\documentclass[a4paper,12pt,titlepage]{book}

\usepackage{amsfonts}
\usepackage{amsmath}
\usepackage[english]{babel}
\usepackage{amssymb}
\usepackage{amstext}
\usepackage{amsthm}
\usepackage{bbold}
\usepackage{cancel}
\usepackage{comment}
\usepackage{easymat}
\usepackage{enumitem}
\usepackage[T1]{fontenc}
\usepackage{frontespizio}
\usepackage[a4paper]{geometry}
\usepackage{graphicx}
\usepackage{hyperref}
\usepackage[utf8]{inputenc}
\usepackage{physics}
\usepackage{simpler-wick}
\usepackage[active]{srcltx}
\usepackage{tikz}
\usepackage[compat=1.0.0]{tikz-feynman}
\newcommand\beq{\begin{equation}}
\newcommand\eeq{\end{equation}}
\newcommand\bcc{\begin{cancel}}
\newcommand\ecc{\end{cancel}}
\addto\appendix{
}
\renewcommand{\theequation}{\thesection.\arabic{equation}}
\theoremstyle{definition}
\newtheorem*{ex}{Text}
\theoremstyle{definition}
\newtheorem*{soluzione}{Solution}
\makeatletter
\newenvironment{sol}
  {\protected@edef\theparentequation{\theequation}%
   \setcounter{parentequation}{\value{equation}}%
   \setcounter{equation}{0}%
   \begin{soluzione}}
  {\end{soluzione}\setcounter{equation}{\value{parentequation}}}
\makeatother

\begin{document}
\selectlanguage{english}

\thispagestyle{empty}
\linespread{1}

\begin{minipage}[c]{\textwidth}
\vspace{9cm}
\begin{center}
\Large\bfseries INTRODUCTIVE EXERCISES TO \\QUANTUM FIELD THEORY
\end{center}
\begin{center}
Stefano Disca, Roberto Demaria, Lorenzo Formaggio
\end{center}
\vspace{0.5cm}
\begin{center}
(last revision on 24/04/2024)
\end{center}
\end{minipage}
\clearpage\null\thispagestyle{empty}\clearpage

\tableofcontents
\setcounter{page}{1}
\clearpage\null\thispagestyle{empty}\clearpage

\chapter*{Preface}\label{preface}
\addcontentsline{toc}{chapter}{Preface}
The following is a collection of 12 exercises picked from the exam tests of the course  \textit{Elements of Quantum Field Theory}, teached by professor Mauro Moretti in the academic year 2021-22 for the Master's Degree in Physics at the University of Ferrara. \\The goal of this review is to provide a general method to approach the main types of calculations in Quantum Field Theory. The topics that will be treated are: applications of Noether's theorem, checking of Ward identities, computations of cross sections and decay rates. In order to be as clear as possible in the solutions we preferred to be very explicit in the calculations, at the risk of being tedious. \\We must thank Thomas Lovo, for the constant help and suggestions given, and dr. Alessandro Greco, whose lecture notes of the course proved to be essential in order to write these ones. \\Needless to say, this text is not free from errors; if you find errors in calculations or conceptual mistakes in general (or even more elegant solutions to the problems), please contact the authors by the following e-mail addresses: \href{mailto: stefano.disca@edu.unife.it}{stefano.disca@edu.unife.it}, \href{mailto: roberto.demaria@edu.unife.it}{roberto.demaria@edu.unife.it}, \href{mailto: lformagg@cougarnet.uh.edu}{lformagg@cougarnet.uh.edu}.
\clearpage\null\thispagestyle{empty}\clearpage

\chapter{Noether's Theorem and Conserved Charges}

\section{Complex Klein-Gordon field}\label{KGC}
\begin{ex}\label{KGC_cariche}
Given a complex Klein-Gordon field, described by the Lagrangian
\[
\mathcal{L} = \partial_{\mu} \phi \partial^{\mu} \phi^* - m^2 \phi \phi^*
\]
determine the conserved charges of the theory.
\end{ex}
\begin{sol}
First of all, remember that for every lagrangian theory we distinguish symmetries in \textit{space-time} symmetries (resulting from transformations of space-time coordinates) and \textit{internal} symmetries (resulting from field transformations which leave the coordinates untouched). The former type is common to all Lagrangians that are invariant under space-time translations, and its associated Noether currents correspond to the stress-energy tensor
\beq
T^{\mu\nu} = \frac{\partial \mathcal{L}}{ \partial(\partial_\mu \phi) } \partial^{\nu} \phi - \mathcal{L} g^{\mu\nu}
\eeq
where $g^{\mu\nu}$ is the metric tensor; the following ones are the conserved charges
\beq
P^{\nu} = \int{ d^3x T^{0\nu} }
\eeq
This Lagrangian has also a \textit{phase} symmetry
\beq\begin{split}
\phi \to &\phi' = e^{i\varepsilon} \phi \sim (1 + i\varepsilon) \phi \\
\phi^* \to &\phi'^{*} = e^{-i\varepsilon} \phi^{*} \sim (1 - i\varepsilon) \phi^*
\end{split}\eeq
where $\varepsilon$ is an arbitrary small parameter. The total variations of the fields are
\beq\begin{split}
&\varepsilon \overline{\delta} \phi \overset{def}{=} \phi' - \phi = i\varepsilon \phi \\
&\varepsilon \overline{\delta} \phi^* \overset{def}{=} \phi'^{*} - \phi^* = -i\varepsilon \phi^*
\end{split}\eeq
so the Noether currents are
\beq
j^{\mu} = \frac{\partial \mathcal{L}}{ \partial(\partial_\mu \phi) } \overline{\delta} \phi + \frac{\partial \mathcal{L}}{ \partial(\partial_\mu \phi^*) } \overline{\delta} \phi^* = i \bigg[ \phi \partial_\mu \phi^* - \phi \partial_\mu \phi \bigg]
\eeq
therefore its associated conserved charge is
\beq
Q = \int{ d^3x j^0 } = i \int{ d^3x \bigg[ \phi \partial_0 \phi^* - \phi^* \partial_0 \phi \bigg] } =: Q_1 + Q_2
\eeq
The \textit{real} Klein-Gordon field can be written as a combination of the ladder operators $a^\dagger$ and $a$; in the case of a \textit{complex} scalar field we must treat $\phi$ and $\phi^*$ as two separate fields, so that we have two distinct types of ladder operators, $a$ and $b$, associated to two different types of particles. Being the two fields simply related by complex conjugation, we can write them as follows
\beq\begin{split}
&\phi(x) = \int{ \frac{d^3p}{(2\pi)^3} \frac{1}{\sqrt{2E_\textbf{p}}} \bigg[ a_\textbf{p} e^{-i E_\textbf{p}t} e^{i\textbf{p}\cdot \textbf{x} }+ b_\textbf{p}^\dagger e^{iE_\textbf{p}t}e^{-i\textbf{p}\cdot\textbf{x}}\bigg] } \\
&\phi^*(x) = \int{ \frac{d^3p}{(2\pi)^3} \frac{1}{\sqrt{2E_\textbf{p}}} \bigg[ a^\dagger_\textbf{p} e^{i E_\textbf{p} t} e^{-i \textbf{p} \cdot \textbf{x} } + b_\textbf{p} e^{-i E_\textbf{p} t} e^{i \textbf{p} \cdot \textbf{x} } \bigg] }
\end{split}\eeq
We reobtain the case of a real Klein-Gordon field by setting $b_\textbf{p} = a_\textbf{p}$. \\Let's now calculate $Q_1 = i \int{ d^3x \phi \partial_0 \phi^*}$
\beq
\partial_0 \phi^*(x) = \int{ \frac{d^3p}{(2\pi)^3} \frac{1}{\sqrt{2E_\textbf{p}}} \bigg[ i E_\textbf{p} a^\dagger_\textbf{p} e^{i E_\textbf{p} t} e^{-i \textbf{p} \cdot \textbf{x} } - i E_\textbf{p} b_\textbf{p} e^{-i E_\textbf{p} t} e^{i \textbf{p} \cdot \textbf{x} } \bigg] }
\eeq
\beq\begin{split}
\phi(x) \partial_0 \phi^*(x) = \int &\frac{d^3p}{(2\pi)^3} \frac{1}{\sqrt{2E_\textbf{p}}} \frac{d^3q}{(2\pi)^3} \frac{1}{\sqrt{2E_\textbf{q}}} (i E_\textbf{p}) \bigg[ a_\textbf{p} e^{-i E_\textbf{p} t} e^{i \textbf{p} \cdot \textbf{x} } + b^\dagger_\textbf{p} e^{i E_\textbf{p} t} e^{-i \textbf{p} \cdot \textbf{x} } \bigg] \\
&\bigg[ i E_\textbf{p} a^\dagger_\textbf{p} e^{i E_\textbf{p} t} e^{-i \textbf{p} \cdot \textbf{x} } - i E_\textbf{p} b_\textbf{p} e^{-i E_\textbf{p} t} e^{i \textbf{p} \cdot \textbf{x} } \bigg]
\end{split}\eeq
so that
\begin{subequations}
\beq\begin{split}
Q_1 = i \frac{1}{(2\pi)^3} \frac{1}{2\sqrt{E_\textbf{q}E_\textbf{p}}} (iE_\textbf{p}) \int &d^3x d^3p d^3q \bigg[ a_\textbf{q} a^\dagger_\textbf{p} e^{ i(E_\textbf{q} - E_\textbf{p})t } e^{ i(\textbf{q}-\textbf{p}) \cdot \textbf{x} } + \\
&- a_\textbf{q} b_\textbf{p} e^{ -i(E_\textbf{q} + E_\textbf{p})t } e^{ i(\textbf{p}+\textbf{q}) \cdot \textbf{x} } + \\
&+b^\dagger_\textbf{q} a^\dagger_\textbf{p} e^{ i(E_\textbf{q} + E_\textbf{p})t } e^{ -i(\textbf{p}+\textbf{q}) \cdot \textbf{x}} + \\
&-b^\dagger_\textbf{q} b_\textbf{p} e^{ i(E_\textbf{q} - E_\textbf{p})t } e^{ i(\textbf{p}-\textbf{q}) \cdot \textbf{x} } \bigg]
\end{split}\eeq
\beq\begin{split}
Q_1 = i \frac{1}{(2\pi)^3} \frac{1}{2\sqrt{E_\textbf{q}E_\textbf{p}}} (iE_\textbf{p}) \int &d^3p d^3q \bigg[ a_\textbf{q} a^\dagger_\textbf{p} e^{ i(E_\textbf{p} - E_\textbf{q})t } (2\pi)^3 \delta(\textbf{q}-\textbf{p}) + \\
&- a_\textbf{q} b_\textbf{p} e^{ -i(E_\textbf{q} + E_\textbf{p})t } (2\pi)^3 \delta(\textbf{p}+\textbf{q}) + \\
&+b^\dagger_\textbf{q} a^\dagger_\textbf{p} e^{ i(E_\textbf{q} + E_\textbf{p})t } (2\pi)^3 \delta(\textbf{p}+\textbf{q}) + \\
&-b^\dagger_\textbf{q} b_\textbf{p} e^{ i(E_\textbf{q} - E_\textbf{p})t } (2\pi)^3 \delta(\textbf{p}-\textbf{q}) \bigg]
\end{split}\eeq
\beq
Q_1 = i \frac{1}{(2\pi)^3} \frac{1}{2\sqrt{E_\textbf{q}E_\textbf{p}}} (iE_\textbf{p}) (2\pi)^3 \int{ d^3p \bigg[ a_\textbf{p} a^\dagger_\textbf{p} - a_{-\textbf{p}} b_\textbf{p} e^{ -2i E_\textbf{p} t } + b^\dagger_{-\textbf{p}} a^\dagger_\textbf{p} e^{ 2i E_\textbf{p} t } - b^\dagger_\textbf{p} b_\textbf{p} \bigg] }
\eeq
\end{subequations}
By naming $\varphi = 2 E_\textbf{p} t$, we have
\beq
Q_1 = - \frac{1}{2} \frac{1}{(2\pi)^3} \int{ d^3p \bigg[ a_\textbf{p} a^\dagger_\textbf{p} - a_{-\textbf{p}} b_\textbf{p} e^{ -i \varphi } + b^\dagger_{-\textbf{p}} a^\dagger_\textbf{p} e^{ i \varphi } - b^\dagger_\textbf{p} b_\textbf{p} \bigg] }
\eeq
An analogous calculation holds for $Q_2 = -i \int{ d^3x \phi^* \partial_0 \phi }$
\beq
\phi^* \partial_0 \phi = (\phi \partial_0 \phi^*)^* = \dots
\eeq
\beq
Q_2 = - \frac{1}{2} \frac{1}{(2\pi)^3} \int{ d^3p \bigg[ a^\dagger_\textbf{p} a_\textbf{p} - a^\dagger_{-\textbf{p}} b^\dagger_\textbf{p} e^{ i \varphi } + b_{-\textbf{p}} a_\textbf{p} e^{ -i \varphi } - b_\textbf{p} b^\dagger_\textbf{p} \bigg] }
\eeq
Finally
\beq\begin{split}\label{Q_KG}
Q = -\frac{1}{2} \frac{1}{(2\pi)^3} \int d^3p &\bigg[ a_\textbf{p} a^\dagger_\textbf{p} + a^\dagger_\textbf{p} a_\textbf{p} - b^\dagger_\textbf{p} b_\textbf{p} - b_\textbf{p} b^\dagger_\textbf{p} + e^{ -i \varphi } \bigg( -a_{-\textbf{p}} b_\textbf{p} + b_{-\textbf{p}} a_\textbf{p} \bigg) + \\
&+e^{i \varphi} \bigg( b^\dagger_{-\textbf{p}} a^\dagger_\textbf{p} - a^\dagger_{-\textbf{p}} b^\dagger_\textbf{p} \bigg) \bigg]
\end{split}\eeq
The first two terms in \eqref{Q_KG} can be written as
\beq
a_\textbf{p} a^\dagger_\textbf{p} + a^\dagger_\textbf{p} a_\textbf{p} = 2 a^\dagger_\textbf{p} a_\textbf{p} + [ a_\textbf{p}, a^\dagger_\textbf{p} ] = 2 N^a_\textbf{p} + (2\pi)^3 \delta(\textbf{0})
\eeq
where we ignored the infinite c-number given by the Dirac delta. For what concerns the term containing the phase we have instead
\beq
I = \int_{-\infty}^{+\infty}{ d^3p \bigg( -a_{-\textbf{p}} b_\textbf{p} + b_{-\textbf{p}} a_\textbf{p} \bigg) }
\eeq
\beq
\begin{cases}
p' = -p \\
d^3p' = d^3p
\end{cases}
\eeq
\beq\begin{split}
I &= \int_{+\infty}^{-\infty}{ -d^3p' \bigg( -a_{\textbf{p}'} b_{-\textbf{p}'} + b_{\textbf{p}'} a_{-\textbf{p}'} \bigg) } = \\
&= \int_{-\infty}^{+\infty}{ d^3p' \bigg( a_{-\textbf{p}'} b_{\textbf{p}'} - b_{-\textbf{p}'} a_{\textbf{p}'} \bigg) } = -I \Rightarrow I=0
\end{split}\eeq
where we used the commutation rules between ladder operators of different particles. \\The conserved charge associated with the phase symmetry therefore is
\beq
Q = \int{ \frac{d^3p}{(2\pi)^3} \bigg[ N_\textbf{p}^b - N_\textbf{p}^a \bigg] }
\eeq
where the number operators $N_\textbf{p}^a$ e $N_\textbf{p}^b$
tell us the number of particles of type $a$ and $b$. It is worth noting that $Q_1 = - Q_2$, so that the fields $\phi$ and $\phi^*$ describe particles with the same mass but opposite charge; we then identify particles of type $b$ as the antiparticles of type $a$ (and viceversa).
\end{sol}

\newpage
\section{Yukawa Lagrangian}
\begin{ex}\label{Yukawa}
Consider the following Lagrangian
\[
\mathcal{L} = i \overline \psi \gamma^\mu \partial_\mu \psi - m \overline \psi \psi + \frac{1}{2} \partial_\mu \phi \partial^\mu \phi - m^2 \phi^2 + y \overline \psi \psi \phi
\]
where $\psi$ is a Dirac spinor field and $\phi$ is a real Klein-Gordon field.
\begin{itemize}
\item Determine the conserved charges of the theory.
\item For each conserved charge determine its associated operator as a function of the ladder operators and verify their action on single particle states.
\end{itemize}
\end{ex}
\begin{sol}
Besides the stress-energy tensor, the Lagrangian produces another Noether current, associated with phase symmetries of the Dirac spinor fields
\beq\begin{split}
&\psi \to \psi' = e^{i \varepsilon} \psi \sim (1 + i \varepsilon) \psi \\
&\overline \psi \to \overline \psi' = e^{- i \varepsilon} \overline \psi \sim (1 - i \varepsilon) \overline \psi
\end{split}\eeq
\beq
\begin{split}
&\varepsilon \overline \delta \psi = \psi' - \psi = i \varepsilon \\
&\varepsilon \overline \delta \; \overline \psi = \overline \psi' - \overline \psi = - i \varepsilon
\end{split} \Rightarrow
\begin{split}
&\overline \delta \psi = i \psi \\
&\overline \delta \; \overline \psi = i \overline \psi
\end{split}
\eeq
The Noether current associated with this symmetry is 
\beq
j^\mu = \frac{\partial \mathcal{L} }{\partial (\partial_\mu \psi) } \overline \delta \psi + \frac{\partial \mathcal{L} }{\partial (\partial_\mu \overline \psi) } \overline \delta \overline \psi = - \overline \psi \gamma^\mu \psi
\eeq
with Noether charge
\beq
Q = \int{ d^3x j^0 } = -\int{ d^3x \overline \psi \gamma^0 \psi }
\eeq
We write the charge $Q$ as a function of the number operators
\beq\begin{split}
Q = -\frac{1}{ (2\pi)^6 } \frac{1}{2 \sqrt{E_\textbf{p} E_\textbf{q}} } \int& d^3x d^3p d^3q \bigg[ e^{-i \textbf{q} \cdot \textbf{x}} \bigg( \sum_r{ \overline u^r(\textbf{q}) a_\textbf{q}^{r^\dagger} } \bigg) + e^{i \textbf{q} \cdot \textbf{x}} \bigg( \sum_r{ \overline v^r(\textbf{q}) b_\textbf{q}^r } \bigg) \bigg] \\
&\gamma^0 \bigg[ e^{i \textbf{p} \cdot \textbf{x}} \bigg( \sum_s{ u^s(\textbf{p}) a_\textbf{p}^s } \bigg) + e^{-i \textbf{p} \cdot \textbf{x}} \bigg( \sum_s{ v^s(\textbf{p}) b_\textbf{p}^{s^\dagger} } \bigg) \bigg]
\end{split}\eeq
Expanding the products we can make use of the Dirac delta in the form 
\beq
\int{ d^3x e^{i (\textbf{p} - \textbf{q}) \cdot \textbf{x}} } = (2\pi)^3 \delta(\textbf{p} - \textbf{q})
\eeq
so the charge $Q$ becomes
\beq\begin{split}
Q = -\frac{1}{ (2\pi)^6 } \frac{1}{2 E_\textbf{p}} \int& d^3p \bigg( \sum_r{ u^{r^\dagger}(\textbf{p}) a_{\textbf{p}}^{r^\dagger} } \bigg) \bigg( \sum_s{ u^s(\textbf{p}) a_{\textbf{p}}^s} \bigg) + \\
&+ \bigg( \sum_r{ u^{r^\dagger}(-\textbf{p}) a_{-\textbf{p}}^{r^\dagger} } \bigg) \bigg( \sum_s{ v^s(\textbf{p}) b_{\textbf{p}}^{s^\dagger} } \bigg) + \\
&+ \bigg( \sum_r{ v^{r^\dagger}(-\textbf{p}) b_{-\textbf{p}}^r } \bigg) \bigg( \sum_s{ u^s(\textbf{p}) a_{\textbf{p}}^s } \bigg) + \\
&+ \bigg( \sum_r{ v^{r^\dagger}(\textbf{p}) b_{\textbf{p}}^r } \bigg) \bigg( \sum_s{ v^s(\textbf{p}) b_{\textbf{p}}^{s^\dagger} } \bigg)
\end{split}\eeq
where we made explicit the dependence on the 3-momenta (since the time components do not get affected by the action of the Dirac delta) and exploited the following identities $\overline u = u^\dagger \gamma^0$, $\overline v = v^\dagger \gamma^0$. Thanks to \eqref{id_spinori_zero} the second and third term are both equal to zero, while the first and the fourth one become
\beq\begin{split}
&\bigg( \sum_r{ u^{r^\dagger}(\textbf{p}) a_{\textbf{p}}^{r^\dagger} } \bigg) \bigg( \sum_s{ u^s(\textbf{p}) a_{\textbf{p}}^s} \bigg) = 2 E_\textbf{p} a_\textbf{p}^{1^\dagger} a_\textbf{p}^1 + 2 E_\textbf{p} a_\textbf{p}^{2^\dagger} a_\textbf{p}^2 \\
&\bigg( \sum_r{ v^{r^\dagger}(\textbf{p}) b_{\textbf{p}}^r } \bigg) \bigg( \sum_s{ v^s(\textbf{p}) b_{\textbf{p}}^{s^\dagger} } \bigg) = 2 E_\textbf{p} b_\textbf{p}^1 b_\textbf{p}^{1^\dagger} + 2 E_\textbf{p} b_\textbf{p}^2 b_\textbf{p}^{2^\dagger} \\
\end{split}\eeq
so that
\beq\begin{split}
Q &= -\frac{1}{ (2\pi)^3 } \int d^3p \bigg( a_\textbf{p}^{1^\dagger} a_\textbf{p}^1 + a_\textbf{p}^{2^\dagger} a_\textbf{p}^2 \bigg) + \bigg( b_\textbf{p}^1 b_\textbf{p}^{1^\dagger} + b_\textbf{p}^2 b_\textbf{p}^{2^\dagger} \bigg) = \\
&= -\frac{1}{ (2\pi)^3 } \int d^3p \bigg( a_\textbf{p}^{1^\dagger} a_\textbf{p}^1 + a_\textbf{p}^{2^\dagger} a_\textbf{p}^2 \bigg) - \bigg( b_\textbf{p}^{1^\dagger} b_\textbf{p}^1 + b_\textbf{p}^{2^\dagger} b_\textbf{p}^2 \bigg) + \{ b_\textbf{p}^1, b_\textbf{p}^{1^\dagger} \} + \{ b_\textbf{p}^2, b_\textbf{p}^{2^\dagger} \}
\end{split}\eeq
As usual we ignore the term proportional to $\delta(\textbf{0})$ (see exercise \ref{KGC_cariche}) and the Noether charge becomes
\beq
Q = \int \frac{d^3p}{ (2\pi)^3 } \bigg[ N_\textbf{p}^b - N_\textbf{p}^a \bigg]
\eeq
where the number operators contain both spin contributions. As expected, we have that the conserved charge is defined through the difference between the numbers of particles and antiparticles.
\end{sol}

\chapter{Cross Sections and Decay Rates}

\section{Higgs boson decay}
\begin{ex}\label{Higgs_dec}
The Higgs boson is a scalar particle whose interaction with the electromagnetic field is described by the following Lagrangian
\[
\mathcal{L} = \frac{1}{2} \partial_{\mu} \phi \partial^{\mu} \phi - m^2 \phi - \frac{1}{4} F_{\alpha\beta} F^{\alpha\beta} + \lambda \phi F_{\alpha\beta} F^{\alpha\beta}
\]
where $F_{\alpha\beta}$ is the electromagnetic stress tensor, defined as
\[
F_{\alpha\beta} = \partial_\alpha A_\beta - \partial_\beta A_\alpha
\]
Compute the decay rate of the Higgs boson $\phi$.
\end{ex}
\begin{sol}
We start by deriving the Feynman rules of the theory. \\The complete Lagrangian is the sum of the Klein-Gordon Lagragian, the electromagnetic Lagrangian and the interaction one, which is given by
\beq
\mathcal{L}_{INT} = \lambda \phi F_{\alpha\beta} F^{\alpha\beta}
\eeq
Explicitly
\beq\begin{split}
\mathcal{L}_{INT} &= \lambda \phi ( \partial_{\alpha} A_{\beta} - \partial_{\beta} A_{\alpha} ) ( \partial^{\alpha} A^{\beta} - \partial^{\beta} A^{\alpha}) = \\
&= \lambda \phi ( \partial_\alpha A_\beta \partial^\alpha A^\beta - \partial_\alpha A_\beta \partial^\beta A^\alpha - \partial_\beta A_\alpha \partial^\alpha A^\beta + \partial_\beta A_\alpha \partial^\beta A^\alpha ) = \\
&=  \lambda \phi ( 2 \partial_\alpha A_\beta \partial^\alpha A^\beta - 2 \partial_\alpha A_\beta \partial^\beta A^\alpha ) = \\
&= 2 \lambda \phi ( \partial_\alpha A_\beta \partial^\alpha g^{\beta\gamma} A_\gamma - \partial_\alpha A_\beta \partial^\beta g^{\alpha\gamma} A_\gamma )
\end{split}\eeq
Moving to the momentum space ($\partial_\mu \to i k_\mu$) we obtain
\beq\begin{split}
\mathcal{L}_{INT} &\to 2 \lambda \phi ( i k_{1\alpha} A_\beta i k_2^\alpha g^{\beta\gamma} A_\gamma - i k_{1\alpha} A_\beta i k_2^\beta g^{\alpha\gamma} A_\gamma ) = \\
&= -2 \lambda \phi ( k_{1\alpha} k_2^\alpha g^{\beta\gamma} - k_1^\gamma k_2^\beta ) A_\beta(k_1) A_\gamma(k_2)
\end{split}\eeq
therefore the Feynman rules of the theory are
\beq
\begin{tikzpicture}
\begin{feynman}
\node at (0.5, 0) {$=$};

	\vertex (a) {\(\phi\)};
	\vertex[left=of a] (b);
	\vertex[above left=of b] (c) {\(A_\beta\)};
	\vertex[below left=of b] (d) {\(A_\gamma\)};

	\diagram {
	(a) -- [dashed] (b),
	(c) -- [boson, momentum={ [arrow style=lightgray] $k_1$ } ] (b),
	(d) -- [boson, momentum={ [arrow style=lightgray] $k_2$ } ] (b)
	};
\end{feynman}
\end{tikzpicture}
\begin{tikzpicture}
\begin{feynman}
\node at (2.5, 0) {$= -4 i \lambda ( k_{1\alpha} k_2^\alpha g^{\beta\gamma} - k_1^\gamma k_2^\beta )$};

	\vertex (a) {\(\phi\)};
	\vertex[left=of a] (b);
	\vertex[above left=of b] (c) {\(A_\beta\)};
	\vertex[below left=of b] (d) {\(A_\gamma\)};

	\diagram {
	(a) -- [dashed] (b),
	(c) -- [boson, reversed momentum={ [arrow style=lightgray] $k_1$ } ] (b),
	(d) -- [boson, reversed momentum={ [arrow style=lightgray] $k_2$ } ] (b)
	};
\end{feynman}
\end{tikzpicture}
\eeq
\beq
\begin{tikzpicture}
\begin{feynman}
\node at (0.5, 0) {$=$};

	\vertex (a) {\(\phi\)};
	\vertex[left=of a] (b);
	\vertex[above left=of b] (c) {\(A_\beta\)};
	\vertex[below left=of b] (d) {\(A_\gamma\)};

	\diagram {
	(a) -- [dashed] (b),
	(c) -- [boson, momentum={ [arrow style=lightgray] $k_1$ } ] (b),
	(d) -- [boson, reversed momentum={ [arrow style=lightgray] $k_2$ } ] (b)
	};
\end{feynman}
\end{tikzpicture}
\begin{tikzpicture}
\begin{feynman}
\node at (2.5, 0) {$= 4 i \lambda ( k_{1\alpha} k_2^\alpha g^{\beta\gamma} - k_1^\gamma k_2^\beta )$};

	\vertex (a) {\(\phi\)};
	\vertex[left=of a] (b);
	\vertex[above left=of b] (c) {\(A_\beta\)};
	\vertex[below left=of b] (d) {\(A_\gamma\)};

	\diagram {
	(a) -- [dashed] (b),
	(c) -- [boson, reversed momentum={ [arrow style=lightgray] $k_1$ } ] (b),
	(d) -- [boson, momentum={ [arrow style=lightgray] $k_2$ } ] (b)
	};
\end{feynman}
\end{tikzpicture}
\eeq
The factor $i$ comes from the first order expansion, while the further factor of $2$ is added in order to consider the indistinguishability of the photons. \\Our goal is to compute the decay rate of the $\phi$ boson: $\phi \to \gamma \gamma$; at first order the process is described by the following diagram
\beq
\begin{tikzpicture}
\begin{feynman}
	\vertex (a){$\phi$};
	\vertex[right=of a] (b);
	\vertex[above right=of b] (c){$\gamma$};
	\vertex[below right=of b] (d){$\gamma$};

	\diagram {
	(a) -- [dashed, momentum={ [arrow style=lightgray] $p$ } ] (b),
	(b) -- [boson, momentum={ [arrow style=lightgray] $k_1$ } ] (c),
	(b) -- [boson, momentum={ [arrow style=lightgray] $k_2$ } ] (d)
	};
\end{feynman}
\end{tikzpicture}
\eeq
The scattering amplitude is given by
\beq\begin{split}
\mathcal{A} =& 1 \cdot (-4 i \lambda) \bigg[ k_1 k_2 g^{\beta\gamma} - k_1^\gamma k_2^\beta \bigg] \bigg( \varepsilon_{\beta}^{\lambda_1}(k_1) \bigg)^* \bigg( \varepsilon_{\gamma}^{\lambda_2}(k_2) \bigg)^* = \\
=& -4 i \lambda \bigg[ k_1 k_2 g^{\beta\gamma} - k_1^\gamma k_2^\beta \bigg] \bigg( \varepsilon_{\beta}^{\lambda_1}(k_1) \bigg)^* \bigg( \varepsilon_{\gamma}^{\lambda_2}(k_2) \bigg)^*
\end{split}\eeq
In order to compute the decay rate, we need the modulus square of the scattering matrix element, defined as
\beq
|\mathcal{M}|^2 = \sum_{\lambda_1, \lambda_2} \mathcal{A} \mathcal{A}^*
\eeq
So we have
\beq\begin{split}
|\mathcal{M}|^2 &= 16 \lambda^2 \sum_{\lambda_1, \lambda_2} \bigg[k_1 k_2 g^{\beta\gamma}-k_1^\gamma k_2^\beta \bigg] \bigg[k_1 k_2 g^{\mu\nu} - k_1^\nu k_2^\mu \bigg] (\varepsilon_\beta^{\lambda_1})^* \varepsilon_\mu^{\lambda_1} (\varepsilon_\gamma^{\lambda_2})^* \varepsilon_\nu^{\lambda_2}
\end{split}\eeq
Summing over the photon polarization we can use the following relations
\beq\begin{split}
&\sum_{\lambda_1}(\varepsilon_\beta^{\lambda_1})^* \varepsilon_\mu^{\lambda_1} = -g_{\beta\mu} \\
&\sum_{\lambda_2}(\varepsilon_\gamma^{\lambda_2})^* \varepsilon_\nu^{\lambda_2} = -g_{\gamma\nu}
\end{split}\eeq
so that
\beq\begin{split}
|\mathcal{M}|^2 &= 16 \lambda^2 \bigg[k_1 k_2 g^{\beta\gamma} - k_1^\gamma k_2^\beta \bigg] \bigg[k_1 k_2 g^{\mu\nu} - k_1^\nu k_2^\mu \bigg](-g_{\beta\mu})(-g_{\gamma\nu}) = \\
&= 16\lambda^2 \bigg[k_1 k_2 g_{\mu\nu} - k_{1\nu} k_{2\mu} \bigg] \bigg[k_1 k_2 g^{\mu\nu} - k_1^\nu k_2^\mu \bigg] = \\
&=16 \lambda^2 \bigg[4 (k_1 k_2)^2 - (k_1 k_2)^2 - (k_1 k_2)^2 + \underbrace{k_1^2}_{=0} \underbrace{k_2^2}_{=0} \bigg] = \\
&= 32\lambda^2 (k_1 k_2)^2
\end{split}\eeq
The 2-body phase space element, after integrating on $d\varphi$, is
\beq
d \Phi_2 = \frac{1}{8 \pi} \frac{ | \textbf{k}_1 | }{ E_1 + E_2 } d \cos{\theta}
\eeq
where $E_1$ and $E_2$ are the photon final energies. \\The physics of the process must be independent on the reference frame chosen, so we can put ourselves into the most simple one. In this case it is useful to use the Higgs boson comoving frame, where the momenta satisfy
\beq\begin{cases}
k_1 = (E_1, 0, |\textbf{k}| \cos{\theta}, |\textbf{k}| \sin{\theta}) \\
k_2 = (E_2, 0, -|\textbf{k}| \cos{\theta}, -|\textbf{k}| \sin{\theta})
\end{cases}\eeq
and the following relations hold
\beq\begin{split}
&E_1 + E_2 = m \Rightarrow E_1 = E_2 = |\textbf{k}| = \frac{m}{2} \\
&k_1k_2 = E_1E_2 + |\textbf{k}|^2 = \frac{m^2}{2}
\end{split}\eeq
We therefore obtain the differential decay rate
\beq\begin{split}
d \Gamma =& \frac{1}{2m} |\mathcal{M}|^2 d \Phi_2 = \bigg(\frac{1}{2m}\bigg) \bigg(32 \lambda^2 \frac{m^4}{4} \bigg) \bigg(\frac{1}{8 \pi} \frac{ | \textbf{k}_1 | }{ E_1 + E_2 } d \cos{\theta} \bigg) = \\
=& \frac{\lambda^2}{4\pi}m^3 d \cos{\theta}
\end{split}\eeq
and finally
\beq
\Gamma = \frac{\lambda^2}{4\pi} m^3 \int_{-1}^{1} d \cos{\theta} = \frac{\lambda^2}{2\pi} m^3
\eeq
\end{sol}

\newpage
\section{Scalar and standard QED: $\psi \overline \psi \to \phi \phi^*$}
\begin{ex}\label{QED_scalare}
Given the following Lagrangian
\[
\mathcal{L} = i \overline \psi \gamma_\mu D^\mu \psi - m \overline \psi \psi + (D_{\mu}\phi)^* (D^{\mu}\phi) - m_\phi^2 \phi \phi^* - \frac{1}{4}F_{\alpha\beta}F^{\alpha\beta} - \frac{1}{2\xi} (\partial_\alpha A^\alpha)^2
\]
Compute the differential cross section for the process
\[
\psi \overline \psi \to \phi \phi^*
\]
\end{ex}
\begin{sol}
First of all, we derive the Feynman rules in the momentum space, taking into account that the gauge-fixing term $\frac{1}{2\xi} (\partial_\alpha A^\alpha)^2$ does not modify the interactions
\beq\begin{split}
\mathcal{L}_{INT} &= ie [ A_\mu \phi^* \partial^\mu \phi - A_\mu \phi \partial^\mu \phi^* ] + e^2 A_\mu A^\mu \phi \phi^* = \\
&= ie [ A_\mu \phi^* \partial^\mu \phi - A^\mu \phi \partial_\mu \phi^* ] + e^2 g_{\mu\nu} A_\mu A_\nu \phi \phi^* \to \\
&\to -e [ p_1^\mu + p_2^\mu ] A_\mu \phi \phi^* + e^2 g^{\mu\nu} A_\mu A_\nu \phi \phi^*
\end{split}\eeq
\beq
\begin{tikzpicture}
\begin{feynman}
\node at (2.5, 0) {$= -i e[ p_1^\mu - p_2^\mu ]$};

	\vertex (a) {\(A_\mu\)};
	\vertex[left=of a] (b);
	\vertex[above left=of b] (c) {\(\phi\)};
	\vertex[below left=of b] (d) {\(\phi^*\)};

	\diagram {
	(a) -- [boson] (b),
	(c) -- [dashed, momentum=\({p_1}\)] (b),
	(d) -- [dashed, momentum=\({p_2}\)] (b)
	};
\end{feynman}
\end{tikzpicture}
\eeq
\beq
\begin{tikzpicture}
\begin{feynman}
\node at (2.5, 0) {$= -i e[ -p_1^\mu + p_2^\mu ]$};

	\vertex (a) {\(A_\mu\)};
	\vertex[left=of a] (b);
	\vertex[above left=of b] (c) {\(\phi\)};
	\vertex[below left=of b] (d) {\(\phi^*\)};

	\diagram {
	(a) -- [boson] (b),
	(b) -- [dashed, momentum=\({p_1}\)] (c),
	(b) -- [dashed, momentum=\({p_2}\)] (d)
	};
\end{feynman}
\end{tikzpicture}
\eeq
\beq
\begin{tikzpicture}
\begin{feynman}
\node at (2.5, 0) {$= -i e[ -p_1^\mu - p_2^\mu ]$};

	\vertex (a) {\(A_\mu\)};
	\vertex[left=of a] (b);
	\vertex[above left=of b] (c) {\(\phi\)};
	\vertex[below left=of b] (d) {\(\phi^*\)};

	\diagram {
	(a) -- [boson] (b),
	(b) -- [dashed, momentum'=\({p_1}\)] (c),
	(d) -- [dashed, momentum=\({p_2}\)] (b)
	};
\end{feynman}
\end{tikzpicture}
\eeq
\beq
\begin{tikzpicture}
\begin{feynman}
\node at (2.5, 0) {$= -i e[ p_1^\mu + p_2^\mu ]$};

	\vertex (a) {\(A_\mu\)};
	\vertex[left=of a] (b);
	\vertex[above left=of b] (c) {\(\phi\)};
	\vertex[below left=of b] (d) {\(\phi^*\)};

	\diagram {
	(a) -- [boson] (b),
	(c) -- [dashed, momentum=\({p_1}\)] (b),
	(b) -- [dashed, momentum=\({p_2}\)] (d)
	};
\end{feynman}
\end{tikzpicture}
\eeq
\beq
\begin{tikzpicture}
\begin{feynman}
\node at (2.5, 0) {$= 2 i e^2 g^{\mu\nu}$};

	\vertex (a);
	\vertex[above right=of a] (b) {\(A_\mu\)};
	\vertex[below right=of a] (c) {\(A_\nu\)};
	\vertex[above left=of a] (d) {\(\phi\)};
	\vertex[below left=of a] (e) {\(\phi^*\)};

	\diagram {
	(a) -- [boson] (b),
	(a) -- [boson] (c),
	(a) -- [dashed] (d),
	(a) -- [dashed] (e)
	};
\end{feynman} 
\end{tikzpicture} 
\eeq
where we have added, as usual, a factor $i$ due to the first order expansion of the S-matrix and a factor $2$ in the last diagram to account for the exchange of the two photons; these rules will be used together with the usual rules for QED and the ones for the Klein-Gordon field (the former interaction is usually known as \textit{scalar QED}). \\The differential cross section of a 2-body interactions is defined as
\beq
d \sigma = \mathcal{J} |M|^2 d \Phi_2
\eeq
where
\beq
\begin{cases}
\mathcal{J} = \frac{1}{ 4 \sqrt{ (p_1p_2)^2 - m_1^2 m_2^2 } } \\
|M|^2 = \sum_{s_j, \lambda_j, \dots}{ |A|^2 } \\
d \Phi_2 = \frac{ |\textbf{q}_1| }{E_1+E_2} \frac{d \cos{\theta} d\varphi}{16 \pi^2} \bigg|_{ \textbf{q}_2 = -\textbf{q}_1 }
\end{cases}
\eeq
Identifying the initial and final states as $\ket{p_1, s_1; p_2, s_2}$ and $\ket{q_1; q_2}$, the process $\psi \overline \psi \to \phi \phi^*$ is described, at first order by the following diagram
\beq
\begin{tikzpicture}
\begin{feynman}
	\vertex (a);
	\vertex[above left=of a] (b){$\overline{\psi}$};
	\vertex[below left=of a] (c){$\psi$};
	\vertex[right=of a] (d);
	\vertex[above right=of d] (e){$\phi^*$};
	\vertex[below right=of d] (f){$\phi$};

	\diagram {
	(a) -- [fermion, momentum={ [arrow shorten=0.85, arrow style=lightgray] $p_2$ } ] (b),
	(c) -- [fermion, momentum={ [arrow style=lightgray] $p_1$ } ] (a),
	(a) -- [boson, momentum={ [arrow style=lightgray] $k$ } ] (d),
	(d) -- [anti charged scalar, momentum={ [arrow style=lightgray] $q_1$ } ] (e),
	(d) -- [charged scalar, momentum={ [arrow style=lightgray] $q_2$ } ] (f),
	};
\end{feynman}
\end{tikzpicture}
\eeq
where the four-momentum $k$ is $k = p_1+p_2 = q_1+q_2$. We start calculating the scattering amplitude
\beq\begin{split}
\mathcal{A} &= \frac{i e^2 g_{\mu\nu} }{k^2} ( -q_1^\nu + q_2^\nu ) \overline v^{s_2}(p_2) \gamma^\mu u^{s_1}(p_1) = \\
&= -\frac{i e^2}{k^2} \overline v^{s_2}(p_2) ( q_{1\mu} - q_{2\mu} ) \gamma^\mu u^{s_1}(p_1) = \\
&= -\frac{i e^2}{k^2} \overline v^{s_2}(p_2) ( {\bcc q \ecc}_1 - {\bcc q \ecc}_2 ) u^{s_1}(p_1)
\end{split}\eeq
where $\bcc q \ecc := q_\mu \gamma^\mu$. We therefore compute the invariant matrix element averaging on the initial particles spins
\beq\begin{split}
|\mathcal{M}|^2 &= \sum_{s_1, s_2} |\mathcal{A}|^2 = \frac{e^4}{ (k^2)^2 } \sum_{s_1, s_2} \overline v^{s_2} ({\bcc q \ecc}_1 - {\bcc q \ecc}_2) u^{s_1} ( u^{s_1} )^\dagger ({\bcc q \ecc}_1 - \bcc {\bcc q \ecc}_2 \ecc)^\dagger ( \overline v^{s_2} )^\dagger = \\
&= \frac{e^4}{ (k^2)^2 } \sum_{s_1, s_2} \overline v^{s_2} ({\bcc q \ecc}_1 - {\bcc q \ecc}_2) u^{s_1} \overline u^{s_1} \gamma^0 (\gamma^0 {\bcc q \ecc}_1 \gamma^0 - \gamma^0 {\bcc q \ecc}_2 \gamma^0) \gamma^0 v^{s_2} = \\
&= \frac{e^4}{ (k^2)^2 } \sum_{s_1, s_2} \overline v^{s_2} ({\bcc q \ecc}_1 - {\bcc q \ecc}_2) u^{s_1} \overline u^{s_1} ({\bcc q \ecc}_1 - {\bcc q \ecc}_2) v^{s_2} = \\
\end{split}\eeq
Now we make explicit the tensorial products
\beq\begin{split}
|\mathcal{M}|^2 &= \frac{e^4}{ (k^2)^2 } \sum_{s_1, s_2} \sum_{a, b, c, d} (\overline v^{s_2} )_a ({\bcc q \ecc}_1 - {\bcc q \ecc}_2)_{ab} (u^{s_1} )_b (\overline u^{s_1} )_c ({\bcc q \ecc}_1 - {\bcc q \ecc}_2)_{cd} (v^{s_2} )_d = \\
&= \frac{e^4}{ (k^2)^2 } \sum_{a, b, c, d} ({\bcc q \ecc}_1 - {\bcc q \ecc}_2)_{ab} ({\bcc p \ecc}_1 + m)_{bc} ({\bcc q \ecc}_1 - {\bcc q \ecc}_2)_{cd} ({\bcc p \ecc}_2 - m)_{da} = \\
\end{split}\eeq
This corresponds to the computation of the following trace
\beq\begin{split}
|\mathcal{M}|^2 &= \frac{e^4}{ (k^2)^2 } Tr \bigg[ ({\bcc q \ecc}_1 - {\bcc q \ecc}_2) ({\bcc p \ecc}_1 + m) ({\bcc q \ecc}_1 - {\bcc q \ecc}_2) ({\bcc p \ecc}_2 - m) \bigg]
\end{split}\eeq
Recall that the trace of an odd number of gamma matrices is null; therefore we can simplify the above expression in
\beq
|\mathcal{M}|^2 = \frac{e^4}{ (k^2)^2 } \bigg\{ Tr \bigg[ ({\bcc q \ecc}_1 - {\bcc q \ecc}_2) {\bcc p \ecc}_1 ({\bcc q \ecc}_1 - {\bcc q \ecc}_2) {\bcc p \ecc}_2 \bigg] - m^2 Tr \bigg[ ({\bcc q \ecc}_1 - {\bcc q \ecc}_2) ({\bcc q \ecc}_1 - {\bcc q \ecc}_2) \bigg] \bigg\}
\eeq
The explicit calculation is quite lengthy. For the first trace we have
\beq
Tr \bigg[ ({\bcc q \ecc}_1 - {\bcc q \ecc}_2) {\bcc p \ecc}_1 ({\bcc q \ecc}_1 - {\bcc q \ecc}_2) {\bcc p \ecc}_2 \bigg] = Tr \bigg[ {\bcc q \ecc}_1 {\bcc p \ecc}_1 {\bcc q \ecc}_1 {\bcc p \ecc}_2 - {\bcc q \ecc}_1 {\bcc p \ecc}_1 {\bcc q \ecc}_2 {\bcc p \ecc}_2 - {\bcc q \ecc}_2 {\bcc p \ecc}_1 {\bcc q \ecc}_1 {\bcc p \ecc}_2 + {\bcc q \ecc}_2 {\bcc p \ecc}_1 {\bcc q \ecc}_2 {\bcc p \ecc}_2 \bigg]
\eeq
so that
\beq\begin{split}
Tr \bigg[ ({\bcc q \ecc}_1 - {\bcc q \ecc}_2) {\bcc p \ecc}_1 ({\bcc q \ecc}_1 - {\bcc q \ecc}_2) {\bcc p \ecc}_2 \bigg] = Tr \bigg[ (& q_{1\mu} p_{1\nu} q_{1\rho} p_{2\sigma} - q_{1\mu} p_{1\nu} q_{2\rho} p_{2\sigma} - q_{2\mu} p_{1\nu} q_{1\rho} p_{2\sigma} + \\
&+ q_{2\mu} p_{1\nu} q_{2\rho} p_{2\sigma} ) \gamma^\mu \gamma^\nu \gamma^\rho \gamma^\sigma \bigg]
\end{split}\eeq
For the trace of the four gamma matrices we have
\beq\begin{split}
Tr \bigg[ ({\bcc q \ecc}_1 - {\bcc q \ecc}_2) {\bcc p \ecc}_1 ({\bcc q \ecc}_1 - {\bcc q \ecc}_2) {\bcc p \ecc}_2 \bigg] =& ( q_{1\mu} p_{1\nu} q_{1\rho} p_{2\sigma} - q_{1\mu} p_{1\nu} q_{2\rho} p_{2\sigma} - \\
&- q_{2\mu} p_{1\nu} q_{1\rho} p_{2\sigma} + q_{2\mu} p_{1\nu} q_{2\rho} p_{2\sigma} ) \cdot \\
&\cdot 4 ( g^{\mu\nu} g^{\rho\sigma} - g^{\mu\rho} g^{\nu\sigma} + g^{\mu\sigma} g^{\nu\rho} )
\end{split}\eeq
With some patience we can carry out the various contractions, to obtain
\beq\begin{split}
Tr \bigg[ ({\bcc q \ecc}_1 - {\bcc q \ecc}_2) {\bcc p \ecc}_1 ({\bcc q \ecc}_1 - {\bcc q \ecc}_2) {\bcc p \ecc}_2 \bigg] = 4 \bigg\{& (p_1 p_2) \bigg[ - (q_1)^2 - (q_2)^2 + 2 (q_1 q_2) \bigg] + \\
&+ (q_1 p_2) \bigg[ (p_1 q_1) - (p_1 q_2) \bigg] + \\
&+ (q_2 p_2) \bigg[ (p_1 q_2) - (p_1 q_1) \bigg] + \\
&+ (q_1 p_1) \bigg[ (q_1 p_2) - (q_2 p_2) \bigg] + \\
&+ (q_2 p_1) \bigg[ (q_2 p_2) - (q_1 p_2) \bigg] \bigg\}
\end{split}\eeq
For the second trace instead we have
\beq\begin{split}
Tr \bigg[ ({\bcc q \ecc}_1 - {\bcc q \ecc}_2) ({\bcc q \ecc}_1 - {\bcc q \ecc}_2) \bigg] &= Tr \bigg[ {\bcc q \ecc}_1 {\bcc q \ecc}_1 - {\bcc q \ecc}_1{\bcc q \ecc}_2 - {\bcc q \ecc}_2 {\bcc q \ecc}_1 + {\bcc q \ecc}_2 {\bcc q \ecc}_2 \bigg] = \\
&= Tr \bigg[ (q_{1\mu} q_{1\nu} + q_{2\mu} q_{2\nu} ) \gamma^\mu \gamma^\nu \bigg] - Tr \bigg[ \{ {\bcc q \ecc}_1, {\bcc q \ecc}_2 \} \bigg] = \\
&= (q_{1\mu} q_{1\nu} + q_{2\mu} q_{2\nu} ) \cdot 4 g^{\mu\nu} - q_{1\mu} q_{2\nu} \cdot 2 g^{\mu\nu} \cdot 4 = \\
&= 4 \bigg[ (q_1)^2 + (q_2)^2 - 2 (q_1 q_2) \bigg]
\end{split}\eeq
Putting all together and making explicit $k^2 = (p_1)^2 + (p_2)^2 + 2 (p_1 p_2) = 2 m^2 + 2 (p_1 p_2)$ and $(q_1)^2 = (q_2)^2 = m_\phi^2$, the final expression for $|\mathcal{M}|^2$ is
\beq\begin{split}
|\mathcal{M}|^2 = \frac{e^4}{ [ m^2 + (p_1 p_2) ]^2 } \bigg\{& \bigg[ (p_1 p_2) - m^2 \bigg] \bigg[ - 2 m_\phi^2 + 2 (q_1 q_2) \bigg] + \\
&+ 2 \bigg[ (p_1 q_1) - (p_1 q_2) \bigg] \bigg[ (q_1 p_2) - (q_2 p_2) \bigg]
\end{split}\eeq
The flux factor is
\beq
\mathcal{J} = \frac{1}{ 4 \sqrt{ (p_1 p_2)^2 - m_1^2 m_2^2 } } = \frac{1}{ 4 \sqrt{ (p_1 p_2)^2 - m^4 } }
\eeq
and the 2-body phase space element is
\beq
d \Phi_2 = \frac{ | \textbf{q}_1 | }{ E_1 + E_2 } \frac{ d \cos{\theta} d \varphi}{16\pi^2} = \frac{1}{8 \pi} \frac{ | \textbf{q}_1 | }{ E_1 + E_2 } d \cos{\theta}
\eeq
where $E_1$ and $E_2$ stand for the final energies. In the centre of mass frame, setting the two final particles 3-momenta equal and opposite, and enforcing 4-momentum conservation we have
\beq\begin{cases}
p_1 = (E_1, 0, 0, |\textbf{p}|) \\
p_2 = (E_2, 0, 0, -|\textbf{p}|) \\
q_1 = (E_3, 0, |\textbf{q}| \sin{\theta}, |\textbf{q}| \cos{\theta}) \\
q_2 = (E_4, 0, -|\textbf{q}| \sin{\theta}, -|\textbf{q}| \cos{\theta}) \\
\end{cases}\eeq
The energies are not independent from each other, since
\beq\begin{split}
&(p_1)^2 = (p_2)^2 = m^2 \implies E_1^2 - m^2 = E_2^2 - m^2 \implies E_1 = E_2 \\
&(q_1)^2 = (q_2)^2 = m_\phi^2 \implies E_2^2 - m_\phi^2 = E_4^2 - m_\phi^2 \implies E_3 = E_4 \\
&E_1 + E_2 = E_3 + E_4 \implies 2 E_1 = 2 E_3 \implies E_1 = E_3
\end{split}\eeq
It follows that $E_1 = E_2 = E_3 = E_4 = E$, so
\beq\begin{cases}
p_1 = (E, 0, 0, |\textbf{p}|) \\
p_2 = (E, 0, 0, -|\textbf{p}|) \\
q_1 = (E, 0, |\textbf{q}| \sin{\theta}, |\textbf{q}| \cos{\theta}) \\
q_2 = (E, 0, -|\textbf{q}| \sin{\theta}, -|\textbf{q}| \cos{\theta}) \\
\end{cases}\eeq
\beq\begin{split}
&p_1 p_2 = 2 E^2 - m^2 \\
&q_1 q_2 = 2 E^2 - m_\phi^2 \\
&p_1 q_1 = q_2 p_2 = E^2 - \sqrt{ (E^2 - m^2) (E^2 - m_\phi^2) } \cos{\theta} \\
&p_1 q_2 = q_1 p_2 = E^2 + \sqrt{ (E^2 - m^2) (E^2 - m_\phi^2) } \cos{\theta}
\end{split}\eeq
and the differential cross-section of the process, after some simplifications becomes
\beq
\frac{d \sigma}{d \cos{\theta} } = \frac{e^4}{2^7 \pi} \frac{ \sqrt{ (E^2 - m_\phi^2)^3 (E^2 - m^2) } }{E^6} (1 - 2 \cos^2{\theta} )
\eeq
\end{sol}

\newpage
\section{Scalar-boson interaction: $\phi \phi \to \gamma \gamma$}
\begin{ex}\label{Higgs_sez1}
Given the Lagrangian
\[
\mathcal{L} = \frac{1}{2} \partial_{\mu} \phi \partial^{\mu} \phi - m^2 \phi - \frac{1}{4} F_{\alpha\beta} F^{\alpha\beta} -\frac{1}{2\xi} (\partial^\alpha A_\alpha)^2 + \lambda \phi F_{\alpha\beta} F^{\alpha\beta}
\]
Compute the differential cross section for the process
\[
\phi \phi \to \gamma \gamma
\]
\end{ex}
\begin{sol}
Since the gauge-fixing term does not affect the scattering amplitude, we can use the Feynman rules already derived in Exercise \ref{Higgs_dec}. \\At first order we have the following Feynman diagram (\textit{t}-channel)
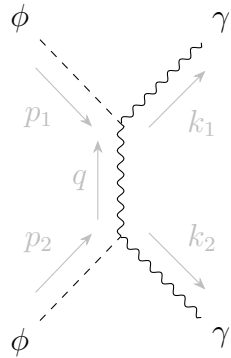
\beq
\begin{tikzpicture}
\begin{feynman}
	\vertex (a);
	\vertex[above left=of a] (b){$\phi$};
	\vertex[above right=of a] (c){$\gamma$};
	\vertex[below=of a] (d);
	\vertex[below left=of d] (e){$\phi$};
	\vertex[below right=of d] (f){$\gamma$};

	\diagram {
	(b) -- [dashed, momentum'={ [arrow style=lightgray] $p_1$ } ] (a),
	(a) -- [boson, momentum'={ [arrow style=lightgray] $k_1$ } ] (c),
	(d) -- [boson, momentum={ [arrow style=lightgray] $q$ } ] (a),
	(e) -- [dashed, momentum={ [arrow style=lightgray] $p_2$ } ] (d),
	(d) -- [boson, momentum={ [arrow style=lightgray] $k_2$ } ] (f),
	};
\end{feynman}
\end{tikzpicture}
\eeq
where  $q = k_1 - p_1 = p_2 - k_2$, and a second one with the final state photons exchanged (\textit{u}-channel)
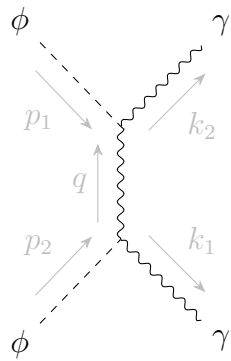
\beq
\begin{tikzpicture}
\begin{feynman}
	\vertex (a);
	\vertex[above left=of a] (b){$\phi$};
	\vertex[above right=of a] (c){$\gamma$};
	\vertex[below=of a] (d);
	\vertex[below left=of d] (e){$\phi$};
	\vertex[below right=of d] (f){$\gamma$};

	\diagram {
	(b) -- [dashed, momentum'={ [arrow style=lightgray] $p_1$ } ] (a),
	(a) -- [boson, momentum'={ [arrow style=lightgray] $k_2$ } ] (c),
	(d) -- [boson, momentum={ [arrow style=lightgray] $q$ } ] (a),
	(e) -- [dashed, momentum={ [arrow style=lightgray] $p_2$ } ] (d),
	(d) -- [boson, momentum={ [arrow style=lightgray] $k_1$ } ] (f),
	};
\end{feynman}
\end{tikzpicture}
\eeq
where $q = k_2 - p_1 = p_2 - k_1$. \\Setting $\ket{p_1; p_2} \to \ket{k_1, \lambda_1; k_2, \lambda_2}$, the scattering amplitude of the first diagram is
\beq\begin{split}
\mathcal{A} =& \bigg( \varepsilon_\beta^1 \bigg)^* \cdot 4 i \lambda \bigg[ (k_1 q) g^{\mu\beta} - k_1^\beta q^\mu \bigg] \cdot \bigg( \frac{-i g_{\mu\nu}}{q^2} \bigg) \cdot \\
&\cdot \bigg( \varepsilon_\gamma^2 \bigg)^* \cdot (-4 i \lambda) \bigg[ (q k_2) g^{\gamma\nu} - q^\gamma k_2^\nu \bigg] = \\
=& \frac{-16 i \lambda^2}{q^2} \bigg( \varepsilon_\beta^1 \bigg)^* \bigg( \varepsilon_\gamma^2 \bigg)^* g_{\mu\nu} \bigg[ (k_1 q) g^{\mu\beta} - k_1^\beta q^\mu ] [ (q k_2) g^{\gamma\nu} - q^\gamma k_2^\nu \bigg] = \\
=& \frac{-16 i \lambda^2}{q^2} \bigg( \varepsilon_\beta^1 \bigg)^* \bigg( \varepsilon_\gamma^2 \bigg)^* g_{\mu\nu} \bigg[ (k_1 q) (k_2 q) g^{\mu\beta} g^{\gamma\nu} - (k_1 q) g^{\mu\beta} q^\gamma k_2^\nu - \\
&\quad\quad\quad\quad\quad\quad\quad\quad\quad\quad\quad - (q k_2) g^{\gamma\nu} k_1^\beta q^\mu + k_1^\beta q^\mu q^\gamma k_2^\nu \bigg] = \\
=& \frac{-16 i \lambda^2}{q^2} \bigg( \varepsilon_\beta^1 \bigg)^* \bigg( \varepsilon_\gamma^2 \bigg)^* \bigg[ (k_1 q) (k_2 q) g^{\beta\gamma} - (k_1 q) q^\gamma k_2^\beta - (q k_2) k_1^\beta q^\gamma + (k_2 q) k_1^\beta q^\gamma \bigg]
\end{split}\eeq
where $\varepsilon_\mu^{\lambda_j}(k_j) =: \varepsilon_\mu^j$. \\We then use \eqref{orto_polarizzazioni} to simplify the contractions of the polarization vectors, so that the amplitude reduces to 
\beq\begin{split}
&\mathcal{A} = \frac{-16 i \lambda^2}{q^2} \bigg[ (k_2 q) g^{\beta\gamma} - q^\gamma k_2^\beta \bigg] (k_1 q) \bigg( \varepsilon_\beta^1 \bigg)^* \bigg( \varepsilon_\gamma^2 \bigg)^* \\
&\mathcal{A}^* = \frac{16 i \lambda^2}{q^2} \bigg[ (k_2 q) g^{\beta\gamma} - q^\gamma k_2^\beta \bigg] (k_1 q) \varepsilon_\beta^1 \varepsilon_\gamma^2
\end{split}\eeq
where $q = k_1 - p_1 = p_2 - k_2$. \\Setting $\tilde q = k_2 - p_1 = p_2 - k_1$, it is easy to convince ourselves that the second amplitude $\tilde A$ can be obtained from the first one with the following substitutions
\beq\begin{cases}\label{da_prima_a_seconda}
k_1 \leftrightarrow k_2 \\
\lambda_1 \leftrightarrow \lambda_2 \\
q \to \tilde q
\end{cases}\eeq
so that
\beq\begin{split}
&\mathcal{\tilde A} = \frac{-16 i \lambda^2}{\tilde q^2} \bigg[ (k_1 \tilde q) g^{\beta\gamma} - \tilde q^\gamma k_1^\beta \bigg] (k_2 \tilde q) \bigg( \varepsilon_\beta^2 \bigg)^* \bigg( \varepsilon_\gamma^1 \bigg)^* \\
&\mathcal{\tilde A}^* = \frac{16 i \lambda^2}{\tilde q^2} \bigg[ (k_1 \tilde q) g^{\beta\gamma} - \tilde q^\gamma k_1^\beta \bigg] (k_2 \tilde q) \varepsilon_\beta^2 \varepsilon_\gamma^1 \\
\end{split}\eeq
We can then sum the square modulus of $\mathcal{A}$ over the polarization indices
\beq
\sum_{\lambda_1, \lambda_2} |\mathcal{A}|^2 = \frac{2^8 \lambda^4}{(q^2)^2} (k_1 q)^2 \bigg[ (k_2 q) g^{\beta\gamma} - q^\gamma k_2^\beta \bigg] \bigg[ (k_2 q) g^{\mu\nu} - q^\nu k_2^\mu \bigg] \sum_{\lambda_1, \lambda_2} \bigg( \varepsilon_\beta^1 \bigg)^* \bigg( \varepsilon_\gamma^2 \bigg)^* \varepsilon_\mu^1 \varepsilon_\nu^2
\eeq
Using \eqref{id_polarizzazioni} we get
\beq\begin{split}
\sum_{\lambda_1, \lambda_2} |\mathcal{A}|^2 =& \frac{2^8 \lambda^4}{(q^2)^2} (k_1 q)^2 \bigg[ (k_2 q) g^{\beta\gamma} - q^\gamma k_2^\beta \bigg] \bigg[ (k_2 q) g^{\mu\nu} - q^\nu k_2^\mu \bigg] g_{\mu\beta} g_{\nu\gamma} = \\
=& \frac{2^8 \lambda^4}{(q^2)^2} (k_1 q)^2 \bigg[ (k_2 q) g_{\mu\nu} - q_\nu k_{2\mu} \bigg] \bigg[ (k_2 q) g^{\mu\nu} - q^\nu k_2^\mu \bigg] = \\
=& \frac{2^8 \lambda^4}{(q^2)^2} (k_1 q)^2 \bigg[ 2 (k_2 q)^2 + q^2 k_2^2 \bigg]
\end{split}\eeq
Therefore, since $k_2^2 = 0$, we have
\beq\begin{split}
\sum_{\lambda_1, \lambda_2} |\mathcal{A}|^2 =& \frac{2^9 \lambda^4}{(q^2)^2} (k_1 q)^2 (k_2 q)^2 = \\
=& \frac{2^9 \lambda^4}{(q^2)^2} (k_1 p_1)^2 (k_2 p_2)^2
\end{split}\eeq
With the substitutions \eqref{da_prima_a_seconda} we immediately obtain
\beq
\sum_{\lambda_1, \lambda_2} |\mathcal{\tilde A}|^2 = \frac{2^9 \lambda^4}{(\tilde q^2)^2} (k_2 \tilde q)^2 (k_1 \tilde q)^2 = \frac{2^9 \lambda^4}{(\tilde q^2)^2} (k_2 p_1)^2 (k_1 p_2)^2
\eeq
The interference terms are obtained with analogous calculations  
\beq\begin{split}
\sum_{\lambda_1, \lambda_2} \mathcal{A} \tilde{\mathcal{A}}^* =& \sum_{\lambda_1, \lambda_2} \mathcal{A}^* \tilde{\mathcal{A}} = \frac{2^8 \lambda^4}{q^2 \tilde q^2} (k_1 q) (k_2 \tilde q) \bigg[ 2 (k_1 \tilde q) (k_2 q) + (q k_1) (\tilde q k_2) \bigg] = \\
=& \frac{2^8 \lambda^4}{q^2 \tilde q^2} (k_1 p_1) (k_2 p_1) \bigg[ 2 (k_1 p_2) (k_2 p_2) + (k_1 p_1) (k_2 p_1) \bigg]
\end{split}\eeq
The invariant matrix element becomes
\beq\begin{split}
\frac{ |\mathcal{M}|^2 }{2^9 \lambda^4} =& \frac{ (k_1 p_1)^2 (k_2 p_2)^2 }{ (m^2 - 2k_1p_1 )^2 } + \frac{ (k_2 p_1)^2 (k_1 p_2)^2 }{ (m^2 - 2k_1p_2 )^2 } + \\
+& \frac{ (k_1 p_1) (k_2 p_1) }{ (m^2 - 2 k_1 p_1) (m^2 - 2k_1p_2 ) } \bigg[ 2 (k_1 p_2) (k_2 p_2) + (k_1 p_1 )(k_2 p_1) \bigg]
\end{split}\eeq
where we substituted  
\beq\begin{split}
q^2 =& k_1^2 + p_1^2 - 2 k_1 p_1 = m^2 - 2 k_1 p_1 = \\
=& p_2^2 + k_2^2 - 2 p_2 k_2 = m^2 - 2 k_2 p_2
\end{split}\eeq
and analogous relations for $\tilde q$. \\The flux factor is
\beq
\mathcal{J} = \frac{1}{4 \sqrt{ (p_1p_2)^2 - m^4 } }
\eeq
while the 2-body phase space element is
\beq
d \Phi_2 = \frac{1}{8 \pi} \frac{ | \textbf{k}_1 | }{ E_1 + E_2 } d \cos{\theta}
\eeq
where $E_1$ and $E_2$ stand for the final state energies. The calculations become simpler in the centre of mass reference frame, where the two couples of particles have 3-momenta with equal modulus and opposite direction; therefore, after imposing the conservation of four-momentum, we have 
\beq\begin{cases}
p_1 = (E, 0, 0, |\textbf{p}|) \\
p_2 = (E, 0, 0, -|\textbf{p}|) \\
k_1 = (E, 0, |\textbf{k}| \sin{\theta}, |\textbf{k}| \cos{\theta}) \\
k_2 = (E, 0, -|\textbf{k}| \sin{\theta}, -|\textbf{k}| \cos{\theta}) \\
\end{cases}\eeq
\beq\begin{split}
&|\textbf{k}| = E \\
&|\textbf{p}| = \sqrt{E^2 - m^2} \\
&p_1 p_2 = 2 E^2 - m^2 \\
&k_1 p_1 = k_2 p_2 = E^2 - E \sqrt{E^2 - m^2} \cos{\theta} \\
&k_1 p_2 = k_2 p_1 = E^2 + E \sqrt{E^2 - m^2} \cos{\theta} \\
\end{split}\eeq
After some algebra we get 
\beq\begin{split}\label{M_2.3}
|\mathcal{M}|^2 = 2^9 \lambda^4 \bigg\{& \frac{ (E^2 - E \sqrt{E^2 - m^2} \cos{\theta})^4 }{[ m^2 - 2(E^2 - E \sqrt{E^2 - m^2} \cos{\theta}) ]^2} + \\
&+ \frac{ (E^2 + E \sqrt{E^2 - m^2} \cos{\theta})^4 }{[ m^2 - 2(E^2 + E \sqrt{E^2 - m^2} \cos{\theta}) ]^2} + \\
&+ \frac{ 3 [E^4 - E^2 (E^2 - m^2) \cos^2{\theta}]^2 }{ [m^2 - 2(E^2 - E \sqrt{E^2 - m^2} \cos{\theta}) ] [m^2 - 2(E^2 + E \sqrt{E^2 - m^2} \cos{\theta}) ] } \bigg\}
\end{split}\eeq
Finally the differential cross section for this process is
\beq
\frac{d \sigma}{d \cos{\theta} } = \frac{1}{ 128 \pi \sqrt{ E^4 - E^2 m^2 } } |\mathcal{M}|^2
\eeq
where $|\mathcal{M}|^2$ is given by \eqref{M_2.3}. \\In the ultra-relativistic limit, where we can neglect the mass of the particles, the cross section simplifies considerably 
\beq
\frac{d \sigma}{d \cos{\theta} } \sim \frac{\lambda^4}{4 \pi} E^2 ( 5 - \cos^2{\theta} )\eeq
\end{sol}

\newpage
\section{Scalar-fermion-boson interaction: $e^+ e^- \to \phi \gamma$}
\begin{ex}\label{Higgs_sez2}
Let $\phi$ be a neutral (i.e. real) Klein-Gordon field which interacts with photons through the following interaction
\[
\mathcal{L}_{INT} = \lambda \phi F_{\alpha\beta} F^{\alpha\beta}
\]
Compute the differential cross section for the process
\[
e^+ e^- \to \phi \gamma
\]
\end{ex}
\begin{sol}
The Lagrangian is
\beq
\mathcal{L} = \mathcal{L}_{QED} + \mathcal{L}_{KG} + \lambda \phi F_{\alpha\beta} F^{\alpha\beta}
\eeq
The Feynman rules for the theory are the usual QED ones, to which we add those derived in the exercise \ref{Higgs_dec}. \\At first order, the process $e^+ e^- \to \phi \gamma$ is represented by the following diagram
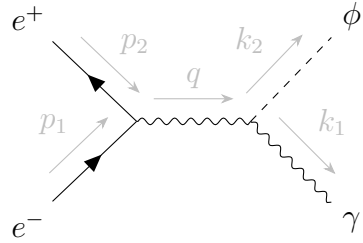
\beq
\begin{tikzpicture}
\begin{feynman}
	\vertex (a);
	\vertex[above left=of a] (b){$e^+$};
	\vertex[below left=of a] (c){$e^-$};
	\vertex[right=of a] (d);
	\vertex[above right=of d] (e){$\phi$};
	\vertex[below right=of d] (f){$\gamma$};

	\diagram {
	(a) -- [fermion, reversed momentum'= { [arrow style=lightgray] $p_2$ } ] (b),
	(c) -- [fermion, momentum={ [arrow style=lightgray] $p_1$ } ] (a),
	(a) -- [boson, momentum={ [arrow style=lightgray] $q$ } ] (d),
	(d) -- [dashed, momentum={ [arrow style=lightgray] $k_2$ } ] (e),
	(d) -- [boson, momentum={ [arrow style=lightgray] $k_1$ } ] (f),
	};
\end{feynman}
\end{tikzpicture}
\eeq
where $q=p_1+p_2=k_1+k_2$. \\Setting $\ket{p_1, s_1; p_2, s_2} \to \ket{k_1, \lambda; k_2}$ the scattering amplitude is
\beq\begin{split}
\mathcal{A} &= \frac{-4 i e \lambda}{q^2} g_{\mu\nu} \bigg[ qk_1 g^{\nu\delta} - q^\delta k_1^\nu \bigg] \bigg( \varepsilon_\delta^\lambda (k_1) \bigg)^* \overline v^{s_2}(p_2) \gamma^\mu u^{s_1}(p_1) = \\
&= \frac{-4 i e \lambda}{q^2} \bigg[ qk_1 g_\mu^\delta - q^\delta k_{1\mu} \bigg] \bigg( \varepsilon_\delta^\lambda (k_1) \bigg)^* \overline v^{s_2}(p_2) \gamma^\mu u^{s_1}(p_1) = \\
&= \frac{-4 i e \lambda}{q^2} \bigg[ qk_1 \bigg( \varepsilon_\mu^\lambda (k_1) \bigg)^* - q \bigg( \varepsilon^\lambda (k_1) \bigg)^* k_{1\mu} \bigg] \overline v^{s_2}(p_2) \gamma^\mu u^{s_1}(p_1) = \\
&= \frac{-4 i e \lambda}{q^2} qk_1 \bigg( \varepsilon_\mu^\lambda (k_1) \bigg)^* \overline v^{s_2}(p_2) \gamma^\mu u^{s_1}(p_1) + \frac{4 i e \lambda}{q^2} q \bigg( \varepsilon^\lambda (k_1) \bigg)^* k_{1\mu} \overline v^{s_2}(p_2) \gamma^\mu u^{s_1}(p_1) = \\
&=: \mathcal{A}_1 + \mathcal{A}_2
\end{split}\eeq
The invariant matrix element, averaging on the initial particles spins and summing over polarizations, is
\beq
|\mathcal{M}|^2 = \frac{1}{4} \sum_{s_1, s_2} \sum_\lambda |\mathcal{A}_1|^2 + |\mathcal{A}_2|^2 + \mathcal{A}_1^* \mathcal{A}_2 + \mathcal{A}_1 \mathcal{A}_2^*
\eeq
We separately compute the various terms
\beq\begin{split}
\sum_{s_1, s_2} \sum_\lambda |\mathcal{A}_1|^2 &= \frac{16 e^2 \lambda^2}{ (q^2)^2 } (qk_1)^2 \sum_{s_1, s_2} \sum_\lambda (\varepsilon_\mu^\lambda)^* \overline v^{s_2} \gamma^\mu u^{s_1} \bigg[ (\varepsilon_\nu^\lambda)^* \overline v^{s_2} \gamma^\nu u^{s_1} \bigg]^* = \\
&= \frac{16 e^2 \lambda^2}{ (q^2)^2 } (qk_1)^2 \sum_{s_1, s_2} \sum_\lambda (\varepsilon_\mu^\lambda)^* \overline v^{s_2} \gamma^\mu u^{s_1} \varepsilon_\nu^\lambda \overline u^{s_1} \gamma^0 \gamma^0 \gamma^\nu \gamma^0 \gamma^0 v^{s_2} = \\
&= \frac{16 e^2 \lambda^2}{ (q^2)^2 } (qk_1)^2 \sum_{s_1, s_2} g_{\nu\mu} \overline v^{s_2} \gamma^\mu u^{s_1} \overline u^{s_1} \gamma^\nu v^{s_2} = \\
&= \frac{16 e^2 \lambda^2}{ (q^2)^2 } (qk_1)^2 g_{\nu\mu} \sum_{s_1, s_2} \sum_{a,b,c,d} ( \overline v^{s_2} )_a (\gamma^\mu)_{ab} ( u^{s_1} )_b ( \overline u^{s_1} )_c (\gamma^\nu)_{cd} ( v^{s_2} )_d = \\
&= \frac{16 e^2 \lambda^2}{ (q^2)^2 } (qk_1)^2 g_{\nu\mu} \sum_{a,b,c,d} (\gamma^\mu)_{ab} ( {\bcc p \ecc}_1 + m )_{bc} (\gamma^\nu)_{cd} ({\bcc p \ecc}_2 - m )_{da} = \\
&= \frac{16 e^2 \lambda^2}{ (q^2)^2 } (qk_1)^2 g_{\nu\mu} Tr \bigg[ \gamma^\mu ( {\bcc p \ecc}_1 + m ) \gamma^\nu ({\bcc p \ecc}_2 - m ) \bigg]
\end{split}\eeq
\beq\begin{split}
\sum_{s_1, s_2} \sum_\lambda |\mathcal{A}_2|^2 &= \frac{16 e^2 \lambda^2}{ (q^2)^2 } \sum_{s_1, s_2} \sum_\lambda q^\alpha (\varepsilon_\alpha^\lambda)^* k_{1\mu} \overline v^{s_2} \gamma^\mu u^{s_1} \bigg[ q^\beta (\varepsilon_\beta^\lambda)^* k_{1\nu} \overline v^{s_2} \gamma^\nu u^{s_1} \bigg]^* = \\
&= \frac{16 e^2 \lambda^2}{ (q^2)^2 } q^\alpha q^\beta k_{1\mu} k_{1\nu} \sum_{s_1, s_2} \sum_\lambda (\varepsilon_\alpha^\lambda)^* \overline v^{s_2} \gamma^\mu u^{s_1} \varepsilon_\beta^\lambda \overline u^{s_1} \gamma^\nu v^{s_2} = \\
&= \frac{16 e^2 \lambda^2}{ (q^2)^2 } q^\alpha q^\beta k_{1\mu} k_{1\nu} g_{\beta\alpha} \sum_{s_1, s_2} \overline v^{s_2} \gamma^\mu u^{s_1} \overline u^{s_1} \gamma^\nu v^{s_2} = \\
&= \frac{16 e^2 \lambda^2}{ (q^2)^2 } q^\alpha q_\alpha k_{1\mu} k_{1\nu} \sum_{s_1, s_2} \sum_{a,b,c,d} ( \overline v^{s_2} )_a (\gamma^\mu)_{ab} ( u^{s_1} )_b ( \overline u^{s_1} )_c (\gamma^\nu)_{cd} ( v^{s_2} )_d = \\
&= \frac{16 e^2 \lambda^2}{ (q^2) } k_{1\mu} k_{1\nu} \sum_{a,b,c,d} (\gamma^\mu)_{ab} ( {\bcc p \ecc}_1 + m )_{bc} (\gamma^\nu)_{cd} ({\bcc p \ecc}_2 - m )_{da} = \\
&=  \frac{16 e^2 \lambda^2}{ (q^2) } k_{1\mu} k_{1\nu} Tr \bigg[ \gamma^\mu ( {\bcc p \ecc}_1 + m ) \gamma^\nu ({\bcc p \ecc}_2 - m ) \bigg]
\end{split}\eeq
\beq\begin{split}
\sum_{s_1, s_2} \sum_\lambda \mathcal{A}_1^* \mathcal{A}_2 &= \frac{-16 e^2 \lambda^2}{ (q^2)^2 } \sum_{s_1, s_2} \sum_\lambda (qk_1) \varepsilon_\mu^\lambda \bigg[ \overline v^{s_2} \gamma^\mu u^{s_1} \bigg]^* q(\varepsilon^\lambda)^* k_{1\nu} \overline v^{s_2} \gamma^\nu u^{s_1} = \\
&= \frac{-16 e^2 \lambda^2}{ (q^2)^2 } (qk_1) q^\alpha k_{1\nu} \sum_{s_1, s_2} \sum_\lambda \varepsilon_\mu^\lambda (\varepsilon_\alpha^\lambda)^* \bigg[ \overline v^{s_2} \gamma^\mu u^{s_1} \bigg]^* \overline v^{s_2} \gamma^\nu u^{s_1} = \\
&= \frac{-16 e^2 \lambda^2}{ (q^2)^2 } (qk_1) q^\alpha k_{1\nu} g_{\mu\alpha} \sum_{s_1, s_2} \overline u^{s_1} \gamma^\mu v^{s_2} \overline v^{s_2} \gamma^\nu u^{s_1} = \\
&= \frac{-16 e^2 \lambda^2}{ (q^2)^2 } (qk_1) q_\mu k_{1\nu} \sum_{s_1, s_2} \sum_{a,b,c,d} ( \overline u^{s_1} )_a (\gamma^\mu)_{ab} ( v^{s_2} )_b ( \overline v^{s_2} )_c (\gamma^\nu)_{cd} ( u^{s_1} )_d = \\
&= \frac{-16 e^2 \lambda^2}{ (q^2)^2 } (qk_1) q_\mu k_{1\nu} \sum_{a,b,c,d} (\gamma^\mu)_{ab} ( {\bcc p \ecc}_2 - m )_{bc} (\gamma^\nu)_{cd} ( {\bcc p \ecc}_1 + m )_{da} = \\
&= \frac{-16 e^2 \lambda^2}{ (q^2)^2 } (qk_1) q_\mu k_{1\nu} Tr \bigg[ \gamma^\mu ( {\bcc p \ecc}_2 - m ) \gamma^\nu ( {\bcc p \ecc}_1 + m ) \bigg]
\end{split}\eeq
\beq\begin{split}
\sum_{s_1, s_2} \sum_\lambda \mathcal{A}_1 \mathcal{A}_2^* &= \frac{-16 e^2 \lambda^2}{ (q^2)^2 } \sum_{s_1, s_2} \sum_\lambda (qk_1) (\varepsilon_\mu^\lambda)^* \overline v^{s_2} \gamma^\mu u^{s_1} q \varepsilon^\lambda k_{1\nu} \bigg[ \overline v^{s_2} \gamma^\nu u^{s_1} \bigg]^* = \\
&= \frac{-16 e^2 \lambda^2}{ (q^2)^2 } (qk_1) q^\alpha k_{1\nu} \sum_{s_1, s_2} \sum_\lambda (\varepsilon_\mu^\lambda)^* \varepsilon_\alpha^\lambda \overline v^{s_2} \gamma^\mu u^{s_1} \bigg[ \overline v^{s_2} \gamma^\nu u^{s_1} \bigg]^* = \\
&= \frac{-16 e^2 \lambda^2}{ (q^2)^2 } (qk_1) q^\alpha k_{1\nu} g_{\alpha\mu} \sum_{s_1, s_2} \overline v^{s_2} \gamma^\mu u^{s_1} \overline u^{s_1} \gamma^\nu v^{s_2} = \\
&= \frac{-16 e^2 \lambda^2}{ (q^2)^2 } (qk_1) q_\mu k_{1\nu} \sum_{s_1, s_2} \sum_{a,b,c,d} ( \overline v^{s_2} )_a (\gamma^\mu)_{ab} ( u^{s_1} )_b ( \overline u^{s_1} )_c (\gamma^\nu)_{cd} ( v^{s_2} )_d = \\
&= \frac{-16 e^2 \lambda^2}{ (q^2)^2 } (qk_1) q_\mu k_{1\nu} \sum_{a,b,c,d} (\gamma^\mu)_{ab} ( {\bcc p \ecc}_1 + m )_{bc} (\gamma^\nu)_{cd} ( {\bcc p \ecc}_2 - m )_{da} = \\
&= \frac{-16 e^2 \lambda^2}{ (q^2)^2 } (qk_1) q_\mu k_{1\nu} Tr \bigg[ \gamma^\mu ( {\bcc p \ecc}_1 + m ) \gamma^\nu ( {\bcc p \ecc}_2 - m ) \bigg]
\end{split}\eeq
Thanks to the cyclic property of the trace \eqref{cyclic}, the two "mixed" terms turn out to be equal
\beq\begin{split}
\sum_{s_1, s_2} \sum_\lambda \mathcal{A}_1^* \mathcal{A}_2 &= \frac{-16 e^2 \lambda^2}{ (q^2)^2 } (qk_1) q_\mu k_{1\nu} Tr \bigg[ \gamma^\mu ( {\bcc p \ecc}_2 - m ) \gamma^\nu ( {\bcc p \ecc}_1 + m ) \bigg] = \\
&= \frac{-16 e^2 \lambda^2}{ (q^2)^2 } (qk_1) q_\mu k_{1\nu} Tr \bigg[ ( {\bcc p \ecc}_2 - m ) \gamma^\nu ( {\bcc p \ecc}_1 + m ) \gamma^\mu \bigg] = \\
&= \frac{-16 e^2 \lambda^2}{ (q^2)^2 } (qk_1) q_\mu k_{1\nu} Tr \bigg[ \gamma^\nu ( {\bcc p \ecc}_1 + m ) \gamma^\mu ( {\bcc p \ecc}_2 - m ) \bigg] = \\
&= \frac{-16 e^2 \lambda^2}{ (q^2)^2 } (qk_1) q_\nu k_{1\mu} Tr \bigg[ \gamma^\mu ( {\bcc p \ecc}_1 + m ) \gamma^\nu ( {\bcc p \ecc}_2 - m ) \bigg] = \\
&= \sum_{s_1, s_2} \sum_\lambda \mathcal{A}_1 \mathcal{A}_2^*
\end{split}\eeq
The computation of the scattering amplitude then reduces to
\beq
|\mathcal{M}|^2 = \frac{4 e^2 \lambda^2}{q^2} \bigg[ \frac{1}{q^2} (qk_1)^2 g_{\nu\mu} + k_{1\mu} k_{2\nu} - \frac{2}{q^2} (qk_1) q_\mu k_{1\nu} \bigg] Tr \bigg[ \gamma^\mu ( {\bcc p \ecc}_1 + m ) \gamma^\nu ( {\bcc p \ecc}_2 - m ) \bigg]
\eeq
that is
\beq\begin{split}
|\mathcal{M}|^2 &= \frac{4 e^2 \lambda^2}{q^2} \bigg[ \frac{1}{q^2} (qk_1)^2 g_{\nu\mu} + k_{1\mu} k_{2\nu} - \frac{2}{q^2} (qk_1) q_\mu k_{1\nu} \bigg] \cdot \\
&\cdot \bigg[ p_1^\mu p_2^\nu - (p_1p_2) g^{\mu\nu} + p_1^\nu p_2^\mu - m^2g^{\mu\nu} \bigg]
\end{split}\eeq
Carrying out the contractions we obtain
\beq\begin{split}
|\mathcal{M}|^2 = \frac{4 e^2 \lambda^2}{q^2} \bigg\{& \frac{1}{q^2} (qk_1)^2 \bigg[ -2(p_1p_2) - 4m^2 \bigg] + \\
&+ \bigg[ (k_1p_1) (k_2p_2) - (k_1k_2) (p_1p_2) + (k_1p_2) (k_2p_1) - m^2(k_1k_2) \bigg] - \\
&- \frac{2}{q^2} (qk_1) \bigg[ (qp_1) (k_1p_2) - (qk_1) (p_1p_2) + (qp_2) (k_1p_1) - m^2 (qk_1) \bigg] \bigg\} 
\end{split}\eeq
Therefore we have
\beq\begin{split}
|\mathcal{M}|^2 = \frac{4 e^2 \lambda^2}{q^4} \bigg\{& -2 (qk_1)^2 (p_1p_2) - 4 m^2 (qk_1)^2 + q^2 (k_1p_1) (k_2p_2) - \\
&- q^2 (k_1k_2) (p_1p_2) + q^2 (k_1p_2) (k_2p_1) - m^2 q^2 (k_1k_2) - \\
&- 2 (qk_1) (qp_1) (k_1p_2) + 2 (qk_1)^2 (p_1p_2) - \\
&- 2 (qk_1) (qp_2) (k_1p_1) + 2 m^2 (qk_1)^2 \bigg\}
\end{split}\eeq
The expressions for the flux factor, the phase space element and the 4-momenta in the centre of mass frame are the usual ones
\beq
\mathcal{J} = \frac{1}{4 \sqrt{ (p_1p_2)^2 - m^4 } }
\eeq
\beq
d \Phi_2 = \frac{1}{8 \pi} \frac{ | \textbf{k}_1 | }{ E_\gamma + E_\phi } d \cos{\theta}
\eeq
\beq\begin{cases}
p_1 = (E, 0, 0, |\textbf{p}|) \\
p_2 = (E, 0, 0, -|\textbf{p}|) \\
k_1 = (E_\gamma, 0, |\textbf{k}| \sin{\theta}, |\textbf{k}| \cos{\theta}) \\
k_2 = (E_\phi, 0, -|\textbf{k}| \sin{\theta}, -|\textbf{k}| \cos{\theta})
\end{cases}\eeq
where $|\textbf{p}| = \sqrt{E^2 - m^2}$ and $|\textbf{k}| = E_\gamma = \sqrt{E_\phi^2 - m_\phi^2}$. \\Recalling that $q = p_1+p_2 = k_1+k_2$, we can make explicit the scalar products
\beq\begin{split}
&(qk_1) = (k_1k_2) = E_\gamma E_\phi + E_\gamma^2 \\
&(qp_1) = (qp_2) = 2 E^2 \\
&(p_1p_2) = 2 E^2 - m^2 \\
&(k_1p_1) = E_\gamma E - E_\gamma \sqrt{E^2 - m^2} \cos{\theta} \\
&(k_2p_2) = E_\phi E - E_\gamma \sqrt{E^2 - m^2} \cos{\theta} \\
&(k_1p_2) = E_\gamma E + E_\gamma \sqrt{E^2 - m^2} \cos{\theta} \\
&(k_2p_1) = E_\phi E + E_\gamma \sqrt{E^2 - m^2} \cos{\theta} \\
&q^2 = 4 E^2 = m_\phi^2 + 2 E_\gamma E_\phi + 2 E_\gamma^2
\end{split}\eeq
After some simplification we get
\beq\begin{split}
|\mathcal{M}|^2 = \frac{4 e^2 \lambda^2}{16 E^4} \bigg\{& -2 m^2 \bigg[ E_\gamma^2 E_\phi^2 + E_\gamma^4 + 2 E_\gamma^3 E_\phi \bigg] - 8 E^3 E_\gamma \bigg[ E_\gamma E_\phi + E_\gamma^2 - E E_\gamma \bigg] + \\
&+ 8 E^2 E_\gamma^2 (E^2 - m^2) \cos^2{\theta} \bigg\}
\end{split}\eeq
and finally
\beq\begin{split}
\frac{d \sigma}{d \Omega} = \frac{e^2 \lambda^2}{256\pi} \frac{E_\gamma}{E^5 \sqrt{E^4 - E^2 m^2} } \bigg\{& -m^2 \bigg[ E_\gamma^2 E_\phi^2 + E_\gamma^4 + 2 E_\gamma^3 E_\phi \bigg] - \\
&- 4 E^3 E_\gamma \bigg[ E_\gamma E_\phi + E_\gamma^2 - E E_\gamma \bigg] + \\
&+ 4 E^2 E_\gamma^2 (E^2 - m^2) \cos^2{\theta} \bigg\}
\end{split}\eeq
\end{sol}

\section{Yukawa Lagrangian: $f \bar f \to \phi \phi$}
\begin{ex}\label{Yukawa}
Consider the following lagrangian density 
\[
\mathcal{L} = i \overline \psi \gamma^\mu \partial_\mu \psi - m \overline \psi \psi + \frac{1}{2} \partial_\mu \phi \partial^\mu \phi - m^2 \phi^2 + y \overline \psi \psi \phi
\]
where $\psi$ is a Dirac spinor field and $\phi$ a real scalar field. Compute the differential cross section for the process
\[
f \bar f \to \phi \phi
\]
where $f$ ($\bar f$) stands for the fermionic particle (antiparticle) and $\phi$ for the scalar one.
\end{ex}
\begin{sol}
Besides the usual Feynman rules for QED, we also have the following one, due to the interaction between Dirac fields and the Klein-Gordon one (\textit{Yukawa interaction} term in the Lagrangian).
\beq
\begin{tikzpicture}
\begin{feynman}
\node at (1, 0) {$= i y$};

	\vertex (a) {\(\phi\)};
	\vertex[left=of a] (b);
	\vertex[above left=of b] (c) {\(\overline \psi\)};
	\vertex[below left=of b] (d) {\(\psi\)};

	\diagram {
	(a) -- [dashed] (b),
	(b) -- [fermion] (c),
	(d) -- [fermion] (b),
	};
\end{feynman}
\end{tikzpicture}
\eeq
The Feynman diagrams associated with the process in question are the following
\beq
\begin{tikzpicture}
\begin{feynman}
	\vertex (a);
	\vertex[above left=of a] (b){$ f$};
	\vertex[above right=of a] (c){$\phi$};
	\vertex[below=of a] (d);
	\vertex[below left=of d] (e){$\overline f$};
	\vertex[below right=of d] (f){$\phi$};

	\diagram {
	(a) -- [anti fermion, reversed momentum'={ [arrow style=lightgray] $p_1$ } ] (b),
	(a) -- [dashed, momentum'={ [arrow style=lightgray] $p_3$ } ] (c),
	(d) -- [anti fermion, reversed momentum={ [arrow style=lightgray] $q_1$ } ] (a),
	(e) -- [anti fermion, momentum={ [arrow style=lightgray] $p_2$ } ] (d),
	(d) -- [dashed, momentum={ [arrow style=lightgray] $p_4$ } ] (f),
	};
\end{feynman}
\end{tikzpicture}
\eeq
where $q_1 = p_1-p_3$, and
\beq
\begin{tikzpicture}
\begin{feynman}
	\vertex (a);
	\vertex[above left=of a] (b){$ f$};
	\vertex[above right=of a] (c){$\phi$};
	\vertex[below=of a] (d);
	\vertex[below left=of d] (e){$\overline f$};
	\vertex[below right=of d] (f){$\phi$};

	\diagram {
	(a) -- [anti fermion, reversed momentum'={ [arrow style=lightgray] $p_1$ } ] (b),
	(a) -- [dashed, momentum'={ [arrow style=lightgray] $p_4$ } ] (c),
	(d) -- [anti fermion, reversed momentum={ [arrow style=lightgray] $q_2$ } ] (a),
	(e) -- [anti fermion, momentum={ [arrow style=lightgray] $p_2$ } ] (d),
	(d) -- [dashed, momentum={ [arrow style=lightgray] $p_3$ } ] (f),
	};
\end{feynman}
\end{tikzpicture}
\eeq
where $q_2 = p_1-p_4$. \\The two corresponding amplitudes are
\beq
\mathcal{A}_1 = \overline v^{s_2}(p_2) \bigg( -i\lambda \bigg) \frac{ i(\bcc q_1 \ecc + m_f)}{q_1^2 - m_f^2} \bigg( -i\lambda \bigg) u^{s_1}(p_1)
\eeq
\beq
\mathcal{A}_2 = \overline v^{s_2}(p_2) \bigg( -i\lambda \bigg) \frac{ i(\bcc q_2 \ecc + m_f)} {q_2^2 - m_f^2} \bigg( -i\lambda \bigg) u^{s_1}(p_1)
\eeq
We now substitute the expressions for $q_1$ and $q_2$ in the respective amplitudes. Regarding $\mathcal{A}_1$, in the denominator we have 
\beq
q_1^2 - m_f^2 = \underbrace{p_1^2}_{= m_f^2} + \underbrace{p_3^2}_{= m_\phi^2} - 2p_1 \cdot p_3 - m_f^2 = m_\phi^2 - 2p_1 \cdot p_3
\eeq
The numerator is instead
\beq
(\bcc q_1 \ecc + m_f) u^{s_1}(p_1) = (\bcc p_1 \ecc - \bcc p_3 \ecc + m_f) u^{s_1}(p_1) = (2m_f - \bcc p_3 \ecc) u^{s_1}(p_1)
\eeq
where we have exploited the Dirac equation $(\cancel{p}-m)u(p)=0$. \\Carrying out the same passages for $\mathcal{A}_2$ one obtains
\beq
\mathcal{A}_1 = \frac{-i\lambda^2} {\underbrace{m_\phi^2 - 2p_1 \cdot p_3}_{=: A}} \overline v^{s_2}(p_2) \bigg( 2m_f - \bcc p_3 \ecc \bigg) u^{s_1}(p_1)
\eeq
\beq
\mathcal{A}_2 = \frac{-i\lambda^2}{\underbrace{m_\phi^2 - 2p_1 \cdot p_4}_{=: B}} \overline v^{s_2}(p_2) \bigg(2m_f - \bcc p_4 \ecc \bigg) u^{s_1}(p_1)
\eeq
The invariant matrix element, averaged on the initial spins, is
\beq\begin{split}
|\mathcal{M}|^2 &= \frac{1}{4} \sum_{s_1,s_2} \big[ \mathcal{A}_1 \mathcal{A}_1^* + \mathcal{A}_2 \mathcal{A}_2^* + \mathcal{A}_1 \mathcal{A}_2^* + \mathcal{A}_1^* \mathcal{A}_2 \big] = \\
&= \frac{\lambda^4}{4} \sum_{s_1,s_2} \bigg\{ \frac{1}{A^2} \overline v^{s_2} \bigg( 2m_f - \bcc p_3 \ecc \bigg) u^{s_1} \overline u^{s_1} \bigg(2m_f - \bcc p_3 \ecc \bigg) v^{s_2} + \\
&\quad\quad\quad\quad + \frac{1}{B^2} \overline v^{s_2} \bigg( 2m_f - \bcc p_4 \ecc \bigg) u^{s_1} \overline u^{s_1} \bigg( 2m_f - \bcc p_4 \ecc \bigg) v^{s_2} + \\
&\quad\quad\quad\quad + \frac{1}{AB} \bigg[ \overline v^{s_2} \bigg( 2m_f - \bcc p_3 \ecc \bigg) u^{s_1} \overline u^{s_1} \bigg( 2m_f - \bcc p_4 \ecc \bigg) v^{s_2} + \\
&\quad\quad\quad\quad\quad + \overline v^{s_2} \bigg( 2m_f - \bcc p_4 \ecc \bigg) u^{s_1} \overline u^{s_1} \bigg( 2m_f - \bcc p_3 \ecc \bigg) v^{s_2} \bigg] \bigg\} = \\
&= \frac{\lambda^4}{4} \bigg\{ \frac{1}{A^2} \underbrace{Tr\big[ (\bcc p_1 \ecc + m_f) (2m_f - \bcc p_3 \ecc) (\bcc p_2 \ecc - m_f) (2m_f - \bcc p_3 \ecc)\big]}_{(1)} + \\
&\quad\quad\quad + \frac{1}{B^2} \underbrace{Tr\big[ (\bcc p_1 \ecc + m_f) (2m_f - \bcc p_4 \ecc) (\bcc p_2 \ecc - m_f) (2m_f - \bcc p_4 \ecc) \big]}_{(2)} + \\
&\quad\quad\quad + \frac{1}{AB} \bigg( \underbrace{Tr\big[(\bcc p_1 \ecc + m_f) (2m_f - \bcc p_4 \ecc) (\bcc p_2 \ecc - m_f) (2m_f - \bcc p_3 \ecc) \big]}_{(3)} + \\
&\quad\quad\quad\quad\quad + \underbrace{Tr\big[ (\bcc p_1 \ecc + m_f) (2m_f - \bcc p_3 \ecc) (\bcc p_2 \ecc - m_f) (2m_f - \bcc p_4 \ecc) \big]}_{(4)}\bigg) \bigg\}
\end{split}\eeq
The four traces share the same structure, therefore it is sufficient to compute only one of them; the other traces will be obtained with appropriate substitutions. Let's see the computation of the trace $(3)$.
\beq\begin{split}
(3) =& Tr\big[ (\bcc p_1 \ecc + m_f) (2m_f - \bcc p_4 \ecc) (\bcc p_2 \ecc - m_f) (2m_f - \bcc p_3 \ecc) \big] = \\
=& Tr\big[ (\bcc p_1 \ecc + m_f) \bcc p_4 \ecc (\bcc p_2 \ecc - m_f) \bcc p_3 \ecc \big] + 4m_f^2 Tr\big[ (\bcc p_1 \ecc + m_f) (\bcc p_2 \ecc - m_f) \big] = \\
=& Tr\big[ \bcc p_1 \ecc \bcc p_4 \ecc \bcc p_2 \ecc \bcc p_3 \ecc \big] - m_f^2 Tr\big[\bcc p_4 \ecc \bcc p_3 \ecc \big] + 4m_f^2 Tr\big[\bcc p_1 \ecc \bcc p_2 \ecc \big] - 16m_f^4 = \\
=& 4\big[ (p_1 p_4) (p_2 p_3) - (p_1 p_2) (p_4 p_3) + (p_1 p_3) (p_4 p_2) \big] - 4m_f^2 (p_4 p_3) + 16m_f^2 (p_1 p_2) - 16m_f^4
\end{split}\eeq
The result of trace number $(3)$ is symmetric under the exchange $p_3 \longleftrightarrow p_4$, so that $(3) = (4)$. \\Computing the first trace we get
\beq
(1) = 4\big[ 2 (p_1 p_3) (p_2 p_3) - (p_1 p_2) p_3^2 \big] - 4m_f^2 p_3^2 + 16m_f^2 (p_1 p_2) - 16m_f^4
\eeq
Finally, trace $(2)$ can be obtained from trace $(1)$ substituting $p_3$ with $p_4$. \\In the centre of mass reference frame the 4-momenta can be written as 
\beq\begin{cases}
p_1 = (E_1, 0, 0, |\textbf{p}| ) \\
p_2 = (E_2, 0, 0, -|\textbf{p}| ) \\
p_3 = (E_3, \textbf{q}) \\
p_4 = (E_4, -\textbf{q})
\end{cases}\eeq
\beq\begin{split}
&E_1 = E_2 =: E \\
&E_3 = E_4 =: E' \\
&p_1^2 = p_2^2 = m_f^2 \\
&p_3^2 = p_4^2 = m_\phi^2
\end{split}\eeq
By introducing the Mandelstam variables
\beq \begin{split}
&s := (p_1 + p_2)^2 \\
&t := (p_1 - p_3)^2 \\
&u := (p_1 - p_4)^2
\end{split}\eeq
the products of the 4-momenta now read
\beq \begin{split}
&p_1 p_2 = \frac{1}{2} (s - 2m_f^2) \\
&p_1 p_3 = p_2 p_4 = \frac{1}{2} (m_f^2 + m_\phi^2 - t) \\
&p_1 p_4 = p_2 p_3 = \frac{1}{2} (m_f^2 + m_\phi^2 - u) \\
&p_3 p_4 = \frac{1}{2} (s - 2m_\phi^2)
\end{split}\eeq
where we made use of four-momentum conservation between initial and final states and also used the relation $s + t + u = \sum_i {m_i^2}$ (which is valid in general), where $m_i$ are the masses of all the external particles involved in the process.
\beq\begin{split}\label{M_yukawa}
|\mathcal{M}|^2 = \frac{\lambda^4}{4} \bigg\{& \bigg( \frac{1}{(t - m_f^2)^2} + \frac{1}{(u - m_f^2)^2} \bigg) \bigg[ 2(m_f^2 + m_\phi^2 - t) (m_f^2 + m_\phi^2 - u) - \\
&- 4m_f^2 m_\phi^2 + 8m_f^2 (s - 4m_f^2) \bigg] + \\
&+ \frac{2}{(t-m_f^2)(u-m_f^2)}\bigg[ (m_f^2 + m_\phi^2 - t)^2 (m_f^2 + m_\phi^2 - u)^2 - \\
&- (s - 2m_f^2) (s - 2m_\phi^2) + 2m_f^2 \bigg( 4(s - 2m_f^2) - (s - 2m_\phi^2) \bigg) - 16m_f^4 \bigg] \bigg\}
\end{split}\eeq
The flux factor is 
\beq
\mathcal{J} = \frac{1}{\sqrt{4(s - 2m_f^2)^2 - 16m_f^4}}
\eeq
while the two-body phase space is
\beq
d \Phi_2 = \frac{\sqrt{\frac{s}{4} - m_\phi^2} }{\sqrt{s}/2} \frac{d \cos{\theta} d \varphi}{16\pi^2}
\eeq
therefore, after integrating over $d\varphi$, the differential cross section for the process is
\beq
\frac{d \sigma}{d \cos{\theta}} = \frac{1}{16\pi} \sqrt{\frac{1 - \frac{4m_\phi^2}{s}}{(s - 2m_f^2)^2 - 4m_f^4}} |\mathcal{M}|^2
\eeq
where $|\mathcal{M}|^2$ is given by \eqref{M_yukawa}.

\end{sol}

\newpage
\section{Fermi Lagrangian: $\nu e^- \to \nu e^-$}
\begin{ex}\label{Fermi}
Consider the standard QED Lagrangian for the electron; add to that a neutral and massless spinor field $\psi_\nu$, described by the following Lagrangian
\[
\mathcal{L} = i \overline \psi_\nu \bcc \partial \ecc \psi_\nu + G \overline \psi_\nu \gamma_\alpha \psi_\nu \overline \psi_e \gamma^\alpha \psi_e + \mu \overline \psi_\nu \sigma_{\alpha\beta} F^{\alpha\beta} \psi_\nu
\]
Compute the differential cross section for the process
\[
\nu e^- \to \nu e^-
\]
\end{ex}
\begin{sol}
The full Lagrangian of the problem is
\beq
\mathcal{L} = \overline{\psi}_e (i \mathcal{\bcc D \ecc} - m_e) \psi_e - \frac{1}{4} F_{\mu\nu} F^{\mu\nu} + i \overline \psi_\nu \bcc \partial \ecc \psi_\nu + G \overline \psi_\nu \gamma_\alpha \psi_\nu \overline \psi_e \gamma^\alpha \psi_e + \mu \overline \psi_\nu \sigma_{\alpha\beta} F^{\alpha\beta} \psi_\nu
\eeq
In addition the the usual QED vertex we also have the following new ones
\beq
\begin{tikzpicture}
\begin{feynman}
\node at (2.5, 0) {$= i G \gamma_\alpha \gamma^\alpha$};

	\vertex (a);
	\vertex[above right=of a] (b) {$\overline \psi_\nu$};
	\vertex[below right=of a] (c) {$\psi_\nu$};
	\vertex[above left=of a] (d) {$\overline \psi_e$};
	\vertex[below left=of a] (e) {$\psi_e$};

	\diagram {
	(a) -- [anti fermion] (b),
	(a) -- [fermion] (c),
	(a) -- [fermion] (d),
	(a) -- [anti fermion] (e)
	};
\end{feynman}
\end{tikzpicture} 
\eeq
\beq
\begin{tikzpicture}
\begin{feynman}
\node at (2, 0) {$=2i\mu(\gamma^\sigma \bcc k \ecc-k^\sigma)$};

	\vertex (a) {\(A_\sigma\)};
	\vertex[left=of a] (b);
	\vertex[above left=of b] (c) {$\psi_\nu$};
	\vertex[below left=of b] (d) {$\overline{\psi}_\nu$};

	\diagram {
	(a) -- [boson] (b),
	(c) -- [fermion] (b),
	(b) -- [fermion] (d)
	};
\end{feynman}
\end{tikzpicture}
\eeq
This last vertex factor was computed as follows
\beq\begin{split}
\mathcal{L}_{INT}=& \mu \overline{\psi}_\nu \sigma^{\alpha\beta} F_{\alpha\beta} \psi_\nu = \mu \overline{\psi}_\nu \frac{i}{2} [\gamma^\alpha,\gamma^\beta] (\partial_\alpha A_\beta - \partial_\beta A_\alpha) \psi_\nu = \\
=& i \frac{\mu}{2} \overline{\psi}_\nu (\gamma^\alpha \gamma^\beta - \gamma^\beta \gamma^\alpha) (\partial_\alpha A_\beta - \partial_\beta A_\alpha)\psi_\nu= i \frac{\mu}{2} \overline{\psi}_\nu 2 (\gamma^\alpha \gamma^\beta \partial_\alpha A_\beta - \gamma^\alpha \gamma^\beta \partial_\beta A_\alpha) = \\
=& i \mu \overline{\psi}_\nu \gamma^\alpha \gamma^\beta (g_\beta^\sigma \partial_\alpha - g_\alpha^\sigma \partial_\beta) A_\sigma= i \mu \overline{\psi}_\nu (\gamma^\alpha \gamma^\sigma \partial_\alpha - \gamma^\sigma \gamma^\beta \partial_\beta) A_\sigma
\end{split}\eeq
In the momentum space we obtain
\beq\begin{split}
\mathcal{L}_{INT} =& i \mu \overline{\psi}_\nu \bigg(\gamma^\alpha \gamma^\sigma (ik_\alpha) - \gamma^\sigma \gamma^\beta (ik_\beta)\bigg) A_\sigma = \mu \overline{\psi}_\nu (\gamma^\sigma \bcc k \ecc - \bcc k \ecc \gamma^\sigma) A_\sigma \\
=& 2 \mu \overline{\psi}_\nu (\gamma^\sigma \bcc k \ecc - k^\sigma) A_\sigma
\end{split}\eeq
The process $\nu e^- \to \nu e^-$ is described by the following diagrams
\beq
\begin{tikzpicture}
\begin{feynman}
        \node at (3.6,-1){$+$};
	\vertex (a);
	\vertex[above left=of a] (b){$e^-$};
	\vertex[above right=of a](c) {$e^-$};
	\vertex[below=of a] (d);
	\vertex[below left=of d] (e){$\nu$};
	\vertex[below right=of d] (f){$\nu$};

	\diagram {
	(a) -- [anti fermion, reversed momentum={ [arrow style=lightgray] $p_1$ } ] (b),
	(a) -- [fermion, momentum'={ [arrow style=lightgray] $p_3$ } ] (c),
	(d) -- [boson, reversed momentum={ [arrow style=lightgray] $k$ } ] (a),
	(e) -- [fermion, momentum={ [arrow style=lightgray] $p_2$ } ] (d),
	(d) -- [fermion, momentum={ [arrow style=lightgray] $p_4$ } ] (f),
	};

	\vertex (A) at (7,-1);
	\vertex[above right=of A] (B) {$e^-$};
	\vertex[below right=of A] (C) {$\nu$};
	\vertex[above left=of A] (D) {$e^-$};
	\vertex[below left=of A] (E) {$\nu$};

	\diagram {
	(A) -- [fermion, momentum={ [arrow style=lightgray] $p_3$ }] (B),
	(A) -- [fermion, momentum={ [arrow style=lightgray] $p_4$ }] (C),
	(A) -- [anti fermion, reversed momentum={ [arrow style=lightgray] $p_1$ }] (D),
	(A) -- [anti fermion, reversed momentum={ [arrow style=lightgray] $p_2$ }] (E)
	};
\end{feynman}
\end{tikzpicture}
\eeq
so that the amplitude is
\beq\begin{split}
\mathcal{A}=& \overline{u}^{s_3}(p_3) \bigg(-ie \gamma^\mu \bigg) u^{s_1}(p_1) \bigg(\frac{-ig_{\mu\nu}}{k^2} \bigg) \overline{u}^{s_4}(p_4)\bigg(2i \mu (\gamma^\nu \bcc k \ecc - k^\nu) \bigg) u^{s_2} + \\
&+ iG \overline{u}^{s_4}(p_4) \gamma_\sigma u^{s_2}(p_2) \overline{u}^{s_3}(p_3) \gamma^\sigma u^{s_1}(p_1)
\end{split}\eeq
which means
\beq
\mathcal{A} = - \frac{2i\mu}{k^2} \overline{u}^{s_3} \gamma_\nu u^{s_1} \overline{u}^{s_4} (\gamma^\nu \bcc k \ecc - k^\nu) u^{s_2} + iG \overline{u}^{s_4} \gamma_\sigma u^{s_2} \overline{u}^{s_3} \gamma^\sigma u^{s_1}
\eeq
\beq
\mathcal{A}^*= \frac{2i\mu}{k^2} \overline{u}^{s_2} (\bcc k \ecc \gamma^\mu - k^\mu) u^{s_4} \overline{u}^{s_1} \gamma_\mu u^{s_3} - iG \overline{u}^{s_1} \gamma^\rho u^{s_3} \overline{u}^{s_2} \gamma_\rho u^{s_4}
\eeq
The matrix element for an unpolarized cross section is obtained averaging over the spins of the initial particles and summing over the spins of the final ones
\beq\begin{split}
|\mathcal{M}|^2 = \frac{1}{4}\,\sum_{s_1,s_2,s_3,s_4}\,\bigg\{& \frac{4\mu^2}{k^4} \overline{u}^{s_3} \gamma_\nu u^{s_1} \overline{u}^{s_4} (\gamma^\nu \bcc k \ecc - k^\nu) u^{s_2} \overline{u}^{s_2} (\bcc k \ecc \gamma^\mu - k^\mu) u^{s_4} \overline{u}^{s_1} \gamma_\mu u^{s_3} - \\
&- \frac{2\mu G}{k^2} \bigg[ \overline{u}^{s_3} \gamma_\nu u^{s_1} \overline{u}^{s_4} (\gamma^\nu \bcc k \ecc - k^\nu) u^{s_2} \overline{u}^{s_1} \gamma^\rho u^{s_3} \overline{u}^{s_2} \gamma_\rho u^{s_4} + \\
&+ \overline{u}^{s_4} \gamma_\sigma u^{s_2} \overline{u}^{s_3} \gamma^\sigma u^{s_1} \overline{u}^{s_2} (\bcc k \ecc \gamma^\mu - k^\mu) u^{s_4} \overline{u}^{s_1} \gamma_\mu u^{s_3} \bigg] + \\
&+ G^2 \overline{u}^{s_4} \gamma_\sigma u^{s_2} \overline{u}^{s_3} \gamma^\sigma u^{s_1} \overline{u}^{s_1} \gamma^\rho u^{s_3} \overline{u}^{s_2}\gamma_\rho u^{s_4} \bigg\}
\end{split}\eeq
Using the Dirac spinors sum rules we obtain
\beq\begin{split}
|\mathcal{M}|^2= \frac{1}{4}\bigg\{& \frac{4\mu^2}{k^4} \underbrace{Tr\bigg[ ({\bcc p \ecc}_1 + m_e) \gamma_\mu ({\bcc p \ecc}_3 + m_e) \gamma_\nu \bigg]}_{(1)} \underbrace{Tr\bigg[ {\bcc p \ecc}_2 (\bcc k \ecc \gamma^\mu - k^\mu){\bcc p \ecc}_4 (\gamma^\nu \bcc k \ecc - k^\nu) \bigg]}_{(2)} + \\
&- \frac{2\mu G}{k^2} \bigg[\underbrace{Tr\bigg[ ({\bcc p \ecc}_1 + m_e) \gamma^\rho ({\bcc p \ecc}_3 + m_e) \gamma_\nu \bigg]}_{(3)} \underbrace{Tr\bigg[ {\bcc p \ecc}_2 \gamma_\rho {\bcc p \ecc}_4 (\gamma^\nu \bcc k \ecc - k^\nu) \bigg]}_{(4)} + \\
&\underbrace{Tr\bigg[ ({\bcc p \ecc}_1 + m_e) \gamma_\mu ({\bcc p \ecc}_3 + m_e) \gamma^\sigma \bigg]}_{(5)} \underbrace{Tr\bigg[ {\bcc p \ecc}_2 (\bcc k \ecc \gamma^\mu - k^\mu){\bcc p \ecc}_4 \gamma_\sigma \bigg]}_{(6)} \bigg] + \\
&+G^2 \underbrace{Tr\bigg[ ({\bcc p \ecc}_1 + m_e) \gamma^\rho ({\bcc p \ecc}_3 + m_e) \gamma^\sigma \bigg]}_{(7)} \underbrace{Tr\bigg[ {\bcc p \ecc}_2 \gamma_\rho {\bcc p \ecc}_4 \gamma_\sigma \bigg]}_{(8)} \bigg\}
\end{split}\eeq
Traces $(4)$ and $(6)$ contain an odd number of gamma matrices, so they are equal to zero; therefore the terms proportional to $-\frac{2\mu G}{k^2}$ are null.
\beq\begin{split}
(1) =& p_{1\alpha} p_{3\gamma} Tr[\gamma^\alpha \gamma_\mu \gamma^\gamma \gamma_\nu] + m_e^2 Tr[\gamma_\mu \gamma_\nu] \\
=& 4 p_{1\alpha} p_{3\gamma} (g_\mu^\alpha g_\nu^\gamma - g^{\alpha\gamma} g_{\mu\nu} + g_\nu^\alpha g_\mu^\gamma) + 4m_e^2 g_{\mu\nu} \\
=& 4 [p_{1\mu} p_{3\nu} + p_{1\nu} p_{3\mu} + (m_e^2 - p_1 p_3) g_{\mu\nu}]
\end{split}\eeq
Trace $(7)$ is equal to trace $(1)$ we have just computed, so we can immediately write
\beq
(7) = 4 [p_1^\rho p_3^\sigma + p_1^\sigma p_3^\rho + (m_e^2 - p_1 p_3) g^{\rho\sigma}]
\eeq
The remaining traces are
\beq\begin{split}
(8) =& p_{2\beta} p_{4\delta} Tr[\gamma^\beta \gamma_\rho \gamma^\delta \gamma_\sigma] \\
=& 4 p_{2\beta} p_{4\delta} (g_\rho^\beta g_\sigma^\delta - g^{\beta\delta} g_{\rho\sigma} + g_\sigma^\beta g_\rho^\delta) \\
=& 4 [p_{2\rho} p_{4\sigma} + p_{2\sigma} p_{4\rho} - p_2 p_4 g_{\rho\sigma}]
\end{split}\eeq
\beq\begin{split}
(2)=& p_{2\beta} p_{4\delta} k_\sigma k_\rho Tr[\gamma^\beta \gamma^\sigma \gamma^\mu \gamma^\delta \gamma^\nu \gamma^\rho] + p_{2\beta} p_{4\delta} k^\mu k^\nu Tr[\gamma^\beta \gamma^\delta] \\
=& 4 p_{2\beta} p_{4\delta} k_\sigma k_\rho \bigg[g^{\beta\sigma} (g^{\mu\delta} g^{\nu\rho} - g^{\mu\nu} g^{\delta\rho} + g^{\mu\rho} g^{\delta\nu}) - g^{\beta\mu} (g^{\sigma\delta} g^{\nu\rho} - g^{\sigma\nu} g^{\delta\rho} + g^{\sigma\rho} g^{\delta\nu}) + \\
&\quad\quad\quad\quad\quad\quad + g^{\beta\delta} (g^{\sigma\mu} g^{\nu\rho} - g^{\sigma\nu} g^{\mu\rho} + g^{\sigma\rho} g^{\mu\nu}) - g^{\beta\nu} (g^{\sigma\mu} g^{\delta\rho} - g^{\sigma\delta} g^{\mu\rho} + g^{\sigma\rho} g^{\mu\delta}) + \\
&\quad\quad\quad\quad\quad\quad + g^{\beta\rho} (g^{\sigma\mu} g^{\delta\nu} - g^{\sigma\delta} g^{\mu\nu} + g^{\sigma\nu} g^{\mu\delta}) \bigg] + 4 p_{2\beta} p_{4\delta} k^\mu k^\nu g^{\beta\delta} \\
=& 4 \bigg[(p_2 k) \bigg[p_4^\mu k^\nu - (p_4 k) g^{\mu\nu} + k^\mu p_4^\nu \bigg] - p_2^\mu \bigg[(p_4 k) k^\nu - (p_4 k) k^\nu + k^2 p_4^\nu \bigg] + \\
&\quad + (p_2 p_4) \bigg[k^\mu k^\nu - k^\mu k^\nu + k^2 g^{\mu\nu} \bigg] - p_2^\nu \bigg[(p_4 k) k^\mu - (p_4 k) k^\mu + k^2 p_4^\mu \bigg] + \\
&\quad + (p_2 k) \bigg[k^\mu p_4^\nu - (p_4 k) g^{\mu\nu} + k^\nu p_4^\mu \bigg] \bigg] + 4 (p_2 p_4) k^\mu k^\nu \\
=& 4\bigg[2 (p_2 k) \bigg[p_4^\mu k^\nu + p_4^\nu k^\mu - (p_4 k) g^{\mu\nu} \bigg] - k^2 \bigg[p_2^\mu p_4^\nu + p_2^\nu p_4^\mu - (p_2 p_4) g^{\mu\nu} \bigg] + (p_2 p_4) k^\mu k^\nu \bigg]
\end{split}\eeq
Carrying out the products $(1) \times (2)$ and $(7) \times (8)$, the matrix element becomes
\beq\begin{split}
|\mathcal{M}|^2=& \frac{16\mu^2}{k^4} \bigg\{ 4 (p_2 k) (p_1 p_4) (p_3 k) + 4 (p_2 k) (p_1 k) (p_3 p_4) + 2 (p_2 p_4) (p_1 k) (p_3 k) + \\
&\quad\quad\quad - k^2 \bigg[2 (p_1 p_2) (p_3 p_4) + 2 (p_1 p_4) (p_2 p_3) + (p_1 p_3) (p_2 p_4) \bigg] + \\
&\quad\quad\quad + m_e^2 \bigg[3 k^2 (p_2 p_4) - 4 (p_2 k) (p_4 k) \bigg] \bigg\} + \\
&+ 8 G^2 \bigg\{ (p_1 p_2) (p_3 p_4) + (p_1 p_4) (p_2 p_3) - m_e^2 (p_2 p_4) \bigg\}
\end{split}\eeq
In the centre of mass frame we can write the 4-momenta as
\beq\begin{cases}
p_1 = (E_1,|\textbf{p}|\hat{z}) \\
p_2 = (E_2,-|\textbf{p}|\hat{z}) \\
p_3 = (E_3, \textbf{q}) \\
p_4 = (E_4,-\textbf{q}) \\
k=p_1-p_3\\
\end{cases}\eeq
Now $E_4 = E_2$, since
\beq\begin{split}
p_3^2 &= m_e^2 = (p_1 + p_2 - p_4)^2 = \underbrace{p_1^2}_{=m_e^2} + \underbrace{p_2^2}_{=0} + \underbrace{p_4^2}_{=0} + 2 p_1 p_2 - 2 p_4 \cdot (p_1 + p_2) = \\
&= m_e^2 + 2 (E_1 E_2 + \underbrace{|\textbf{p}|^2}_{= E_2^2} ) - 2 E_4 (E_1 + E_2) \\
&\implies 0 = (E_2 - E_4) (E_1 + E_2) \implies E_4 = E_2
\end{split}\eeq
Because of conservation of energy it also follows that $E_3 = E_1$; moreover, $\abs{\textbf{p}}=\abs{\textbf{q}}=E_2$. The scalar products are
\beq\begin{split}
&p_1 p_4 = E_2 (E_1 + E_2 \cos{\theta}) \\
&p_3 p_4 = E_2 (E_1 + E_2) \\
&p_2  p_4 = E_2^2 (1 - \cos{\theta}) \\
&p_1 p_2 = E_2 (E_1 + E_2) \\
&p_2 p_3 = E_2 (E_1 + E_2 \cos{\theta}) \\
&p_1 p_3 = E_1^2 - E_2^2 \cos{\theta} \\
&p_1 k = -E_2^2 (1 - \cos{\theta}) \\
&p_2 k = E_2^2 (1 - \cos{\theta}) \\
&p_3 k = E_2^2 (1 - \cos{\theta}) \\
&p_4 k = -E_2^2 (1 - \cos{\theta}) \\
&k^2 = 2 (E_1^2 - E_2^2) (1 - \cos{\theta})
\end{split}\eeq
and, by defining $\chi:=\frac{E_1}{E_2}$, the matrix element becomes
\beq\begin{split}\label{M_fermi}
|\mathcal{M}|^2=& 8 \mu^2 E_2^2 \frac{\chi (2 \chi^3 + 4 \chi^2 + \chi - 4) + 2 (\chi^4 + 2 \chi^3 - 2 \chi^2 - 2 \chi - 2) \cos{\theta} + 3 \chi^2 \cos^2{\theta}}{(1 - \chi^2) (1 - \cos{\theta})} + \\
&+ 8 G^2 E_2^4 \bigg[1 + (1 + \chi)^2 + [(1 + \chi)^2 - 2] \cos{\theta} + \cos^2{\theta} \bigg]
\end{split}\eeq
The flux factor is
\beq
\mathcal{J} = \frac{1}{4E_2^2 (\chi + 1)}
\eeq
while the 2-body phase space element is
\beq
d \Phi_2 = \frac{1}{\chi + 1} \frac{d \cos{\theta} d \varphi}{16\pi^2}
\eeq
Given that, once having integrated over $d\varphi$, the differential cross section for the process becomes
\beq
\frac{d \sigma}{d \cos{\theta}} = \frac{1}{32\pi E_2^2 (\chi + 1)^2} |\mathcal{M}|^2
\eeq
where $|\mathcal{M}|^2$ is given by \eqref{M_fermi}.
\end{sol}

\newpage
\section{QED with massless electron: $\gamma e^- \to \gamma e^-$}
\begin{ex}\label{e_massless}
Consider the standard QED Lagrangian and assume the electron to be massless. Compute the differential cross section for the process
\[
\gamma e^- \to \gamma e^-
\]
\end{ex}
\begin{sol}
Let's consider the initial particles 4-momenta $p_j$'s and the final ones $q_j$'s in the centre of mass reference frame.
\beq
p_1 = \begin{pmatrix} E_e \\ \textbf{p} \end{pmatrix}
\eeq
\beq
p_2 = \begin{pmatrix} E_\gamma \\ -\textbf{p} \end{pmatrix}
\eeq
\beq
q_1 = \begin{pmatrix} E_{e'} \\ -\textbf{q} \end{pmatrix}
\eeq
\beq
q_2 = \begin{pmatrix} E_{\gamma'} \\ \textbf{q} \end{pmatrix}
\eeq
The assumption of a massless electron results in the square of every 4-momentum to be null: $p_1^2 = 0$, $p_2^2 = 0$, $q_1^2 = 0$, $q_2^2 = 0$. Moreover, because of the mass-shell relation, $E_e = E_\gamma = E$, $E_{e'} = E_{\gamma'} = E'$, so that
\beq
p_1 = \begin{pmatrix} E \\ \textbf{p} \end{pmatrix}
\eeq
\beq
p_2 = \begin{pmatrix} E\\ -\textbf{p} \end{pmatrix}
\eeq
\beq
q_1 = \begin{pmatrix} E' \\ -\textbf{q} \end{pmatrix}
\eeq
\beq
q_2 = \begin{pmatrix} E' \\ \textbf{q} \end{pmatrix}
\eeq
The Feynman rules are the usual ones for QED. We proceed building the relevant diagrams for the process
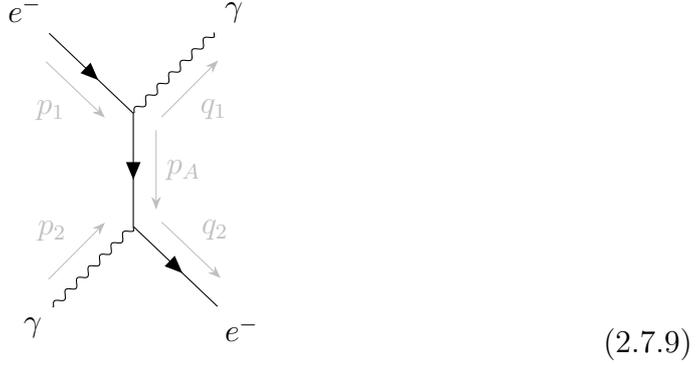
\beq
\begin{tikzpicture}
\begin{feynman}
	\vertex (a);
	\vertex[above left=of a] (b){$e^-$};
	\vertex[above right=of a] (c){$\gamma$};
	\vertex[below=of a] (d);
	\vertex[below left=of d] (e){$\gamma$};
	\vertex[below right=of d] (f){$e^-$};

	\diagram {
	(b) -- [fermion, momentum'={ [arrow style=lightgray] $p_1$ } ] (a),
	(a) -- [boson, momentum'={ [arrow style=lightgray] $q_1$ } ] (c),
	(a) -- [fermion, momentum={ [arrow style=lightgray] $p_A$ } ] (d),
	(e) -- [boson, momentum={ [arrow style=lightgray] $p_2$ } ] (d),
	(d) -- [fermion, momentum={ [arrow style=lightgray] $q_2$ } ] (f),
	};
\end{feynman}
\end{tikzpicture}
\eeq
\beq
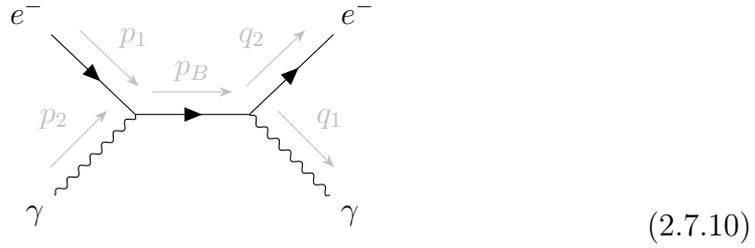
\begin{tikzpicture}
\begin{feynman}
	\vertex (a);
	\vertex[above left=of a] (b){$e^-$};
	\vertex[below left=of a] (c){$\gamma$};
	\vertex[right=of a] (d);
	\vertex[above right=of d] (e){$e^-$};
	\vertex[below right=of d] (f){$\gamma$};

	\diagram {
	(b) -- [fermion,  momentum= { [arrow style=lightgray] $p_1$ } ] (a),
	(c) -- [boson, momentum={ [arrow style=lightgray] $p_2$ } ] (a),
	(a) -- [fermion, momentum={ [arrow style=lightgray] $p_B$ } ] (d),
	(d) -- [fermion, momentum={ [arrow style=lightgray] $q_2$ } ] (e),
	(d) -- [boson, momentum={ [arrow style=lightgray] $q_1$ } ] (f),
	};
\end{feynman}
\end{tikzpicture}
\eeq
Thanks to 4-momentum conservation we have
\beq
p_A = p_1 - q_1 = q_2 - p_2
\eeq
\beq
p_B = p_1 + p_2 = q_1 + q_2
\eeq
Calling $\mathcal{A}$ the amplitude of the first diagram and $\mathcal{B}$ the amplitude of the second one, we have
\beq
\mathcal{A} = ie^2 \overline u^{s_2} \gamma^\nu \frac{\bcc p_A \ecc}{p_A ^2} \gamma^\mu u^{s_1} \varepsilon^{\lambda_2}_\nu \bigg( \varepsilon^{\lambda_1}_\mu \bigg)^*
\eeq
\beq
\mathcal{B} = ie^2 \overline u^{s_2} \gamma^\nu \frac{\bcc p_B \ecc}{p_B^2} \gamma^\mu u^{s_1} \varepsilon^{\lambda_2}_\mu \bigg( \varepsilon^{\lambda_1}_\nu \bigg)
\eeq
The momenta $p_A^2$ and $p_B^2$ are
\beq
p_A^2=p_1^2+q_1^2-2p_1 q_1=-2p_1q_1
\eeq
\beq
p_B^2=p_1^2+p_2^2+2p_1 p_2=2p_1p_2
\eeq
where $p_1^2 = p_2^2 = q_1^2 = 0$ since we're dealing with massless particles. Given that the amplitudes become
\beq
\mathcal{A} = -ie^2 \overline u^{s_2} \gamma^\nu \frac{{\bcc p \ecc}_1 - {\bcc q \ecc}_1}{2p_1 q_1} \gamma^\mu u^{s_1} \varepsilon^{\lambda_2}_\nu \bigg( \varepsilon^{\lambda_1}_\mu \bigg)
\eeq
\beq
\mathcal{B} = ie^2 \overline u^{s_2} \gamma^\nu \frac{{\bcc p \ecc}_1 + {\bcc p \ecc}_2}{2 p_1 p_2} \gamma^\mu u^{s_1} \varepsilon^{\lambda_2}_\mu \bigg( \varepsilon^{\lambda_1}_\nu \bigg)
\eeq
For the total scattering amplitude we have
\beq
|\mathcal{M}|^2 = \frac{1}{4} \sum_{s_1, s_2, \lambda_1, \lambda_2} |\mathcal{A} + \mathcal{B}|^2 = \frac{1}{4} \sum_{s_1, s_2, \lambda_1, \lambda_2} |\mathcal{A}|^2 + |\mathcal{B}|^2 + \mathcal{A}^*\mathcal{B} + \mathcal{A}\mathcal{B}^*
\eeq
where the factor $\frac{1}{4}$ comes from averaging over the initial spins. \\Let's start by computing the term $\sum_{s_1, s_2, \lambda_1, \lambda_2} |\mathcal{A}|^2$
\beq
\sum_{s_1, s_2, \lambda_1, \lambda_2} |\mathcal{A}|^2 = \frac{e^4}{4 (p_1q_1)^2} g_{\nu \beta} g_{\mu \alpha} \sum_{s_1, s_2} \overline u^{s_2} \gamma^\nu ({\bcc p \ecc}_1 - {\bcc q \ecc}_1) \gamma^\mu u^{s_1} \overline u^{s_1} \gamma^\alpha ({\bcc p \ecc}_1 - {\bcc q \ecc}_1) \gamma^\beta u^{s_2}
\eeq
where we have already summed over the polarizations. Contracting the metric tensors we obtain the following trace
\beq
\sum_{s_1, s_2, \lambda_1, \lambda_2} |\mathcal{A}|^2 = \frac{e^4}{4 (p_1q_1)^2} Tr\bigg[ {\bcc q \ecc}_2 \gamma^\nu ({\bcc p \ecc}_1 - {\bcc q \ecc}_1) \gamma^\mu {\bcc p \ecc}_1 \gamma_\mu ({\bcc p \ecc}_1 - {\bcc q \ecc}_1) \gamma_\nu \bigg]
\eeq
Thanks to equation \eqref{id_2gamma} we obtain the sum of four traces, three of which are proportional to terms of the type ${\bcc p \ecc}_1 {\bcc p \ecc}_1 = p_1^2 = 0$; we are then left with
\beq
\sum_{s_1, s_2, \lambda_1, \lambda_2} |\mathcal{A}|^2 = \frac{-2e^4}{4 (p_1q_1)^2} Tr\bigg[ {\bcc q \ecc}_2 \gamma^\nu {\bcc q \ecc}_1 {\bcc p \ecc}_1 {\bcc q \ecc}_1 \gamma_\nu \bigg]
\eeq
Thanks to identity \eqref{id_4gamma} we get
\beq\begin{split}
\sum_{s_1, s_2, \lambda_1, \lambda_2} |\mathcal{A}|^2 &= \frac{4e^4}{4 (p_1q_1)^2} Tr\bigg[ \bcc q_2\ecc {\bcc q \ecc}_1 {\bcc p \ecc}_1 {\bcc q \ecc}_1 \bigg] = \\
&= \frac{4 e^4}{(p_1 q_1)^2} q_{2 \mu}q_{1 \nu} p_{1\alpha} q_{1\beta}(g^{\mu\nu} g^{\alpha\beta} - g^{\mu\alpha} g^{\nu\beta} + g^{\mu\beta} g^{\nu\beta})
\end{split}\eeq
Therefore
\beq\begin{split}
\sum_{s_1,s_2, \lambda_1, \lambda_2} |\mathcal{A}|^2 &= \frac{4e^4}{(p_1q_1)^2} \bigg( (q_1 q_2) (p_1q_1) - (q_2p_1) q_1^2 + (q_2q_1) (q_1p_1) \bigg) = \\
&= \frac{8e^4}{(p_1q_1)^2} (q_1 q_2) (p_1q_1) = \frac{8e^4}{(p_1q_1)} (q_1 q_2)
\end{split}\eeq
Note that the summation for $|\mathcal{B}|^2$ can be obtained from the above by simply exchanging $-q_1$ with $+p_2$
\beq
\sum_{s_1, s_2, \lambda_1, \lambda_2} |\mathcal{B}|^2 = \frac{8e^4}{(p_1 p_2)} (p_2 q_2)
\eeq
All that is left to do is to compute the terms $\mathcal{A}^*\mathcal{B}$ e $\mathcal{A}\mathcal{B}^*$
\beq
\mathcal{A}\mathcal{B}^* = -e^4 \overline u^{s_2} \gamma^\nu \frac{{\bcc p \ecc}_1 - {\bcc q \ecc}_1}{2 p_1 q_1} \gamma^\mu u^{s_1} \varepsilon^{\lambda_2}_\nu \bigg( \varepsilon^{\lambda_1} _\mu \bigg)^* \overline u^{s_1} \gamma^\alpha \frac{{\bcc p \ecc}_1 + {\bcc p \ecc}_2}{2 p_1 p_2} \gamma^\beta u^{s_2} \bigg( \varepsilon^{\lambda_2}_\alpha \bigg)^* \varepsilon^{\lambda_1}_\beta
\eeq
\beq
\mathcal{A}^*\mathcal{B} = -e^4 \overline u^{s_2} \gamma^\nu \frac{{\bcc p \ecc}_1 + {\bcc p \ecc}_2}{2 p_1 p_2} \gamma^\mu u^{s_1} \varepsilon^{\lambda_2}_\mu \bigg( \varepsilon^{\lambda_1}_\nu \bigg)^* \overline u^{s_1} \gamma^\alpha \frac{{\bcc p \ecc}_1 - {\bcc q \ecc}_1}{2 p_1 q_1} \gamma^\beta u^{s_2} \bigg( \varepsilon^{\lambda_2}_\beta \bigg)^* \varepsilon^{\lambda_1}_\alpha
\eeq
\beq\begin{split}
\sum_{s_1, s_2, \lambda_1, \lambda_2} \mathcal{A}\mathcal{B}^* = -e^4 \sum_{s_1, s_2, \lambda_1, \lambda_2}& \overline u^{s_2} \gamma^\nu \frac{{\bcc p \ecc}_1 - {\bcc q \ecc}_1}{2 p_1 q_1} \gamma^\mu u^{s_1} \varepsilon^{\lambda_2}_\nu \bigg( \varepsilon^{\lambda_1} _\mu \bigg)^* \\
&\overline u^{s_1} \gamma^\alpha \frac{{\bcc p \ecc}_1 + {\bcc p \ecc}_2}{2 p_1 p_2} \gamma^\beta u^{s_2} \bigg( \varepsilon^{\lambda_2}_\alpha \bigg)^* \varepsilon^{\lambda_1}_\beta
\end{split}\eeq
Making explicit the summations on the polarization indices we get two metric tensors
\beq\begin{split}
\sum_{s_1, s_2, \lambda_1, \lambda_2} \mathcal{A}\mathcal{B}^* &= -e^4 \sum_{s_1, s_2} \overline u^{s_2} \gamma^\nu \frac{{\bcc p \ecc}_1 - {\bcc q \ecc}_1}{2 p_1 q_1} \gamma^\mu u^{s_1} \overline u^{s_1} \gamma^\alpha \frac{{\bcc p \ecc}_1 + {\bcc p \ecc}_2}{2 p_1 p_2} \gamma^\beta u^{s_2} g_{\alpha \nu} g_{\beta \mu} \\
&= -\frac{e^4}{4 (p_1q_1)(p_1 p_2)} \sum_{s_1, s_2} \overline u^{s_2} \gamma^\nu ({\bcc p \ecc}_1 - {\bcc q \ecc}_1) \gamma^\mu u^{s_1} \overline u^{s_1} \gamma_\nu ({\bcc p \ecc}_1 + {\bcc p \ecc}_2) \gamma_\mu u^{s_2} 
\end{split}\eeq
and summing over the spins we arrive to the following trace
\beq
\sum_{s_1, s_2, \lambda_1, \lambda_2} \mathcal{A}\mathcal{B}^* = \frac{-e^4}{4 (p_1q_1) (p_1p_2)} Tr\bigg[ {\bcc q \ecc}_2 \gamma^\nu ({\bcc p \ecc}_1 - {\bcc q \ecc}_1) \gamma^\mu {\bcc p \ecc}_1 \gamma_\nu ({\bcc p \ecc}_1 + {\bcc p \ecc}_2) \gamma_\mu \bigg]
\eeq
Using again \eqref{id_4gamma} we obtain
\beq
\sum_{s_1, s_2, \lambda_1, \lambda_2} \mathcal{A}\mathcal{B}^* = \frac{2e^4}{4 (p_1q_1) (p_1p_2)} Tr\bigg[ {\bcc q \ecc}_2 {\bcc p \ecc}_1 \gamma^\mu ({\bcc p \ecc}_1 - {\bcc q \ecc}_1) ({\bcc p \ecc}_1 + {\bcc p \ecc}_2) \gamma_\mu \bigg]
\eeq
Applying now \eqref{id_3gamma} we have
\beq\begin{split}
\sum_{s_1, s_2, \lambda_1, \lambda_2} \mathcal{A}\mathcal{B}^* &= \frac{2e^4}{4 (p_1 q_1) (p_1 p_2)} Tr\bigg[ {\bcc q \ecc}_2 {\bcc p \ecc}_1 \cdot 4 (p_1 - q_1) (p_1 + p_2) \bigg] = \\
&= \frac{8e^4}{(p_1 q_1) (p_1 p_2)}(q_2 p_1) (p_1 - q_1) (p_1 + p_2) = \\
&= \frac{8e^4}{(p_1 q_1) (p_1 p_2)} (p_1 q_2)\bigg[ (p_1 p_2) - (q_1 p_1) - (q_1 p_2) \bigg]
\end{split}\eeq
In order to compute $\sum \mathcal{A}^*\mathcal{B}$ it is sufficient to make the following substitutions: $-q_1 \rightarrow p_2$, $p_2 \rightarrow -q_1$. We then obtain
\beq
\sum_{s_1, s_2, \lambda_1, \lambda_2} \mathcal{A}^*\mathcal{B} = \sum_{s_1, s_2, \lambda_1, \lambda_2} \mathcal{A}\mathcal{B}^* = \frac{8e^4}{(p_1 q_1) (p_1 p_2)} (p_1 q_2) \bigg[ (p_1 p_2) - (q_1 p_1) - (q_1 p_2) \bigg]
\eeq
Putting all the pieces together, and remembering the factor $\frac{1}{4}$ in front of the summation, we have
\beq\begin{split}
|\mathcal{M}|^2 = \frac{1}{4} \bigg\{& \frac{8e^4}{(p_1 q_1)} (q_1 q_2) + \frac{8e^4}{(p_1 p_2)} (p_2 q_2) + \\
&+ \frac{16e^4}{(p_1 q_1) (p_1 p_2)} (p_1 q_2) \big[ (p_1 p_2) - (q_1 p_1) - (q_1 p_2) \big] \bigg\}
\end{split}\eeq
and finally
\beq
|\mathcal{M}|^2 = \frac{2e^4}{(p_1 q_1) (p_1 p_2)} \bigg\{ (q_1 q_2) (p_1 p_2) + (p_2 q_2) (p_1 q_1) + 2 (p_1 q_2) [ (p_1 p_2) - (q_1 p_1) - (q_1 p_2) ] \bigg\}
\eeq
As anticipated we put ourselves in the centre of mass frame; imposing 4-momentum conservation, the moduli of the initial and final 3-momenta must be equal to each other, and so thanks to the mass shell relations we have $E' = E$. Our calculations simplify considerably, so we can write
\beq
|\mathcal{M}|^2 = \frac{e^4}{1 + \cos{\theta}} \big[ 4 + (1 + \cos{\theta})^2 \big]
\eeq
The expression for the flux factor $\mathcal{J}$ is
\beq
\mathcal{J} = \frac{1}{4\sqrt{(p_1 p_2)^2} } = \frac{1}{8E^2}
\eeq
and the 2-body phase space element is
\beq\begin{split}
d\Phi_2 &= \frac{|\Vec{q_1}|}{2E} \frac{d \cos{\theta} d\mathcal{\phi} }{16 \pi^2} = \frac{1}{2} \frac{d \cos{\theta} d\varphi}{16 \pi^2} = \\
&= \frac{1}{2} \frac{d \cos{\theta} }{8\pi}
\end{split}\eeq
Therefore, the differential cross section of the process becomes
\beq
\frac{d\sigma}{d\cos{\theta}} = \frac{1}{128} \frac{e^4}{1 + \cos{\theta} } [4 + (1 + \cos{\theta})^2]
\eeq
\end{sol}

\newpage
\section{Chiral lagrangian: $\psi_a \psi_b \to \psi_a \psi_b$}
\begin{ex}\label{gamma5}
Consider the following lagrangian density
\[
\mathcal{L} = \overline \psi_a ( \partial_\mu \gamma^\mu - m_a ) \psi_a + \overline \psi_b \partial_\mu \gamma^\mu \psi_b + \Lambda \overline \psi_a \gamma_\mu ( g_V \mathbb{1} + g_A \gamma_5 ) \psi_a \overline \psi_b \gamma^\mu ( \mathbb{1} - \gamma_5 ) \psi_b
\]
where $\psi_a$ and $\psi_b$ are associated to two different fermionic species. Derive the Feynman rules of the theory and compute the cross section for the process
\[
\psi_a \psi_b \to \psi_a \psi_b
\]
\end{ex}
\begin{sol}
The Lagrangian in question is made up by two Dirac Lagrangians, one for the massive fermions $a$ and the other for the massless fermions $b$, plus the following interaction term
\beq
\mathcal{L}_{INT} = \Lambda \overline \psi_a \gamma^\mu (g_V \mathbb{1} + g_A \gamma^5) \psi_a \overline \psi_b \gamma_\mu (\mathbb{1} - \gamma^5) \psi_b
\eeq
The interaction Lagrangian represents a contact term, since there's no propagator there. To the usual QED Feynman rules we add the following vertex factor
\beq
\begin{tikzpicture}
\begin{feynman}
	\node at (5.5, -1.5) {$=i\Lambda\gamma^\mu(g_V\,\mathbb{1}+g_A\gamma^5)\gamma_\mu(\mathbb{1}-\gamma^5)$};
	\vertex(a) at (0,0) {$\psi_a$} ;
	\vertex(b) at (0,-3) {$\psi_b$};
	\vertex(c) at (1.5,-1.5);
	\vertex(d) at (3,0) {$\psi_a$};
	\vertex(e) at (3,-3) {$\psi_b$};

	\diagram* {
	(a)--[fermion] (c),
	(b)--[fermion] (c),
	(c)--[fermion] (d),
	(c)--[fermion] (e),
	};

\end{feynman}
\end{tikzpicture}
\eeq
where, as usual, we added the factor $i$. At first order the process is uniquely described by the following diagram
\beq
\begin{tikzpicture}
\begin{feynman}
	\node at (5.5, -1.5);
	\vertex(a) at (0,0){$\psi_a$};
	\vertex(b) at (0,-3){$\psi_b$};
	\vertex(c) at (1.5,-1.5);
	\vertex(d) at (3,0){$\psi_a$};
	\vertex(e) at (3,-3){$\psi_b$};

	\diagram* {
	(a)--[fermion, momentum={[arrow shorten=0.3, arrow style=lightgray] \small$p_1$ } ] (c),
	(b)--[fermion,  momentum={[arrow shorten=0.3, arrow style=lightgray] \small$p_2$ } ] (c),
	(c)--[fermion,  momentum'={[arrow shorten=0.3, arrow style=lightgray] \small$p_3$ } ] (d),
	(c)--[fermion,  momentum'={[arrow shorten=0.3, arrow style=lightgray] \small$p_4$ } ] (e),
	};

\end{feynman}
\end{tikzpicture}
\eeq
The scattering amplitude is
\beq
\mathcal{A} = i \Lambda \bar u^{s_3}(p_3) \gamma^\mu (g_V \mathbb{1} + g_A \gamma^5) u^{s_1}(p_1) \bar u^{s_4}(p_4) \gamma_\mu (\mathbb{1} - \gamma^5) u^{s_2}(p_2)
\eeq
with complex conjugate equal to
\beq
\mathcal{A}^* = -i \Lambda \bar u^{s_2}(p_2) (\mathbb{1} + \gamma^5) \gamma_\nu u^{s_4}(p_4) \bar u^{s_1}(p_1) (g_V\mathbb{1} - g_A\gamma^5) \gamma^\nu u^{s_3}(p_3)
\eeq
As usual, in computing unpolarized cross sections, we have to average on the initial spins and sum over the final ones
\beq\begin{split}
|\mathcal{M}|^2 = \frac{\Lambda^2}{4} \sum_{s_1, s_2, s_3, s_4}& \bar u^{s_3} \gamma^\mu (g_V \mathbb{1} + g_A \gamma^5) u^{s_1} \bar u^{s_4} \gamma_\mu (\mathbb{1} - \gamma^5) u^{s_2} \\
&\bar u^{s_2} (\mathbb{1} + \gamma^5) \gamma_\nu u^{s_4} \bar u^{s_1} (g_V \mathbb{1} - g_A \gamma^5) \gamma^\nu u^{s_3}
\end{split}\eeq
Using the Dirac spinors sum rules
\beq\begin{split}
|\mathcal{M}|^2 =& \frac{\Lambda^2}{4} Tr \bigg[ ({\bcc p \ecc}_1 + m_a)(g_V - g_A \gamma^5) \gamma^\nu ({\bcc p \ecc}_3 + m_a) \gamma^\mu (g_V + g_A \gamma^5) \bigg] \\ 
&Tr \bigg[ {\bcc p \ecc}_2 (1 + \gamma_5) \gamma_\nu \bcc p_4\ecc \gamma_\mu (1 - \gamma_5) \bigg]
\end{split}\eeq
An explicit calculation of all the above traces is excessively long and wasteful; our aim is to manipulate the former expression exploiting some properties of the Dirac matrices. \\Let us consider the first trace
\beq
Tr \bigg[ ({\bcc p \ecc}_1 + m_a)(g_V - g_A \gamma^5) \gamma^\nu ({\bcc p \ecc}_3 + m_a) \gamma^\mu (g_V + g_A \gamma^5) \bigg] =: T_1
\eeq
Using the cyclic property of the trace \eqref{cyclic}, together with \eqref{anticom_gamma5}, we can write $T_1$ as follows
\beq\begin{split}
T_1 =& Tr \bigg[ (g_V + g_A \gamma^5)({\bcc p \ecc}_1 + m_a)(g_V - g_A \gamma^5) \gamma^\nu ({\bcc p \ecc}_3 + m_a) \gamma^\mu \bigg] = \\
=& Tr \bigg[ \bigg( m_a (g_V + g_A \gamma^5) + {\bcc p \ecc}_1 (g_V - g_A \gamma^5) \bigg) (g_V - g_A \gamma^5) \gamma^\nu ({\bcc p \ecc}_3 + m_a \gamma^\mu ) \bigg] = \\
=& m_a Tr \bigg[ (g_V + g_A \gamma^5) (g_V - g_A \gamma^5) \gamma^\nu ({\bcc p \ecc}_3 + m_a) \gamma^\mu \bigg] + \\
&\quad + Tr \bigg[ {\bcc p \ecc}_1 (g_V - g_A \gamma^5)^2 \gamma^\nu ({\bcc p \ecc}_3 + m_a) \gamma^\mu \bigg] = \\
=& (g_V^2 - g_A^2) m_a Tr \bigg[ \gamma^\nu ({\bcc p \ecc}_3 + m_a) \gamma^\mu \bigg] + \\
&\quad + Tr \bigg[ {\bcc p \ecc}_1 (g_V^2 + g_A^2 - 2 g_V g_A \gamma^5) \gamma^\nu ({\bcc p \ecc}_3 + m_a) \gamma^\mu \bigg]
\end{split}\eeq
Defining the following constants
\beq\begin{cases}
G^+ := g_V^2+g_A^2 \\
G^- := g_V^2-g_A^2 \\
G_{VA} := g_V g_A \\
\end{cases}\eeq
we obtain a more compact expression
\beq\begin{split}
T_1 =& G^- m_a Tr \bigg[ \gamma^\nu ({\bcc p \ecc}_3 + m_a) \gamma^\mu \bigg] + Tr \bigg[ {\bcc p \ecc}_1 (G^+ - 2 G_{VA} \gamma^5) \gamma^\nu ({\bcc p \ecc}_3 + m_a) \gamma^\mu \bigg] = \\
=& G^- m_a^2 Tr \bigg[ \gamma^\mu \gamma^\nu \bigg] + G^+ Tr \bigg[{\bcc p \ecc}_1 \gamma^\nu ({\bcc p \ecc}_3 + m_a) \gamma^\mu \bigg] - \\
&\quad - 2 G_{VA} Tr \bigg[ {\bcc p \ecc}_1 \gamma^5 \gamma^\nu ({\bcc p \ecc}_3 + m_a) \gamma^\mu \bigg] = \\
=& 4G^- m_a^2 g^{\mu\nu} + G^+ p_{1\alpha} p_{3\gamma} Tr \bigg[ \gamma^\alpha \gamma^\nu \gamma^\gamma \gamma^\mu \bigg] - 2G_{VA} p_{1\alpha} p_{3\gamma} Tr \bigg[ \gamma^\alpha \gamma^5 \gamma^\nu \gamma^\gamma \gamma^\mu \bigg]
\end{split}\eeq
The trace of an odd number of gamma matrices is null, as well as a trace containing $\gamma^5$ multiplied by an odd number of gamma matrices, so that
\beq
T_1 = 4G^- m_a^2 g^{\mu\nu} + 4G^+ p_{1\alpha} p_{3\gamma} (g^{\alpha\nu} g^{\gamma\mu} - g^{\alpha\gamma} g^{\nu\mu} + g^{\alpha\mu} g^{\nu\gamma} ) - 2G_{VA} p_{1\alpha} p_{3\gamma} Tr \bigg[ \gamma^5 \gamma^\nu \gamma^\gamma \gamma^\mu \gamma^\alpha \bigg]
\eeq
and using $Tr\bigg[ \gamma^5 \gamma^\nu \gamma^\gamma \gamma^\mu \gamma^\alpha \bigg] = -4i \varepsilon^{\nu \gamma \mu \alpha}$ we obtain
\beq\begin{split}
T_1 =& 4G^- m_a^2 g^{\mu\nu} + 4G^+ \bigg( p_1^\nu p_3^\mu - (p_1 p_3) g^{\mu\nu} + p_1^\mu p_3^\nu \bigg) - 8 i G_{VA}  p_{1\alpha} p_{3\gamma} \varepsilon^{\nu \gamma \mu \alpha} = \\
=& 4 \bigg[ \bigg(G^- m_a^2 - G^+ (p_1 p_3) \bigg) g^{\mu\nu} + G^+ \bigg(p_1^\nu p_3^\mu + p_1^\mu p_3^\nu \bigg) \bigg] - 8 i G_{VA} p_{1\alpha} p_{3\gamma} \varepsilon^{\nu \gamma \mu \alpha}
\end{split}\eeq
In the expression above we can notice that the first term in square brackets is symmetric in the indices $\mu$ and $\nu$, while the second term is antisymmetric in the same indices (due to the presence of the totally antisymmetric Levi-Civita symbol). This allows us to define the following tensors
\beq\begin{cases}
S^{\mu\nu} := \bigg[ \bigg(G^- m_a^2 - G^+ (p_1 p_3) \bigg) g^{\mu\nu} + G^+ \bigg(p_1^\nu p_3^\mu + p_1^\mu p_3^\nu \bigg) \bigg] \\
A^{\mu\nu} := G_{VA} p_{1\alpha} p_{3\gamma} \varepsilon^{\nu \gamma \mu \alpha}
\end{cases}\eeq
The trace $T_1$ now takes the form
\beq
T_1 = 4S^{\mu\nu} - 8 i A^{\mu\nu}
\eeq
We can handle the second trace with similar manipulations
\beq\begin{split}
T_2 :=& Tr \bigg[ {\bcc p \ecc}_2 (1 + \gamma^5) \gamma_\nu {\bcc p \ecc}_4 \gamma_\mu (1 - \gamma^5) \bigg] = Tr \bigg[ (1 - \gamma^5) {\bcc p \ecc}_2 (1 + \gamma^5) \gamma_\nu {\bcc p \ecc}_4 \gamma_\mu \bigg] = \\
=& Tr \bigg[ {\bcc p \ecc}_2 (1 + \gamma^5) (1 + \gamma^5) \gamma_\nu {\bcc p \ecc}_4 \gamma_\mu \bigg] = 2 Tr \bigg[ {\bcc p \ecc}_2 \gamma_\nu {\bcc p \ecc}_4 \gamma_\mu \bigg] + 2 Tr \bigg[ {\bcc p \ecc}_2 \gamma^5 \gamma_\nu {\bcc p \ecc}_4 \gamma_\mu \bigg] = \\
=& 2 p_2^\beta p_4^\delta Tr \bigg[ \gamma_\beta \gamma_\nu \gamma_\delta \gamma_\mu \bigg] + 2 p_2^\beta p_4^\delta Tr \bigg[ \gamma^5 \gamma_\nu \gamma_\delta \gamma_\mu \gamma_\beta \bigg] = \\
=& 8 p_2^\beta p_4^\delta \bigg(g_{\beta\nu} g_{\delta\mu} - g_{\beta\delta} g_{\nu\mu} + g_{\beta\mu} g_{\nu\delta} \bigg) - 8 i p_2^\beta p_4^\delta \varepsilon_{\nu \delta \mu \beta} = \\
=& 8 \bigg[ p_{2\nu} p_{4\mu} - (p_2 p_4) g_{\mu\nu} + p_{2\mu} p_{4\nu} \bigg] - 8 i p_2^\beta p_4^\delta \varepsilon_{\nu \delta \mu \beta}
\end{split}\eeq
By defining
\beq\begin{cases}
S'_{\mu\nu} := p_{2\nu} p_{4\mu} - (p_2 p_4) g_{\mu\nu} + p_{2\mu} p_{4\nu} \\
A'_{\mu\nu} := p_2^\beta p_4^\delta \varepsilon_{\nu \delta \mu \beta} \\
\end{cases}\eeq
we obtain
\beq
T_2 = Tr \bigg[{\bcc p \ecc}_2 (1 + \gamma^5) \gamma_\nu {\bcc p \ecc}_4 \gamma_\mu (1 - \gamma^5) \bigg] = 8 S'_{\mu\nu} - 8 i A'_{\mu\nu}
\eeq
Therefore we can make explicit the matrix element $|\mathcal{M}|^2$
\beq\begin{split}
|\mathcal{M}|^2 =& \frac{\Lambda^2}{4} \bigg[ 4 S^{\mu\nu} - 8 i A^{\mu\nu} \bigg] \bigg[8 S'_{\mu\nu} - 8 i A'_{\mu\nu} \bigg] = \\
=& \frac{\Lambda^2}{4} \bigg[ 32 S^{\mu\nu} S'_{\mu\nu} - 32 i S^{\mu\nu} A'_{\mu\nu} - 64 i A^{\mu\nu} S'_{\mu\nu} - 64 A^{\mu\nu} A'_{\mu\nu} \bigg]
\end{split}\eeq
The second and third term are both null, since they are products of a symmetric tensor which multiplies an antisymmetric one in the same indices. It follows that
\beq
|\mathcal{M}|^2 = 8 \Lambda^2 \bigg[ S^{\mu\nu} S'_{\mu\nu} - 2 A^{\mu\nu} A'_{\mu\nu} \bigg]
\eeq
After some simple algebraic passages, the first term can be rewritten as
\beq\begin{split}
S^{\mu\nu} S'_{\mu\nu} =& -2 \bigg[ G^- m_a^2 - G^+ (p_1 p_3) \bigg] (p_2 p_4) + \\
&+ 2G^+ \bigg[ (p_1 p_2) (p_3 p_4) - (p_1 p_3) (p_2 p_4) + (p_1 p_4) (p_2 p_3) \bigg]
\end{split}\eeq
For the second term we have
\beq\begin{split}
-2 A^{\mu\nu} A'_{\mu\nu} =& -2 \bigg[ G_{VA} p_{1\alpha} p_{3\gamma} \varepsilon^{\nu \gamma \mu \alpha} \bigg] \bigg[ p_2^\beta p_4^\delta \varepsilon_{\nu \delta \mu \beta} \bigg] = \\
=& -2 G_{VA} p_{1\alpha} p_2^\beta p_{3\gamma} p_4^\delta \varepsilon^{\nu \gamma \mu \alpha} \varepsilon_{\nu \delta \mu \beta}
\end{split}\eeq
The product of two Levi-Civita symbols can be simplified in the following way
\beq
\begin{cases}
\varepsilon^{\nu \gamma \mu \alpha} = -\varepsilon^{\nu \mu \gamma \alpha} \\
\varepsilon_{\nu \delta \mu \beta} = -\varepsilon_{\nu \mu \delta \beta}
\end{cases} \Rightarrow \varepsilon^{\nu \gamma \mu \alpha} \varepsilon_{\nu \delta \mu \beta} = \varepsilon^{\nu \mu \gamma \alpha} \varepsilon_{\nu \mu \delta \beta} = -2 (\delta_\delta^\gamma \delta_\beta^\alpha - \delta_\beta^\gamma \delta_\delta^\alpha)
\eeq
so that
\beq\begin{split}
-2 A^{\mu\nu} A'_{\mu\nu} =& 4 G_{VA} p_{1\alpha} p_2^\beta p_{3\gamma} p_4^\delta (\delta_\delta^\gamma \delta_\beta^\alpha - \delta_\beta^\gamma \delta_\delta^\alpha) = \\
=& 4 G_{VA} \bigg[ (p_1 p_2) (p_3 p_4) - (p_1 p_4) (p_2 p_3) \bigg]
\end{split}\eeq
Putting the previous results together we finally obtain 
\beq
|\mathcal{M}|^2 = 16 \Lambda^2 \bigg[ -G^- m_a^2 (p_2 p_4) + (G^+ + 2 G_{VA} ) (p_1 p_2) (p_3 p_4) +(G^+-2G_{VA}) (p_1 p_4) (p_2 p_3) \bigg]
\eeq
In the centre of mass frame the 4-momenta are
\beq\begin{cases}
p_1 = (E_1,|\textbf{p}|\hat{z}) \\
p_2 = (E_2,-|\textbf{p}|\hat{z}) \\
p_3 = (E_3, \textbf{q}) \\
p_4 = (E_4,-\textbf{q}) \\
\end{cases}\eeq
Notice that $E_4 = E_2$, since
\beq\begin{split}
p_3^2 &= m_a^2 = (p_1 + p_2 - p_4)^2 = \underbrace{p_1^2}_{=m_a^2} + \underbrace{p_2^2}_{=0} + \underbrace{p_4^2}_{=0} + 2 p_1 p_2 - 2 p_4 \cdot (p_1 + p_2) = \\
&= m_a^2 + 2 (E_1 E_2 + \underbrace{|\textbf{p}|^2}_{= E_2^2} ) - 2 E_4 (E_1 + E_2) \implies \\
&\implies 0 = (E_2 - E_4) (E_1 + E_2) \implies E_4 = E_2
\end{split}\eeq
Thanks to energy conservation we also have $E_3 = E_1$. The various scalar products are
\beq\begin{split}
&p_2 p_4 = E_2^2 (1 - \cos{\theta} ) \\
&p_1 p_2 =p_3 p_4 = E_2 (E_1 + E_2) \\
&p_1 p_4 = p_2 p_3 = E_2 (E_1 + E_2 \cos{\theta} )
\end{split}\eeq
The scattering amplitude then results
\beq\begin{split}
|\mathcal{M}|^2 = 16 \Lambda^2 E_2^2 \bigg[& -G^- m_a^2 (1 - \cos{\theta} ) + (G^+ + 2G_{VA} ) (E_1 + E_2)^2 + \\
&+ (G^+ - 2 G_{VA} ) (E_1 + E_2 \cos{\theta} )^2 \bigg]
\end{split}\eeq
Going back to the constants previously defined, we have
\beq\begin{split}
|\mathcal{M}|^2 = 16 \Lambda^2 E_2^2 \bigg[& (g_A^2 - g_V^2) m_a^2 (1 - \cos{\theta} ) + (g_V + g_A)^2 (E_1 + E_2)^2 + \\
&+ (g_V - g_A)^2 (E_1 + E_2 \cos{\theta} )^2 \bigg]
\end{split}\eeq
The flux factor is
\beq
\mathcal{\mathcal{J}} = \frac{1}{4 \sqrt{ (p_1 p_2)^2} } = \frac{1}{4 E_2 (E_1 + E_2) }
\eeq
while the 2-body phase space is
\beq
d\Phi_2 = \frac{|\textbf{q}|}{E_1 + E_2} \frac{d \cos{\theta} d\varphi}{16 \pi^2} = \frac{E_2}{E_1 + E_2} \frac{d \cos{\theta} d\varphi}{16 \pi^2}
\eeq
once having integrated over $d \varphi$, the differential cross section for the process finally becomes
\beq\begin{split}
\frac{d\sigma}{d \cos{\theta} } = \frac{\Lambda^2}{2 \pi} \frac{E_2^2}{(E_1 + E_2)^2} \bigg[& (g_A^2 - g_V^2) m_a^2 (1 - \cos{\theta} ) + \\
&+ (g_V + g_A)^2 (E_1 + E_2)^2 + \\
&+ (g_V - g_A)^2 (E_1 + E_2 \cos{\theta} )^2 \bigg]
\end{split}\eeq
\end{sol}

\chapter{Ward Indentities}

\section{Radiative corrections for $\phi \phi^* \to \phi \phi^* \gamma$}
\begin{ex}\label{Ward1}
Given the following Lagrangian
\[
\mathcal{L} = (D_{\mu}\phi)^* (D^{\mu}\phi) - m^2 |\phi|^2 + \overline \psi (i 	\bcc D \ecc - m_e) \psi - \frac{1}{4}F_{\alpha\beta}F^{\alpha\beta}
\]
Show that the amplitude for the process
\[
\phi \phi^* \to \phi \phi^* \gamma
\]
vanishes after replacing $\varepsilon(k) \to k$.
\end{ex}
\begin{sol}
Using the vertices derived the exercise \ref{QED_scalare} we can build the following diagrams
\beq
\begin{tikzpicture}
\begin{feynman}
\node at (1.6,-1.3){\fontsize{8}{8}\selectfont$\tiny\mu$};
\node at (3.9,-1.3){\fontsize{8}{8}\selectfont$\tiny\nu$};
\node at (8,-1.5) {$=\mathcal{A}_1$};
	\vertex(a) at (0,0){$\phi$};
	\vertex(b) at (0,-3){$\phi^*$};
	\vertex(c) at (1.5,-1.5);
	\vertex(d) at (4,-1.5);
	\vertex(e) at (5.5,0){$\phi$};
	\vertex(f) at (5.5,-3){$\phi^*$};
	\vertex(g) at (4.3,-3.2){$\gamma$};

	\diagram* {
	(a)--[charged scalar, momentum={[arrow shorten=0.3, arrow style=lightgray] $p_1$ } ] (c),
	(b)--[anti charged scalar,  momentum'={[arrow shorten=0.3, arrow style=lightgray] $p_2$ } ] (c),
	(c)--[boson,  momentum={[arrow shorten=0.3, arrow style=lightgray] $k$ } ] (d),
	(d)--[charged scalar,  momentum={[arrow shorten=0.3, arrow style=lightgray] $p_3$ } ] (e),
	(d)--[anti charged scalar,  momentum={[arrow shorten=0.3, arrow style=lightgray] $p_4$ } ] (f),
	(d)--[boson,  momentum'={[arrow shorten=0.3, arrow style=lightgray] $q$ } ] (g),
	};
    	
\end{feynman}
\end{tikzpicture}
\eeq
\beq
\begin{tikzpicture}
\begin{feynman}
\node at (1.6,-1.7){\fontsize{8}{8}\selectfont$\tiny\mu$};
\node at (3.9,-1.7){\fontsize{8}{8}\selectfont$\tiny\nu$};
\node at (8,-1.5) {$=\mathcal{A}_2$};
	\vertex(a) at (0,0){$\phi$};
	\vertex(b) at (0,-3){$\phi^*$};
	\vertex(c) at (1.5,-1.5);
	\vertex(d) at (4,-1.5);
	\vertex(e) at (5.5,0){$\phi$};
	\vertex(f) at (5.5,-3){$\phi^*$};
	\vertex(g) at (2.5,0){$\gamma$};

	\diagram* {
	(a)--[charged scalar, momentum={[arrow shorten=0.3, arrow style=lightgray] $p_1$ } ] (c),
	(b)--[anti charged scalar,  momentum'={[arrow shorten=0.3, arrow style=lightgray] $p_2$ } ] (c),
	(c)--[boson, momentum'={[arrow shorten=0.3, arrow style=lightgray] $k$ } ] (d),
	(d)--[charged scalar, momentum={[arrow shorten=0.3, arrow style=lightgray] $p_3$ } ] (e),
	(d)--[anti charged scalar, momentum={[arrow shorten=0.3, arrow style=lightgray] $p_4$ } ] (f),
	(c)--[boson, momentum'={[arrow shorten=0.3, arrow style=lightgray] $q$ } ] (g),
	};

\end{feynman}
\end{tikzpicture}
\eeq
Notice that the amplitudes are proportional to $e^3$, where $e$ is the coupling constant of the interaction. \\In order to verify the Ward identity we need to include all possible Feynman diagrams proportional to $e^3$. By attaching a 3-lines vertex to external lines, which means taking into account the \textit{radiative corrections}, we can build the following diagrams
\beq
\begin{tikzpicture}
\begin{feynman}
\node at(4.65,-0.65){\fontsize{8}{8}\selectfont$\sigma$};
\node at (1.6,-1.3){\fontsize{8}{8}\selectfont$\tiny\mu$};
\node at (3.9,-1.3){\fontsize{8}{8}\selectfont$\tiny\nu$};
\node at (8,-1.5) {$=\mathcal{A}_3$};
	\vertex(a) at (0,0){$\phi$};
	\vertex(b) at (0,-3){$\phi^*$};
	\vertex(c) at (1.5,-1.5);
	\vertex(d) at (4,-1.5);
	\vertex(e) at (5.5,0){$\phi$};
	\vertex(f) at (5.5,-3){$\phi^*$};
	\vertex(h) at (4.75,-0.75);
	\vertex(g) at (6.25,-0.6){$\gamma$};

	\diagram* {
	(a)--[charged scalar, momentum={[arrow shorten=0.3, arrow style=lightgray] $p_1$ } ] (c),
	(b)--[anti charged scalar,  momentum'={[arrow shorten=0.3, arrow style=lightgray] $p_2$ } ] (c),
	(c)--[boson, momentum'={[arrow shorten=0.3, arrow style=lightgray] $k$ } ] (d),
	(d)--[charged scalar, momentum={[arrow shorten=0.3, arrow style=lightgray] $p'$ } ] (h),
	(h)--[charged scalar, momentum={[arrow shorten=0.2, arrow style=lightgray] $p_3$ } ] (e),
	(d)--[anti charged scalar, momentum={[arrow shorten=0.3, arrow style=lightgray] $p_4$ } ] (f),
	(h)--[boson, momentum'={[arrow shorten=0.3, arrow style=lightgray] $q$ } ] (g),
	};

\end{feynman}
\end{tikzpicture}
\eeq
\beq
\begin{tikzpicture}
\begin{feynman}
\node at(4.55,-2.3){\fontsize{8}{8}\selectfont$\sigma$};
\node at (1.6,-1.3){\fontsize{8}{8}\selectfont$\tiny\mu$};
\node at (3.9,-1.3){\fontsize{8}{8}\selectfont$\tiny\nu$};
\node at (8,-1.5) {$=\mathcal{A}_4$};
	\vertex(a) at (0,0){$\phi$};
	\vertex(b) at (0,-3){$\phi^*$};
	\vertex(c) at (1.5,-1.5);
	\vertex(d) at (4,-1.5);
	\vertex(e) at (5.5,0){$\phi$};
	\vertex(f) at (5.5,-3){$\phi^*$};
	\vertex(h) at (4.75,-2.25);
	\vertex(g) at (6.25,-2.4){$\gamma$};

	\diagram* {
	(a)--[charged scalar, momentum={[arrow shorten=0.3, arrow style=lightgray] $p_1$ } ] (c),
	(b)--[anti charged scalar,  momentum'={[arrow shorten=0.3, arrow style=lightgray] $p_2$ } ] (c),
	(c)--[boson,  momentum={[arrow shorten=0.3, arrow style=lightgray] $k$ } ] (d),
	(d)--[charged scalar, momentum={[arrow shorten=0.3, arrow style=lightgray] $p_3$ } ] (e),
	(h)--[anti charged scalar, momentum'={[arrow shorten=0.2, arrow style=lightgray] $p_4$ } ] (f),
	(d)--[anti charged scalar, momentum'={[arrow shorten=0.3, arrow style=lightgray] $p'$ } ] (h),
	(h)--[boson, momentum={[arrow shorten=0.3, arrow style=lightgray] $q$ } ] (g),
	};
    	
\end{feynman}
\end{tikzpicture}
\eeq
\beq
\begin{tikzpicture}
\begin{feynman}
\node at(0.6,-0.8){\fontsize{8}{8}\selectfont$\sigma$};
\node at (1.6,-1.3){\fontsize{8}{8}\selectfont$\tiny\mu$};
\node at (3.9,-1.3){\fontsize{8}{8}\selectfont$\tiny\nu$};
\node at (8,-1.5) {$=\mathcal{A}_5$};
	\vertex(a) at (0,0){$\phi$};
	\vertex(b) at (0,-3){$\phi^*$};
	\vertex(c) at (1.5,-1.5);
	\vertex(d) at (4,-1.5);
	\vertex(e) at (5.5,0){$\phi$};
	\vertex(f) at (5.5,-3){$\phi^*$};
	\vertex(h) at (0.75,-0.75);
	\vertex(g) at (2.25,-0.5){$\gamma$};

	\diagram* {
	(a)--[charged scalar, momentum'={[arrow shorten=0.2, arrow style=lightgray] $p_1$ } ] (h),
	(h)--[charged scalar, momentum'={[arrow shorten=0.3, arrow style=lightgray] $p'$ } ] (c),
	(b)--[anti charged scalar, momentum'={[arrow shorten=0.3, arrow style=lightgray] $p_2$ } ] (c),
	(c)--[boson,  momentum'={[arrow shorten=0.3, arrow style=lightgray] $k$ } ] (d),
	(d)--[charged scalar, momentum={[arrow shorten=0.3, arrow style=lightgray] $p_3$ } ] (e),
	(d)--[anti charged scalar, momentum'={[arrow shorten=0.3, arrow style=lightgray] $p_4$ } ] (f),
	(h)--[boson, momentum={[arrow shorten=0.3, arrow style=lightgray] $q$ } ] (g),
	};

\end{feynman}
\end{tikzpicture}
\eeq
\beq
\begin{tikzpicture}
\begin{feynman}
\node at(0.65,-2.2){\fontsize{8}{8}\selectfont$\sigma$};
\node at (1.6,-1.3){\fontsize{8}{8}\selectfont$\tiny\mu$};
\node at (3.9,-1.3){\fontsize{8}{8}\selectfont$\tiny\nu$};
\node at (8,-1.5) {$=\mathcal{A}_6$};
	\vertex(a) at (0,0){$\phi$};
	\vertex(b) at (0,-3){$\phi^*$};
	\vertex(c) at (1.5,-1.5);
	\vertex(d) at (4,-1.5);
	\vertex(e) at (5.5,0){$\phi$};
	\vertex(f) at (5.5,-3){$\phi^*$};
	\vertex(h) at (0.75,-2.25);
	\vertex(g) at (2.25,-2.5){$\gamma$};

	\diagram* {
	(a)--[charged scalar, momentum={[arrow shorten=0.3, arrow style=lightgray] $p_1$ } ] (c),
	(b)--[anti charged scalar,  momentum={[arrow shorten=0.2, arrow style=lightgray] $p_2$ } ] (h),
	(h)--[anti charged scalar, momentum={[arrow shorten=0.3, arrow style=lightgray] $p'$ } ] (c),
	(c)--[boson, momentum={[arrow shorten=0.3, arrow style=lightgray] $k$ } ] (d),
	(d)--[charged scalar, momentum={[arrow shorten=0.3, arrow style=lightgray] $p_3$ } ] (e),
	(d)--[anti charged scalar, momentum'={[arrow shorten=0.3, arrow style=lightgray] $p_4$ } ] (f),
	(h)--[boson, momentum'={[arrow shorten=0.3, arrow style=lightgray] $q$ } ] (g),
	};

\end{feynman}
\end{tikzpicture}
\eeq
The total scattering amplitude is the sum of these six amplitudes; the Ward identity predicts that once we make the substitution $\varepsilon^*_\sigma(q) \to q_\sigma$, the total amplitude vanishes.
\\Explicitely the amplitudes are
\beq
\mathcal{A}_1 = -ie (p_1 - p_2)^\mu \frac{-i g_{\mu\nu} }{k^2} 2 ie^2 g^{\nu\sigma} \varepsilon^*_\sigma(q)
\eeq
\beq
\mathcal{A}_2 = 2ie^2 \varepsilon^*_\sigma(q) g^{\sigma\mu} \frac{-i g_{\mu\nu} }{k^2} (-ie)(p_3 - p_4)^\nu
\eeq
\beq
\mathcal{A}_3 = -ie (p_1-p_2)^\mu \frac{-i g_{\mu\nu} }{k^2} (-ie)(p' - p_4)^\nu \frac{i}{p'^2 - m^2} (-ie)(p' + p_3)^\sigma \varepsilon^*_\sigma(q)
\eeq
\beq
\mathcal{A}_4 = -ie (p_1 - p_2)^\mu \frac{-i g_{\mu\nu} }{k^2} (-ie)(p_3 - p')^\nu \frac{i}{p'^2 - m^2} (-ie)(-p' - p_4)^\sigma \varepsilon^*_\sigma(q)
\eeq
\beq
\mathcal{A}_5 = -ie (p_1 + p')^\sigma \varepsilon^*_\sigma(q) \frac{i}{p'^2 - m^2} (-ie)(p' - p_2)^\mu \frac{-i g_{\mu\nu} }{k^2} (-ie)(p_3 - p_4)^\nu
\eeq
\beq
\mathcal{A}_6 = -ie (-p_2 - p')^\sigma \varepsilon^*_\sigma(q) (-ie)(-p' + p_1)^\mu \frac{-i g_{\mu\nu} }{k^2} (-ie)(p_3 - p_4)^\nu
\eeq
Firstly, we want to simplify each amplitude. Starting from $\mathcal{A}_1$ we have
\beq
\mathcal{A}_1 = -ie (p_1 - p_2)^\mu \frac{-i g_{\mu\nu} }{k^2} 2ie^2 g^{\nu\sigma} q_\sigma =  -\frac{2ie^3 (p_1 - p_2)^\mu q_\mu}{k^2}
\eeq
Notice that
\beq\begin{split}
k = p_1+p_2 \implies k^2 = &(p_1 + p_2)^2 = p_1^2 + p_2^2 + 2p_1 p_2 \\
=& m^2 + m^2 + 2p_1p_2 \\
=& 2 (m^2 + p_1 p_2)
\end{split}\eeq
therefore
\beq
\mathcal{A}_1 = -\frac{ie^3 (p_1 - p_2) \cdot q}{m^2 + p_1 p_2}
\eeq
Elaborating in a similar fashion $\mathcal{A}_2$, by using $k = p_3 + p_4$, we get
\beq
\mathcal{A}_2 = -\frac{ie^3 (p_3 - p_4) \cdot q}{m^2 + p_3 p_4}
\eeq
Exploiting once more the conservation of 4-momentum at the vertices, the amplitude $\mathcal{A}_3$ can be written as follows
\beq
\mathcal{A}_3 = -ie (p_1 - p_2)^\mu \frac{-i g_{\mu\nu} }{k^2} (-ie)(p' - p_4)^\nu \frac{i}{p'^2 - m^2} (-ie)(p' + p_3)^\sigma q_\sigma
\eeq
By using $p' = q + p_3$ e $p' = p_1 + p_2 - p_4$, the products in the amplitude simplify in the following way
\beq\begin{split}
(p_1 - p_2)^\mu g_{\mu\nu} (p' - p_4)^\nu =& (p_1 - p_2) \cdot (p' - p_4) \\
=& (p_1 - p_2) \cdot (p_1 + p_2 - 2p_4) \\
=& p_1^2 - p_2^2 + 2p_1 p_2 - 2p_1 p_2 - 2p_4 \cdot (p_1 - p_2) \\
=& m^2 - m^2 - 2p_4 \cdot (p_1 - p_2) \\
=& -2p_4 \cdot (p_1 - p_2)
\end{split}\eeq
and
\beq\begin{split}
(p' + p_3)^\sigma q_\sigma =& (q + p_3 + p_3) \cdot q \\
=& q^2 + 2p_3 q \\
=& 2p_3 q
\end{split}\eeq
At the denominator we have
\beq\begin{split}
k^2 (p'^2 - m^2)= &(p_1 + p_2)^2 \bigg( (p_3 + q)^2 - m^2 \bigg) \\
=& 2(m^2 + p_1 p_2) (p_3^2 + q^2 + 2p_3 q - m^2) \\
=& 2(m^2 + p_1 p_2) (2p_3 q)
\end{split}\eeq
where we have exploited the fact that external particles are on-shell, so that $p_1^2 = p_2^2 = p_3^2 = p_4^2 = m^2$ and $q^2 = 0$ for a real photon. In the end $\mathcal{A}_3$ becomes
\beq
\mathcal{A}_3 = \frac{ie^3}{2(m^2 + p_1 p_2) (2p_3 q)} \bigg(-2p_4 \cdot (p_1 - p_2) \bigg) (2p_3 q) = -\frac{ie^3 (p_1 - p_2) \cdot p_4}{m^2 + p_1 p_2}
\eeq
With similar calculations, substituting each time the more suited expressions for $k$ and $p'$, for the remaining amplitudes we get the following expressions
\beq
\mathcal{A}_4 = -\frac{ie^3 (p_1 - p_2) \cdot p_3}{m^2 + p_1 p_2}
\eeq
\beq
\mathcal{A}_5 = \frac{ie^3 (p_3 - p_4) \cdot p_2}{m^2 + p_3 p_4}
\eeq
\beq
\mathcal{A}_6 = \frac{ie^3 (p_3 - p_4) \cdot p_1}{m^2 + p_3 p_4}
\eeq
We can then sum all the amplitudes:
\beq\begin{split}
\mathcal{A} = ie^3 \bigg[& -\frac{ (p_1 - p_2) \cdot q}{m^2 + p_1 p_2}-\frac{ (p_3 - p_4) \cdot q}{m^2 + p_3 p_4} -\frac{ (p_1 - p_2) \cdot p_4}{m^2 + p_1 p_2} -\frac{ (p_1 - p_2) \cdot p_3}{m^2 + p_1 p_2} + \\
&+ \frac{ (p_3 - p_4) \cdot p_2}{m^2 + p_3 p_4} + \frac{ (p_3 - p_4) \cdot p_1}{m^2 + p_3 p_4} \bigg] \\
\end{split}\eeq
therefore
\beq
\mathcal{A} = ie^3 \bigg[\frac{ (p_1 - p_2) \cdot (-q - p_4 - p_3)}{m^2 + p_1 p_2} + \frac{ (p_3 - p_4) \cdot (-q + p_1 + p_2)}{m^2 + p_3 p_4} \bigg]
\eeq
Using once again 4-momentum conservation
\beq
p_1 + p_2 = p_3 + p_4 + q
\eeq
we can eliminate every dependence on $q$, finally obtaining
\beq\begin{split}
\mathcal{A} =& ie^3 \bigg[\frac{ -(p_1 - p_2) \cdot (p_1 + p_2)}{m^2 + p_1 p_2} + \frac{ (p_3 - p_4) \cdot (p_3 + p_4)}{m^2 + p_3 p_4} \bigg] \\
=& ie^3\bigg[ \frac{-( p_1^2 + p_1 p_2 - p_1 p_2 - p_2^2 ) }{m^2 + p_1 p_2} + \frac{p_3^2 + p_3 p_4 - p_3 p_4 - p_4^2}{m^2 + p_3 p_4} \bigg] = 0
\end{split}\eeq
since $p_1^2 = p_2^2 = p_3^2 = p_4^2 = m^2$, so the Ward identity is satisfied.
\end{sol}

\newpage
\section{Radiative corrections for $\psi \overline \psi \rightarrow \phi \phi^* \gamma $}
\begin{ex}\label{Ward2}
Consider the interaction between a complex scalar field, a fermionic field and the electromagnetic one, described by the following Lagrangian
\[
\mathcal{L} = \overline \psi (i \bcc D \ecc - m_\psi) \psi + (D_\mu \phi)^* (D^\mu \phi) - m_{\phi}^2 \phi^* \phi - \frac{1}{4} F_{\mu\nu} F^{\mu\nu}
\]
Show that the scattering amplitude for the process
\[
\psi \overline \psi \rightarrow \phi \phi^* \gamma
\]
vanishes after the substitution $\varepsilon_\mu(q) \rightarrow q_\mu$, where $\varepsilon_\mu$ is the polarization wave vector of the photon and $q_\mu$ its 4-momentum.
\end{ex}
\begin{sol}
The Feynman rules for this Lagrangian are given by the usual rules for QED and for scalar QED (see exercise \ref{QED_scalare}). \\For this process, at first order ($\propto e^3$), we have the following diagrams
\\
\beq
\begin{tikzpicture}
  \begin{feynman}
   	\node at (1.6,-1.3){\fontsize{8}{8}\selectfont$\tiny\mu$};
  	\node at (3.9,-1.3){\fontsize{8}{8}\selectfont$\tiny\nu$};
  	\node at (8,-1.5) {$=\mathcal{A}_1$};
  	\vertex(a) at (0,0) {\(\psi\)} ;
  	\vertex(b) at (0,-3) {$\overline \psi$};
  	\vertex(c) at (1.5,-1.5);
  	\vertex(d) at (4,-1.5);
  	\vertex(e) at (5.5,0) {\(\phi\)};
  	\vertex(f) at (5.5,-3) {\(\phi^*\)};
	\vertex(g) at (4.3,-3.2) {$\gamma$};

	\diagram* {
	(a)--[fermion, momentum={[arrow shorten=0.3, arrow style=lightgray]$p_1$}] (c),
	(b)--[anti fermion,  momentum'={[arrow shorten=0.3, arrow style=lightgray]$p_2$}] (c),
	(c)--[boson,  momentum={[arrow shorten=0.3, arrow style=lightgray]$k$}] (d),
	(d)--[charged scalar,  momentum={[arrow shorten=0.3, arrow style=lightgray]$p_3$}] (e),
	(d)--[anti charged scalar,  momentum={[arrow shorten=0.3, arrow style=lightgray]$p_4$}] (f),
	(d)--[boson,  momentum'={[arrow shorten=0.3, arrow style=lightgray]$q$}] (g),

    };
    	
  \end{feynman}
\end{tikzpicture}
\eeq
\beq
\begin{tikzpicture}
  \begin{feynman}
  	\node at(4.65,-0.65){\fontsize{8}{8}\selectfont$\sigma$};
  	\node at (1.6,-1.3){\fontsize{8}{8}\selectfont$\tiny\mu$};
  	\node at (3.9,-1.3){\fontsize{8}{8}\selectfont$\tiny\nu$};
  	\node at (8,-1.5) {$=\mathcal{A}_2$};
  	\vertex(a) at (0,0) {$\psi$} ;
  	\vertex(b) at (0,-3) {$\overline \psi$};
  	\vertex(c) at (1.5,-1.5);
  	\vertex(d) at (4,-1.5);
  	\vertex(e) at (5.5,0) {\(\phi\)};
  	\vertex(f) at (5.5,-3) {\(\phi^*\)};
  	
  	\vertex(h) at (4.75,-0.75);
	\vertex(g) at (6.25,-0.6) {$\gamma$};

	\diagram* {
	(a)--[fermion, momentum={[arrow shorten=0.3, arrow style=lightgray]$p_1$}] (c),
	(b)--[anti fermion,  momentum'={[arrow shorten=0.3, arrow style=lightgray]$p_2$}] (c),
	(c)--[boson, momentum'={[arrow shorten=0.3, arrow style=lightgray]$k$}] (d),
	(d)--[charged scalar, momentum={[arrow shorten=0.3, arrow style=lightgray]$p'$}] (h),
	(h)--[charged scalar, momentum={[arrow shorten=0.2, arrow style=lightgray]$p_3$}] (e),
	(d)--[anti charged scalar, momentum={[arrow shorten=0.3, arrow style=lightgray]$p_4$}] (f),
	(h)--[boson, momentum'={[arrow shorten=0.3, arrow style=lightgray]$q$}] (g),

    };
    	
  \end{feynman}
\end{tikzpicture}
\eeq
\beq
\begin{tikzpicture}
  \begin{feynman}
 	\node at(4.55,-2.3){\fontsize{8}{8}\selectfont$\sigma$};
  	\node at (1.6,-1.3){\fontsize{8}{8}\selectfont$\tiny\mu$};
  	\node at (3.9,-1.3){\fontsize{8}{8}\selectfont$\tiny\nu$};
  	\node at (8,-1.5) {$=\mathcal{A}_3$};
  	\vertex(a) at (0,0) {$\psi$} ;
  	\vertex(b) at (0,-3) {$\overline \psi$};
  	\vertex(c) at (1.5,-1.5);
  	\vertex(d) at (4,-1.5);
  	\vertex(e) at (5.5,0) {\(\phi\)};
  	\vertex(f) at (5.5,-3) {\(\phi^*\)};
  	
  	\vertex(h) at (4.75,-2.25);
	\vertex(g) at (6.25,-2.4){$\gamma$};

	\diagram* {
	(a)--[fermion, momentum={[arrow shorten=0.3, arrow style=lightgray]$p_1$}] (c),
	(b)--[anti fermion,  momentum'={[arrow shorten=0.3, arrow style=lightgray]$p_2$}] (c),
	(c)--[boson,  momentum={[arrow shorten=0.3, arrow style=lightgray]$k$}] (d),
	(d)--[charged scalar, momentum={[arrow shorten=0.3, arrow style=lightgray]$p_3$}] (e),
	(h)--[anti charged scalar, momentum'={[arrow shorten=0.2, arrow style=lightgray]$p_4$}] (f),
	(d)--[anti charged scalar, momentum'={[arrow shorten=0.3, arrow style=lightgray]$p'$}] (h),
	(h)--[boson, momentum={[arrow shorten=0.3, arrow style=lightgray]$q$}] (g),

    };
    	
  \end{feynman}
\end{tikzpicture}
\eeq
\beq
\begin{tikzpicture}
  \begin{feynman}
  	\node at(0.6,-0.8){\fontsize{8}{8}\selectfont$\sigma$};
  	\node at (1.6,-1.3){\fontsize{8}{8}\selectfont$\tiny\mu$};
  	\node at (3.9,-1.3){\fontsize{8}{8}\selectfont$\tiny\nu$};
  	\node at (8,-1.5) {$=\mathcal{A}_4$};
  	\vertex(a) at (0,0) {$\psi$} ;
  	\vertex(b) at (0,-3) {$\overline \psi$};
  	\vertex(c) at (1.5,-1.5);
  	\vertex(d) at (4,-1.5);
  	\vertex(e) at (5.5,0) {\(\phi\)};
  	\vertex(f) at (5.5,-3) {\(\phi^*\)};
  	
  	\vertex(h) at (0.75,-0.75);
	\vertex(g) at (2.25,-0.5) {$\gamma$};

	\diagram* {
	(a)--[fermion, momentum'={[arrow shorten=0.2, arrow style=lightgray]$p_1$}] (h),
	(h)--[fermion, momentum'={[arrow shorten=0.3, arrow style=lightgray]$p'$}] (c),
	(b)--[anti fermion, momentum'={[arrow shorten=0.3, arrow style=lightgray]$p_2$}] (c),
	(c)--[boson,  momentum'={[arrow shorten=0.3, arrow style=lightgray]$k$}] (d),
	(d)--[charged scalar, momentum={[arrow shorten=0.3, arrow style=lightgray]$p_3$}] (e),
	(d)--[anti charged scalar, momentum'={[arrow shorten=0.3, arrow style=lightgray]$p_4$}] (f),
	(h)--[boson, momentum={[arrow shorten=0.3, arrow style=lightgray]$q$}] (g),

    };
    	
  \end{feynman}
\end{tikzpicture}
\eeq
\beq
\begin{tikzpicture}
  \begin{feynman}
    	\node at(0.65,-2.2){\fontsize{8}{8}\selectfont$\sigma$};
  	\node at (1.6,-1.3){\fontsize{8}{8}\selectfont$\tiny\mu$};
  	\node at (3.9,-1.3){\fontsize{8}{8}\selectfont$\tiny\nu$};
  	\node at (8,-1.5) {$=\mathcal{A}_5$};
  	\vertex(a) at (0,0) {$\psi$} ;
  	\vertex(b) at (0,-3) {$\overline \psi$};
  	\vertex(c) at (1.5,-1.5);
  	\vertex(d) at (4,-1.5);
  	\vertex(e) at (5.5,0) {\(\phi\)};
  	\vertex(f) at (5.5,-3) {\(\phi^*\)};
  	
  	\vertex(h) at (0.75,-2.25);
	\vertex(g) at (2.25,-2.5) {$\gamma$};

	\diagram* {
	(a)--[fermion, momentum={[arrow shorten=0.3, arrow style=lightgray]$p_1$}] (c),
	(b)--[anti fermion,  momentum={[arrow shorten=0.2, arrow style=lightgray]$p_2$}] (h),
	(h)--[anti fermion, momentum={[arrow shorten=0.3, arrow style=lightgray]$p'$}] (c),
	(c)--[boson, momentum={[arrow shorten=0.3, arrow style=lightgray]$k$}] (d),
	(d)--[charged scalar, momentum={[arrow shorten=0.3, arrow style=lightgray]$p_3$}] (e),
	(d)--[anti charged scalar, momentum'={[arrow shorten=0.3, arrow style=lightgray]$p_4$}] (f),
	(h)--[boson, momentum'={[arrow shorten=0.3, arrow style=lightgray]$q$}] (g),

    };
    	
  \end{feynman}
\end{tikzpicture}
\eeq
Explicitely the amplitudes are
\beq
\mathcal{A}_1= \overline v^{s_2}(p_2) (-ie \gamma^\mu) u^{s_1}(p_1) \frac{-i g_{\mu\nu}}{k^2} 2ie^2 g^{\nu\sigma} \varepsilon^*_\sigma(q)
\eeq
\beq
\mathcal{A}_2= \overline v^{s_2}(p_2) (-ie \gamma^\mu) u^{s_1}(p_1) \frac{-i g_{\mu\nu}}{k^2} (-ie) (p' - p_4)^\nu \frac{i}{p'^2 - m_\phi^2} (-ie)(p' + p_3)^\sigma \varepsilon^*_\sigma(q)
\eeq
\beq
\mathcal{A}_3 = \overline v^{s_2}(p_2) (-ie \gamma^\mu) u^{s_1}(p_1) \frac{-i g_{\mu\nu}}{k^2} (-ie)(p_3 - p')^\nu \frac{i}{p'^2 - m_\phi^2} (-ie)(-p' - p_4)^\sigma \varepsilon^*_\sigma(q)
\eeq
\beq
\mathcal{A}_4= \overline v^{s_2}(p_2) (-ie \gamma^\mu) \frac{i (\bcc p' \ecc + m_\psi)}{p'^2 - m_\psi^2} (- ie \gamma^\sigma) \varepsilon^*_\sigma(q) u^{s_1}(p_1) \frac{-i g_{\mu\nu}}{k^2} (-ie) (p_3 - p_4)^\nu
\eeq
\beq
\mathcal{A}_5 = \overline v^{s_2}(p_2) (-ie \gamma^\sigma) \varepsilon^*_\sigma(q) \frac{i (-\bcc p' \ecc + m_\psi)}{p'^2 - m_\psi^2} (-ie \gamma^\mu) u^{s_1}(p_1) \frac{-i g_{\mu\nu}}{k^2} (-ie) (p_3 - p_4)^\nu
\eeq
It is worth noting that in $\mathcal{A}_5$ the term $\bcc p' \ecc$ in the numerator of the fermionic propagator has negative sign, since the fermionic current flows in the opposite direction to the flow of momentum. \\Substituting the photon polarization with its momentum, that is $\varepsilon^*_\sigma(q)\rightarrow q_\sigma$, we can simplify the amplitudes as follows
\beq\begin{split}
\mathcal{A}_1 =& \overline v^{s_2}(p_2) (-ie \gamma^\mu) u^{s_1}(p_1) \frac{-i g_{\mu\nu}}{k^2} 2ie^2 g^{\nu\sigma} \varepsilon^*_\sigma(q) = \\
=& \frac{-2ie^3}{(p_1+p_2)^2} \overline v^{s_2}(p_2) \gamma^\mu u^{s_1}(p_1) q_\mu = \\
=& \frac{-ie^3}{m_\psi^2 + p_1\cdot p_2} \overline v^{s_2}(p_2) \bcc q \ecc u^{s_1}(p_1)
\end{split}\eeq
\beq\begin{split}
\mathcal{A}_2=& \overline v^{s_2}(p_2) (-ie \gamma^\mu) u^{s_1}(p_1) \frac{-i g_{\mu\nu}}{k^2} (-ie) (p' - p_4)^\nu \frac{i}{p'^2 - m_\phi^2} (-ie) (p' + p_3)^\sigma \varepsilon^*_\sigma(q) = \\
=& \frac{ie^3}{(p_1 + p_2)^2 \bigg[(p_3 + q)^2 - m_\phi^2 \bigg]} \overline v^{s_2}(p_2) (\bcc q \ecc + {\bcc p \ecc}_3 - {\bcc p \ecc}_4) u^{s_1}(p_1) (q + 2p_3)^\sigma q_\sigma = \\
=& \frac{ie^3}{2 (m_\psi^2 + p_1 \cdot p_2) (2p_3 \cdot q)} \overline v^{s_2}(p_2) (\bcc q \ecc + {\bcc p \ecc}_3 - {\bcc p \ecc}_4) u^{s_1}(p_1) (2p_3 \cdot q) = \\
=& \frac{ie^3}{2 (m_\psi^2 + p_1 \cdot p_2)} \overline v^{s_2}(p_2) (\bcc q \ecc + {\bcc p \ecc}_3 - {\bcc p \ecc}_4) u^{s_1}(p_1)
\end{split}\eeq
\beq\begin{split}
\mathcal{A}_3 =& \overline v^{s_2}(p_2) (-ie \gamma^\mu) u^{s_1}(p_1) \frac{-i g_{\mu\nu}}{k^2} (-ie) (p_3 - p')^\nu \frac{i}{p'^2 - m_\phi^2} (-ie) (-p' - p_4)^\sigma \varepsilon^*_\sigma(q) = \\
=& \frac{ie^3}{(p_1 + p_2)^2 \bigg[(p_4 + q)^2 - m_\phi^2 \bigg]} \overline v^{s_2}(p_2) ({\bcc p \ecc}_3 - \bcc q \ecc - {\bcc p \ecc}_4) u^{s_1}(p_1) (-q - 2p_4)^\sigma q_\sigma = \\
=& \frac{-ie^3}{2 (m_\psi^2 + p_1 \cdot p_2) (2p_4 \cdot q)} \overline v^{s_2}(p_2) ({\bcc p \ecc}_3 - \bcc q \ecc - {\bcc p \ecc}_4) u^{s_1}(p_1) (2p_4 \cdot q) = \\
=& \frac{-ie^3}{2 (m_\psi^2 + p_1 \cdot p_2)} \overline v^{s_2}(p_2) ({\bcc p \ecc}_3 - \bcc q \ecc - {\bcc p \ecc}_4) u^{s_1}(p_1)
\end{split}\eeq
\beq\begin{split}
\mathcal{A}_4=& \overline v^{s_2}(p_2) (-ie\gamma^\mu) \frac{i (\bcc p' \ecc + m_\psi)}{p'^2 - m_\psi^2} (-ie \gamma^\sigma) \varepsilon^*_\sigma(q) u^{s_1}(p_1) \frac{-i g_{\mu\nu}}{k^2} (-ie)(p_3 - p_4)^\nu = \\
=& \frac{ie^3}{(p_3 + p_4)^2 \bigg[(p_1 - q)^2 - m_\phi^2 \bigg]} \overline v^{s_2}(p_2) \bigg[({\bcc p \ecc}_3 - {\bcc p \ecc}_4) ({\bcc p \ecc}_1 - \bcc q \ecc + m_\psi) \bcc q \ecc \bigg] u^{s_1}(p_1) = \\
=& \frac{ie^3}{2 (m_\phi^2 + p_3 \cdot p_4) (- 2p_1 \cdot q)} \overline v^{s_2}(p_2) \bigg[({\bcc p \ecc}_3 - {\bcc p \ecc}_4) ({\bcc p \ecc}_1 \bcc q \ecc - \bcc q \ecc \bcc q \ecc + m_\psi \bcc q \ecc) \bigg] u^{s_1}(p_1)
\end{split}\eeq
The spinors product can be simplified as
\beq
\overline v^{s_2}(p_2)\bigg[({\bcc p \ecc}_3 - {\bcc p \ecc}_4)({\bcc p \ecc}_1 \bcc q \ecc - \underbrace{\bcc q \ecc \bcc q \ecc}_{=0} + m_\psi \bcc q \ecc)\bigg] u^{s_1}(p_1)
\eeq
The term ${\bcc p \ecc}_1 \bcc q \ecc $ can be rewritten as ${\bcc p \ecc}_1 \bcc q \ecc = 2p_1 q - \bcc q \ecc {\bcc p \ecc}_1$, so that
\beq
\mathcal{A}_4= \frac{ie^3}{2 (m_\phi^2 + p_3 p_4) (-2 p_1 q)} \overline v^{s_2}(p_2) \bigg[({\bcc p \ecc}_3 - {\bcc p \ecc}_4) \bigg(2 p_1 q - \bcc q \ecc ({\bcc p \ecc}_1 - m_\psi) \bigg) \bigg] u^{s_1}(p_1)
\eeq
Thanks to the Dirac equation $({\bcc p \ecc}_1 - m_\psi) u^{s_1}(p_1) = 0$, we have
\beq\begin{split}
\mathcal{A}_4=& \frac{-ie^3}{2 (m_\phi^2 + p_3 p_4) (2 p_1 q)} \overline v^{s_2}(p_2) \bigg[({\bcc p \ecc}_3 - {\bcc p \ecc}_4) (2p_1 q) \bigg] u^{s_1}(p_1) = \\
=& \frac{-ie^3}{2 (m_\phi^2 + p_3 p_4)} \overline v^{s_2}(p_2) \bigg[{\bcc p \ecc}_3 - {\bcc p \ecc}_4 \bigg] u^{s_1}(p_1)
\end{split}\eeq
\beq\begin{split}
\mathcal{A}_5=& \overline v^{s_2}(p_2) (-ie \gamma^\sigma) \varepsilon^*_\sigma(q) \frac{i (-\bcc p' \ecc + m_\psi)}{p'^2 - m_\psi^2} (-ie \gamma^\mu) u^{s_1}(p_1) \frac{-i g_{\mu\nu}}{k^2} (-ie) (p_3 - p_4)^\nu = \\
=& \frac{ie^3}{(p_3 + p_4)^2 \bigg[(p_2 - q)^2 - m_\phi^2 \bigg]} \overline v^{s_2}(p_2) \bigg[\bcc q \ecc(-{\bcc p \ecc}_2 + \bcc q \ecc + m_\psi)({\bcc p \ecc}_3 - {\bcc p \ecc}_4) \bigg] u^{s_1}(p_1) = \\
=& \frac{ie^3}{2 (m_\phi^2 + p_3 p_4) (-2 p_2 q)} \overline v^{s_2}(p_2) \bigg[(-\bcc q \ecc {\bcc p \ecc}_2 + \underbrace{\bcc q \ecc \bcc q \ecc}_{=0} + \bcc q \ecc m_\psi) ({\bcc p \ecc}_3 - {\bcc p \ecc}_4) \bigg] u^{s_1}(p_1)
\end{split}\eeq
As did for $\mathcal{A}_4$, we simplify the spinors product by moving $\bcc q \ecc$ to the right of ${\bcc p \ecc}_2$. We then let $({\bcc p \ecc}_2 + m_\psi) \bcc q \ecc$ act on the spinor $\overline v^{s_2}(p_2)$ from the right, which simply gives us zero thanks again to the Dirac equation $\overline v^{s_2}(p_2) ({\bcc p \ecc}_2 + m_\psi) = 0$. \\It follows that
\beq\begin{split}
\mathcal{A}_5=& \frac{ie^3}{2 (m_\phi^2 + p_3 p_4) (-2 p_2 q)} \overline v^{s_2}(p_2) \bigg[ (-2 p_2 q) ({\bcc p \ecc}_3 - {\bcc p \ecc}_4) \bigg] u^{s_1}(p_1) = \\
=& \frac{ie^3}{2 (m_\phi^2 + p_3 p_4)} \overline v^{s_2}(p_2) \bigg[{\bcc p \ecc}_3 - {\bcc p \ecc}_4 \bigg] u^{s_1}(p_1)
\end{split}\eeq
and by summing all the amplitudes we can check the validity of the Ward identity
\begin{subequations}
\beq\begin{split}
\mathcal{A}= \sum_{j=1}^5{ \mathcal{A}_j } =& \frac{-ie^3}{m_\psi^2 + p_1 p_2} \overline v^{s_2}(p_2) \bcc q \ecc u^{s_1}(p_1) + \\
&+ \frac{ie^3}{2 (m_\psi^2 + p_1 p_2)} \overline v^{s_2}(p_2) (\bcc q \ecc + {\bcc p \ecc}_3 - {\bcc p \ecc}_4) u^{s_1}(p_1) + \\
&+ \frac{-ie^3}{2 (m_\psi^2 + p_1 p_2)} \overline v^{s_2}(p_2) ({\bcc p \ecc}_3 - \bcc q \ecc - {\bcc p \ecc}_4) u^{s_1}(p_1) + \\
&+ \frac{-ie^3}{2 (m_\phi^2 + p_3 p_4)} \overline v^{s_2}(p_2) \bigg[{\bcc p \ecc}_3 - {\bcc p \ecc}_4 \bigg] u^{s_1}(p_1) + \\
&+ \frac{ie^3}{2 (m_\phi^2 + p_3 p_4)} \overline v^{s_2}(p_2) \bigg[{\bcc p \ecc}_3 - {\bcc p \ecc}_4 \bigg] u^{s_1}(p_1)
\end{split}\eeq
\beq\begin{split}
\mathcal{A} =& \frac{ie^3} {m_\psi^2 + p_1 p_2} \overline v^{s_2}(p_2) \bigg[-\bcc q \ecc + \frac{1}{2} \bigg(\bcc q \ecc + {\bcc p \ecc}_3 - {\bcc p \ecc}_4 \bigg) - \frac{1}{2} \bigg({\bcc p \ecc}_3 - \bcc q \ecc - {\bcc p \ecc}_4 \bigg) \bigg] u^{s_1}(p_1) + \\
&+ \frac{ie^3}{2 (m_\phi^2 + p_3 p_4)} \overline v^{s_2}(p_2) \bigg[-({\bcc p \ecc}_3 - {\bcc p \ecc}_4) + ({\bcc p \ecc}_3 - {\bcc p \ecc}_4) \bigg] u^{s_1}(p_1) = 0
\end{split}\eeq
\end{subequations}
\end{sol}

\appendix
\renewcommand{\theequation}{A.\arabic{equation}}
\setcounter{equation}{0}
\chapter{Useful formulae}
{\footnotesize
\textsc{Euler-Lagrange equations}
\beq
\partial_\mu \frac{\partial \mathcal{L} }{\partial( \partial_\mu \phi ) } - \frac{\partial \mathcal{L} }{\partial \phi} = 0
\eeq
\\ \textsc{Hamiltonian}
\beq
\mathcal{H} = \sum_{l} { \frac{\partial \mathcal{L} }{\partial \dot \phi_l} \dot \phi_l } - \mathcal{L}
\eeq
\\ \textsc{Noether currents and conserved charges}
\\ \textbf{Internal symmetries}
\beq
\varepsilon \overline \delta \phi := \phi' - \phi
\eeq
\beq
\mathcal{J}^\mu = \sum \frac{\partial \mathcal{L} }{\partial (\partial_\mu \phi) } \overline \delta \phi, \qquad \partial_\mu \mathcal{J}^\mu = 0
\eeq
\beq
Q(t) = \int d^3x \mathcal{J}^0(\overline x, t)
\eeq
\textbf{Space-time translations symmetry}
\beq
T^{\mu\nu} = \frac{\partial \mathcal{L} }{\partial (\partial_\mu \phi) } \partial^\nu \phi - \mathcal{L} g^{\mu\nu}, \qquad \partial_\mu T^{\mu\nu} = 0
\eeq
\beq
P^\nu = \int d^3x T^{0\nu}
\eeq
\\ \textsc{Klein-Gordon Theory}
\beq
\mathcal{L}_{KG} = \partial_\mu \phi \partial^\mu \phi - m^2 \phi^2
\eeq
\beq
( \Box + m^2 ) \phi = 0
\eeq
\textbf{Field expansion}
\beq
\phi(x) = \int \frac{d^3p}{ (2\pi)^3 } \frac{1}{\sqrt{2E_p}} \bigg( a_p e^{-ipx} + a^\dagger_p e^{ipx} \bigg)
\eeq
\textbf{Ladder operators}
\beq
[ a_p, a^\dagger_{p'} ] = (2\pi)^3 \delta^{(3)}( \textbf{p} - \textbf{p}' ), \qquad [ a_p, a_{p'} ] = [ a^\dagger_p, a^\dagger_{p'} ] = 0
\eeq
\beq
\ket{ \textbf{p} } = \sqrt{2E_p} a^\dagger_p \ket{0}
\eeq
\textsc{Pauli matrices}
\beq
\sigma_1 = 
\begin{pmatrix}
0 & 1 \\
1 & 0
\end{pmatrix}, \qquad
\sigma_2 = 
\begin{pmatrix}
0 & -i \\
i & 0
\end{pmatrix}, \qquad
\sigma_3 = 
\begin{pmatrix}
1 & 0 \\
0 & -1
\end{pmatrix}
\eeq
\\ \textsc{Dirac matrices (chiral representation)}
\beq\begin{split}
&\gamma^0 = 
\begin{pmatrix}
0 & \mathbb{1}_{2x2} \\
\mathbb{1}_{2x2} & 0
\end{pmatrix}, \qquad
\gamma^i = 
\begin{pmatrix}
0 & \sigma^i \\
-\sigma^i & 0
\end{pmatrix} \\
&\gamma^5 = i \gamma^0 \gamma^1 \gamma^2 \gamma^3 =
\begin{pmatrix}
-\mathbb{1}_{2x2} & 0 \\
0 & \mathbb{1}_{2x2}
\end{pmatrix}
\end{split}\eeq
\textbf{$\gamma$ matrices properties}
\beq
\{ \gamma^\mu, \gamma^\nu \} = 2 g^{\mu\nu} \mathbb{1}
\eeq
\beq
(\gamma^\mu)^\dagger = \gamma^0 \gamma^\mu \gamma^0
\eeq
\beq
\gamma^\mu \gamma_\mu = 4 \mathbb{1}
\eeq
\beq
(\gamma^5)^\dagger = \gamma^5, \qquad (\gamma^5)^2 = \mathbb{1}
\eeq
\beq\label{anticom_gamma5}
\{ \gamma^5, \gamma^\mu \} = 0
\eeq
\beq\label{id_2gamma}
\gamma_\mu \gamma^\nu \gamma^\mu = -2 \gamma^\nu
\eeq
\beq\label{id_3gamma}
\gamma^\mu \gamma^\nu \gamma^\rho \gamma_\mu = 4g^{\nu \rho} \mathbb{1}_4
\eeq
\beq\label{id_4gamma}
\gamma^\mu \gamma^\nu \gamma^\rho \gamma^\sigma \gamma_\mu = -2 \gamma^\sigma \gamma^\rho \gamma^\nu
\eeq
\beq
P_\mu \gamma^\mu P_\nu \gamma^\nu = P^2
\eeq
\textbf{Traces of $\gamma$ matrices}
\beq
Tr[ \text{odd \# of $\gamma$} ] = 0, \qquad Tr[ \gamma^5 \cdot \text{odd \# of $\gamma$} ] = 0
\eeq
\beq
Tr[ \gamma^\mu \gamma^\nu ] = 4 g^{\mu\nu}
\eeq
\beq
Tr[ \gamma^\mu \gamma^\nu \gamma^\rho \gamma^\sigma ] = 4( g^{\mu\nu} g^{\rho\sigma} - g^{\mu\rho}g^{\nu\sigma} + g^{\mu\sigma} g^{\nu\rho} )
\eeq
\beq\begin{split}
Tr[ \gamma^\mu \gamma^\nu \gamma^\rho \gamma^\sigma \gamma^\alpha \gamma^\beta ] &= 4 g^{\mu\nu} ( g^{\rho\sigma} g^{\alpha\beta} - g^{\rho\alpha} g^{\sigma\beta} + g^{\rho\beta} g^{\sigma\alpha} ) - \\
&-4 g^{\mu\rho} ( g^{\nu\sigma} g^{\alpha\beta} - g^{\nu\alpha} g^{\sigma\beta} + g^{\nu\beta} g^{\sigma\alpha} ) + \\
&+4 g^{\mu\sigma} ( g^{\nu\rho} g^{\alpha\beta} - g^{\nu\alpha} g^{\rho\beta} + g^{\nu\beta} g^{\rho\alpha} ) - \\
&-4 g^{\mu\alpha} ( g^{\nu\rho} g^{\sigma\beta} - g^{\nu\sigma} g^{\rho\beta} + g^{\nu\beta} g^{\rho\sigma} ) + \\
&+4 g^{\mu\beta} ( g^{\nu\rho} g^{\sigma\alpha} - g^{\nu\sigma} g^{\rho\alpha} + g^{\nu\alpha} g^{\rho\sigma} )
\end{split}\eeq
\beq
Tr[\gamma^\mu \gamma^\nu \gamma^5] = 0
\eeq
\textbf{Cyclic property of the trace}
\beq\label{cyclic}
Tr[ABC] = Tr[CAB] = Tr[BCA]
\eeq
\newpage
\noindent
\textsc{Dirac theory}
\beq
\mathcal{L}_D = i \overline \psi \begin{cancel}\partial\end{cancel} \psi - m \overline \psi \psi
\eeq
\beq
\begin{cases}
i \partial_\mu \overline \psi \gamma^\mu + m \overline \psi = 0 \\
-i \bcc \partial \ecc \psi + m \psi = 0
\end{cases}, \qquad \overline \psi = \psi^\dagger \gamma^0, \qquad \bcc \partial \ecc = \partial_\mu \gamma^\mu
\eeq
\textbf{Dirac spinors}
\beq
\begin{cases}
[i \gamma^\mu p_\mu - m] u(p) = 0 \implies u(p) = 
\begin{pmatrix}
\sqrt{p_\mu \sigma^\mu} \xi \\ \sqrt{p_\mu \overline \sigma^\mu} \xi
\end{pmatrix} \\
[i \gamma^\mu p_\mu + m] v(p) = 0 \implies v(p) = 
\begin{pmatrix}
\sqrt{p_\mu \sigma^\mu} \chi \\ -\sqrt{p_\mu \overline \sigma^\mu} \chi
\end{pmatrix}
\end{cases}, \quad \sigma^\mu = (\mathbb{1}, \textbf{$\sigma$} )^T, \quad \overline \sigma^\mu = (\mathbb{1}, -\textbf{$\sigma$} )^T
\eeq
\textbf{Dirac spinor identities}
\beq
u^{\tau \dagger}(p) u^\rho(p) = 2 E_p \delta^{\tau\rho}
\eeq
\beq
v^{\tau \dagger}(p) v^\rho(p) = 2 E_p \delta^{\tau\rho}
\eeq
\beq
\overline u^\tau (p) v^\rho(p) = \overline v^\tau (p) u^\rho(p) = 0
\eeq
\beq\label{id_spinori_zero}
u^{\tau \dagger}(\textbf{p} ) v^\rho(-\textbf{p} ) = v^{\tau \dagger}(- \textbf{p} ) u^\rho(\textbf{p} ) = 0
\eeq
\beq
\sum_{s=1}^2 { u^s(p) \overline u^s(p) } = \bcc p \ecc + m
\eeq
\beq
\sum_{s=1}^2 { v^s(p) \overline v^s(p) } = \bcc p \ecc - m
\eeq
\textbf{Fields expansions}
\beq
\psi(x) = \sum_{s=1}^2 \int \frac{d^3p}{ (2\pi)^3 } \frac{1}{\sqrt{2E_p}} \bigg[ a^s_p u^s(p) e^{-ipx} + b^{s \dagger}_p v^s(p) e^{ipx} \bigg]
\eeq
\beq
\psi^\dagger(x) = \sum_{s=1}^2 \int \frac{d^3p}{ (2\pi)^3 } \frac{1}{\sqrt{2E_p}} \bigg[ a^{s \dagger}_p u^{s \dagger}(p) e^{ipx} + b^s_p v^{s \dagger}(p) e^{-ipx} \bigg]
\eeq
\textbf{Ladder operators}
\beq\begin{split}
&\{ a^s_p, a^{r \dagger}_q \} = \{ b^s_p, b^{r \dagger}_q \} = (2\pi)^3 \delta^{rs} \delta^{(3)}( \textbf{p} - \textbf{q} ) \\
&\{ a^s_p, a^r_q \} = \{ b^s_p, b^r_q \} = 0 \\
&\{ a^{s \dagger}_p, a^{r \dagger}_q \} = \{ b^{s \dagger}_p, b^{r \dagger}_q \} = 0 \\
&\{ a^s_p, b^r_q \} = \{ a^{s \dagger}_p, b^{r \dagger}_q \} = 0 \\
&\{ a^s_p, b^{r \dagger}_q \} = \{ a^{s \dagger}_p, b^r_q \} = 0
\end{split}\eeq
\beq
\ket{ \textbf{p},s } = \sqrt{2E_p} a^{s^\dagger}_p \ket{0}
\eeq
\newpage
\noindent
\textsc{Electromagnetic theory}
\beq
\mathcal{L}_{EM} = -\frac{1}{4} F^{\mu\nu} F_{\mu\nu}
\eeq
\beq
\partial_\mu F^{\mu\nu} = 0, \qquad F^{\mu\nu} = \partial^\mu A^\nu - \partial^\nu A^\mu
\eeq
\textbf{Gauge transformations}
\beq
A_\mu(x) \to A_\mu(x) + \partial_\mu \lambda(x)
\eeq
\textbf{Lorentz gauge and gauge-fixing term}
\beq
\partial_\mu A^\mu = 0
\eeq
\beq
\mathcal{L} = -\frac{1}{4} F^{\mu\nu} F_{\mu\nu} - \frac{1}{2\xi} (\partial_\mu A^\mu)^2
\eeq
\textbf{Field expansion}
\beq
A_\mu(x) = \int \frac{d^3p}{ (2\pi)^3 } \frac{1}{\sqrt{2 |\textbf{p}| }} \sum_{\lambda=0}^3 \bigg[ \varepsilon_\mu^{(\lambda)}(p) a^{(\lambda)}_p e^{-ipx} + \varepsilon_\mu^{(\lambda)^*}(p) a^{(\lambda)^\dagger}_p e^{ipx} \bigg]
\eeq
\beq\label{orto_polarizzazioni}
p^\mu \varepsilon_\mu^{(1,2)} = 0
\eeq
\beq\label{id_polarizzazioni}
\sum_{\lambda=1}^2 \varepsilon_{(\lambda)}^\mu [\varepsilon_{(\lambda)}^\nu]^* = g^{\mu\nu}
\eeq
\textbf{Ladder operators}
\beq
[ a_p^{(\lambda)}, a_q^{(\lambda')^\dagger} ] = -g^{\lambda \lambda'} (2\pi)^3 \delta^{(3)}( \textbf{p} - \textbf{q} )
\eeq
\beq
\ket{ \textbf{p}, \lambda} = a_p^{(\lambda)^\dagger} \ket{0}
\eeq
\\ \textsc{Interacting fields}
\\ \textbf{Wick's theorem}: let $T$ be the time-ordering operator and $N$ the normal-ordering operator, then:
\beq\begin{split}
\phi(x) &= \int \frac{d^3p}{ (2\pi)^3 } \frac{1}{ \sqrt{2E_p} } \bigg[ a_p e^{-ipx} + a_p^\dagger e^{ipx} \bigg] := \phi^+(x) + \phi^-(x) \\
\bra{0} T \big\{ \phi(x) \phi(y) \big\} \ket{0} &=
\begin{cases}
\bra{0} [ \phi^+(x), \phi^-(y) ] \ket{0}, \quad x^0 > y^0 \\
\bra{0} [ \phi^+(y), \phi^-(x) ] \ket{0}, \quad x^0 < y^0 \\
\end{cases} := \wick{\c \phi(x) \c \phi(y)}
\end{split}\eeq
\beq\begin{split}
\psi(x) &= \sum_{s=1}^2 \int \frac{d^3p}{ (2\pi)^3 } \frac{1}{\sqrt{2E_p}} \bigg[ a^s_p u^s(p) e^{-ipx} + b^{s \dagger}_p v^s(p) e^{ipx} \bigg] := \psi^+(x) + \psi^-(x) \\
\bra{0} T \big\{ \psi(x) \overline \psi(y) \big\} \ket{0} &=
\begin{cases}
\bra{0} \{ \psi^+(x), \psi^-(y) \} \ket{0}, \quad x^0 > y^0 \\
\bra{0} - \{ \overline \psi^+(y), \psi^-(x) \} \ket{0}, \quad x^0 < y^0 \\
\end{cases} := \wick{\c \psi(x) \overline {\c \psi}(y) }
\end{split}\eeq
\beq
T\bigg\{ A B C \dots Z \bigg\} = N \bigg\{ A B C \dots Z + \text{all possible contractions of $A B C \dots Z$ } \bigg\}
\eeq
\textbf{S-matrix and scattering amplitude}
\beq
S = T \bigg\{ exp \bigg[ -i \int d^4x \mathcal{H}_{INT} \bigg] \bigg\}, \qquad A = \bra{in} S \ket{out}
\eeq
\textbf{Differential cross section} $\ket{ \textbf{p}_1, \textbf{p}_2 } \to \ket{ \textbf{q}_1, \textbf{q}_2}$
\beq
d \sigma = \mathcal{J} |M|^2 d \Phi_2, \qquad
\begin{cases}
\mathcal{J} = \frac{1}{ 4 \sqrt{ (p_1p_2)^2 - m_1^2 m_2^2 } } \\
|M|^2 = \sum_{s_\mathcal{j}, \lambda_\mathcal{j}, \dots}{ |A|^2 } \\
d \Phi_2 = \frac{ |\textbf{q}_1| }{E_1+E_2} \frac{d \cos{\theta} d\varphi}{16 \pi^2} \bigg|_{ \textbf{q}_2 = -\textbf{q}_1 }
\end{cases}
\eeq
where $m_1$, $m_2$ are the initial masses, $E_1$, $E_2$ are the final energies and $|\textbf{p}_2| = - |\textbf{p}_1|$ \\ \textbf{Feynman rules for $\phi^4$-theory}
\beq
\begin{tikzpicture}
\begin{feynman}
\node at (3,0) {$= \frac{i}{ p^2 - m^2 + i\varepsilon }$};

	\vertex (a);
	\vertex[right=of a] (b);

	\diagram {
	(a) -- [dashed, momentum=\({p}\)] (b)
	};
\end{feynman}
\end{tikzpicture}
\eeq
\beq
\begin{tikzpicture}
\begin{feynman}
\node at (2,0) {$= (-i \lambda)$};

	\vertex (a);
	\vertex[above left=of a] (b);
	\vertex[below left=of a] (c);
	\vertex[above right=of a] (d);
	\vertex[below right=of a] (e);

	\diagram {
	(b) -- [dashed] (a),
	(d) -- [dashed] (a),
	(c) -- [dashed] (a),
	(e) -- [dashed] (a),
	};
\end{feynman}
\end{tikzpicture}
\eeq
\beq
\begin{tikzpicture}
\begin{feynman}
\node at (2.5,0) {$=1$};

	\vertex (a) {\(x\)};
	\vertex[right=of a] (b);

	\diagram {
	(b) -- [dashed, momentum=\({p}\)] (a),
	};
\end{feynman}
\end{tikzpicture}
\eeq
\\ \textsc{Quantum Electrodynamics}
\beq
\mathcal{L}_{QED} = i \overline \psi \bcc D \ecc \psi - m \overline \psi \psi - \frac{1}{4} F^{\mu\nu} F_{\mu\nu}, \qquad D_\mu = \partial_\mu + i e A_\mu
\eeq
\textbf{Feynman rules for QED}
\beq
\begin{tikzpicture}
\begin{feynman}
\node at (3,0) {$= u^s(p)$};

	\vertex (a);
	\vertex[right=of a] (b);
	\vertex[above right=of b] (c);
	\vertex[below right=of b] (d);

	\diagram {
	(a) -- [fermion, momentum=\({p}\)] (b),
	(b) -- [fermion] (c),
	(d) -- [dashed] (b)
	};
\end{feynman}
\end{tikzpicture} \qquad \qquad
\begin{tikzpicture}
\begin{feynman}
\node at (1,0) {$= \overline u^s(p)$};

	\vertex (a);
	\vertex[left=of a] (b);
	\vertex[above left=of b] (c);
	\vertex[below left=of b] (d);

	\diagram {
	(b) -- [fermion, momentum=\({p}\)] (a),
	(c) -- [fermion] (b),
	(d) -- [dashed] (b)
	};
\end{feynman}
\end{tikzpicture}
\eeq
\beq
\begin{tikzpicture}
\begin{feynman}
\node at (3,0) {$= \overline v^s(p)$};

	\vertex (a);
	\vertex[right=of a] (b);
	\vertex[above right=of b] (c);
	\vertex[below right=of b] (d);

	\diagram {
	(b) -- [fermion] (a),
	(a) -- [momentum=\({p}\)] (b),
	(c) -- [fermion] (b),
	(d) -- [dashed] (b)
	};
\end{feynman}
\end{tikzpicture} \qquad \qquad
\begin{tikzpicture}
\begin{feynman}
\node at (1,0) {$= v^s(p)$};

	\vertex (a);
	\vertex[left=of a] (b);
	\vertex[above left=of b] (c);
	\vertex[below left=of b] (d);

	\diagram {
	(a) -- [fermion] (b),
	(b) -- [momentum=\({p}\)] (a),
	(b) -- [fermion] (c),
	(d) -- [dashed] (b)
	};
\end{feynman}
\end{tikzpicture}
\eeq
\beq
\begin{tikzpicture}
\begin{feynman}
\node at (3,0) {$= \varepsilon_\mu(p)$};

	\vertex (a) {\(\mu\)};
	\vertex[right=of a] (b);

	\diagram {
	(a) -- [boson, momentum=\({p}\)] (b)
	};
\end{feynman}
\end{tikzpicture} \qquad \qquad
\begin{tikzpicture}
\begin{feynman}
\node at (3,0) {$= \varepsilon_\mu^*(p)$};

	\vertex (a);
	\vertex[right=of a] (b) {\(\mu\)};

	\diagram {
	(a) -- [boson, momentum=\({p}\)] (b)
	};
\end{feynman}
\end{tikzpicture}
\eeq
\beq
\begin{tikzpicture}
\begin{feynman}
\node at (1,0) {$= -i e \gamma^\mu$};

	\vertex (a) {\(\mu\)};
	\vertex[left=of a] (b);
	\vertex[above left=of b] (c);
	\vertex[below left=of b] (d);

	\diagram {
	(a) -- [boson] (b),
	(b) -- [fermion] (c),
	(d) -- [fermion] (b)
	};
\end{feynman}
\end{tikzpicture}
\eeq
\beq
\begin{tikzpicture}
\begin{feynman}
\node at (3,0) {$= \frac{i (\bcc p \ecc + m) }{p^2 - m^2 + i\varepsilon}$};

	\vertex (a);
	\vertex[right=of a] (b);

	\diagram {
	(a) -- [fermion, momentum=\({p}\)] (b)
	};
\end{feynman}
\end{tikzpicture}
\eeq
\beq
\begin{tikzpicture}
\begin{feynman}
\node at (3,0) {$= \frac{-i g_{\mu\nu} }{p^2 + i\varepsilon}$};

	\vertex (a) {\(\mu\) };
	\vertex[right=of a] (b) {\(\nu\)};

	\diagram {
	(b) -- [boson, momentum=\({p}\)] (a),
	};
\end{feynman}
\end{tikzpicture}
\eeq
\\ \textbf{Fundamental processes in QED} ($m_e \sim 0$)
\begin{itemize}
\item $e^+ e^- \to \mu^+ \mu^-$
\beq
\ket{p_1, s_1; p_2, s_2} \to \ket{k_1, s_3; k_2, s_4}
\eeq
\beq
|M|^2 = \frac{8 e^4}{ (p_1+p_2)^4 } \bigg[ (p_1k_1) (p_2k_2) + (p_1k_2) (p_2k_1) + m_\mu^2 (p_1p_2) \bigg]
\eeq
\beq
\frac{d \sigma}{ d \cos{\theta} } = \frac{e^4}{128\pi} \frac{1}{E^2} \sqrt{ 1 - \frac{m_\mu^2}{E^2} } \bigg[ \bigg( 1 + \frac{m_\mu^2}{E^2} \bigg) + \bigg( 1 - \frac{m_\mu^2}{E^2} \bigg) \cos{\theta} \bigg]
\eeq
\item $e^+ e^- \to e^+ e^-$
\beq
\ket{p_1, s_1; p_2, s_2} \to \ket{k_1, s_3; k_2, s_4}
\eeq
\beq\begin{split}
|M|^2 = 8 e^4 \bigg\{& \frac{1}{ (p_1+p_2)^4 } \bigg[ (p_1k_1) (p_2k_2) + (p_1k_2) (p_2k_1) \bigg] + \\
&+\frac{1}{ (p_1-k_1)^4 } \bigg[ (p_1p_2) (k_1k_2) + (p_1k_2) (p_2k_1) \bigg] + \frac{ 2 (p_1k_2) (p_2k_1) }{ (p_1+p_2)^2 (p_1-k_2)^2 } \bigg\}
\end{split}\eeq
\beq
\frac{d \sigma}{d \cos{\theta} } = \frac{e^4}{64\pi} \frac{1}{E^2} \bigg[ \frac{ (1 - \cos^2{\theta} )^2 + (1 + \cos^2{\theta} )^2 }{4} + \frac{ 4 + (1 + \cos^2{\theta} )^2 }{ (1 - \cos^2{\theta} )^2 } - (1 + \cos^2{\theta} ) \bigg]
\eeq
\item $e^+ e^- \to \gamma\gamma$
\beq
\ket{p_1, s_1; p_2, s_2} \to \ket{k_1, \lambda_1; k_2, \lambda_2}
\eeq
\beq
|M|^2 = e^4 \bigg[ \frac{1}{ (p_2k_2)^2 } + \frac{1}{ (p_2k_1)^2 } \bigg] \bigg[ (p_1p_2) (k_1k_2) - (p_1k_1) (p_2k_2) - (p_1k_2) (p_2k_1) \bigg] 
\eeq
\beq
\frac{d \sigma}{d \cos{\theta} } = \frac{e^4}{16\pi} \frac{1}{E^2} \bigg( \frac{1 + \cos^2{\theta} }{ \sin^2{\theta} } \bigg)
\eeq
\end{itemize}
}

\addcontentsline{toc}{chapter}{Bibliography}


\begin{thebibliography}{9}
\bibitem[1]{Schwartz} M. D. Schwartz. \textit{Quantum Field Theory and the Standard Model}. Cambridge University Press (2018).
\bibitem[2]{Peskin} M. E. Peskin, D. V. Schroeder. \textit{An Introduction to Quantum Field Theory}. Westview Press (1995).
\bibitem[3]{Greco} A. Greco. \textit{Elements of Quantum Field Theory}. Lectures notes for the Master’s Degree in Physics, University of Ferrara (2019).
\end{thebibliography}
\end{document}